\documentclass[usenatbib]{mn2e}
\usepackage{apjfonts,amsfonts,amsmath,amssymb,bm,ctable,verbatim}
\usepackage[export]{adjustbox}

\newcommand{\paperone}{Paper {\small I}}

\newcommand{\gizmourl}{\href{http://www.tapir.caltech.edu/~phopkins/Site/GIZMO.html}{\url{http://www.tapir.caltech.edu/~phopkins/Site/GIZMO.html}}}
\newcommand{\acknowledgments}[1]{\begin{small}\section*{Acknowledgments}\end{small}{\noindent #1}\vspace{5pt}}
\newcommand{\datastatement}[1]{\begin{small}\section*{Data Availability Statement}\end{small}{\noindent #1}\vspace{5pt}}
\newcommand{\microGauss}{\mu{\rm G}}
\newcommand{\Bangle}{\theta_{B}}
\newcommand{\Alf}{{Alfv\'en}}

\newcommand{\fref}[1]{Fig.~\ref{#1}}

\newcommand{\tref}[1]{Table~\ref{#1}}

\newcommand{\Dt}[1]{\frac{\mathrm{d} #1}{\mathrm{dt}}}
\newcommand{\initvalupper}[1]{#1^{0}}
\newcommand{\initvallower}[1]{#1_{0}}
\newcommand{\driftvel}{{\bf w}_{s}}
\newcommand{\driftvelmag}{w_{s}}
\newcommand{\dustvel}{{\bf v}_{d}}
\newcommand{\gasvel}{{\bf u}_{g}}
\newcommand{\gasden}{\rho_{g}}
\newcommand{\rhobase}{\rho_{\rm base}}
\newcommand{\gaspressure}{P}
\newcommand{\dustden}{\rho_{d}}
\newcommand{\rhodust}{\dustden}
\newcommand{\rhogas}{\gasden}

\newcommand{\opticaldepth}{\tau_{\rm ext}}
\newcommand{\tauparam}{\tau_{\rm SL}}
\newcommand{\ts}{t_{s}}
\newcommand{\cs}{c_{s}}
\newcommand{\vA}{v_{A}}
\newcommand{\tL}{t_{L}}
\newcommand{\grainsuff}{_{\rm grain}}
\newcommand{\internaldensity}{\bar{\rho}\grainsuff^{\,i}}
\newcommand{\grainsize}{\epsilon\grainsuff}
\newcommand{\grainsizebar}{\bar{\epsilon}\grainsuff}
\newcommand{\grainmass}{m\grainsuff}
\newcommand{\graincharge}{q\grainsuff}
\newcommand{\grainchargeZ}{Z\grainsuff}
\newcommand{\grainsizemax}{\grainsize^{\rm max}}
\newcommand{\grainsizemin}{\grainsize^{\rm min}}
\newcommand{\B}{{\bf B}}
\newcommand{\Bmag}{|\B|}
\newcommand{\Bhat}{\hat\B}
\newcommand{\bhat}{\Bhat}
\newcommand{\acc}{{\bf a}}
\newcommand{\Lbox}{L_{\rm box}}
\newcommand{\Lscale}{H_{\rm gas}}
\newcommand{\sizeparam}{\tilde{\alpha}}
\newcommand{\sizeparammax}{\sizeparam_{\rm m}}
\newcommand{\chargeparam}{\tilde{\phi}}
\newcommand{\chargeparammax}{\chargeparam_{\rm m}}
\newcommand{\accparam}{\tilde{a}_{d}}
\newcommand{\accparammax}{\tilde{a}_{\rm d,m}}
\newcommand{\accabsmax}{{a}_{\rm d,m}}
\newcommand{\accsizedep}{\psi_{a}}
\newcommand{\gravparam}{\tilde{g}}
\newcommand{\dustgas}{\mu^{\rm dg}}
\newcommand{\dustgashat}{\hat{\mu}^{\rm dg}}

\title[Dust-Driven Winds: GMCs \&\ HII Regions]{Dust in the Wind with Resonant Drag Instabilities: I. The Dynamics of Dust-Driven Outflows in GMCs and HII Regions}

\author[Hopkins et al.]{
\parbox[t]{\textwidth}{Philip F.~Hopkins$^1$, 
Anna L.~Rosen$^2$, 
Jonathan Squire$^3$,
Georgia V.~Panopoulou$^1$, \\
Nadine H.~Soliman$^1$, 
Darryl Seligman$^4$, 
Ulrich P.~Steinwandel$^5$
} \vspace*{4pt} \\
$^1$ TAPIR, Mailcode 350-17, California Institute of Technology, Pasadena, CA 91125, USA \\
$^2$ Center for Astrophysics, Harvard \&\ Smithsonian, 60 Garden St, Cambridge, MA 02138, USA \\
$^3$ Physics Department, University of Otago, 730 Cumberland St., Dunedin 9016, New Zealand \\
$^4$ Dept. of the Geophysical Sciences, University of Chicago, Chicago, IL 60637 \\
$^5$ Center for Computational Astrophysics, Flatiron Institute, 162 5th Ave., New York, NY 10010 USA
}

\date{}
\begin{document}
\maketitle
\mathchardef\mhyphen="2D 

\begin{abstract}
Radiation-dust driven outflows, where radiation pressure on dust grains accelerates gas, occur in many astrophysical environments. Almost all previous numerical studies of these systems have assumed that the dust was perfectly-coupled to the gas. However, it has recently been shown that the dust in these systems is unstable to a large class of ``resonant drag instabilities'' (RDIs) which de-couple the dust and gas dynamics and could qualitatively change the nonlinear outcome of these outflows. We present the first simulations of radiation-dust driven outflows in stratified, inhomogeneous media, including explicit grain dynamics and a realistic spectrum of grain sizes and charge, magnetic fields and Lorentz forces on grains (which dramatically enhance the RDIs), Coulomb and Epstein drag forces, and explicit radiation transport allowing for different grain absorption and scattering properties. In this paper we consider conditions resembling giant molecular clouds (GMCs), HII regions, and distributed starbursts, where optical depths are modest ($\lesssim 1$), single-scattering effects dominate radiation-dust coupling, Lorentz forces dominate over drag on grains, and the fastest-growing RDIs are similar, such as magnetosonic and fast-gyro RDIs. These RDIs generically produce strong size-dependent dust clustering, growing nonlinear on timescales that are much shorter than the characteristic  times of the outflow. The instabilities produce filamentary and plume-like or ``horsehead'' nebular morphologies that are remarkably similar to observed dust structures in GMCs and HII regions. Additionally, in some cases they strongly alter the magnetic field structure and topology relative to filaments. Despite driving strong micro-scale dust clumping which leaves some gas ``behind,'' an order-unity fraction of the gas is always efficiently entrained by dust. 
\end{abstract}


\section{Introduction}

Almost all astrophysical fluids are laden with grains of dust, which play a central role in planet and star formation; attenuation and extinction; cool-star, brown-dwarf, and planetary evolution; astro-chemistry and heating/cooling of the interstellar medium (ISM); and feedback and outflow-launching from star-forming regions, cool stars, and active galactic nuclei (AGN) \citep[see][for reviews]{draine:2003.dust.review,dorschner:dust.mineralogy.review,apai:dust.review,2018A&ARv..26....1H}. Therefore, the dynamical interactions between dust and gas are of fundamental importance in a broad range of astrophysical environments. Of particular interest are radiation ``dust-driven'' outflows, in systems such as cool stellar atmospheres of giant stars, AGN, and starburst/GMC environments. In these systems, radiation pressure from photons absorbed or scattered by dust grains (which dominate the opacity) may launch outflows, in which gas is entrained along with the dust via collisional, electrostatic, and magnetic interactions \citep{1999isw..book.....L}. These outflows can have dramatic impacts on processes ranging from stellar evolution through star and galaxy formation. 

There has been considerable theoretical work to understand these outflows over the last several decades \citep{sandford:radiatively.driven.dust.bounded.gmc.globules,chang:qso.rad.feedback,franco:dust.rad.pressure.galactic.fountain,berruyer:dust.wind.unstable.pressure.gradient,1999isw..book.....L}. In the ISM (in particular in GMCs, starburst galaxies, and HII regions),  radiation pressure on grains provides a potential acceleration mechanism for outflows and driver of turbulence \citep{heckman:1990.sb.superwinds,scoville:2001.dust.pressure.in.sb.regions,thompson:rad.pressure}. There has been significant controversy about the relative importance of radiation pressure as compared to other feedback mechanisms (e.g.\ stellar outflows, photoionization, and supernovae [SNe]; see \citealt{raskutti:2016.m1.cloud.sims}). Many recent studies  have focused on how differences in radiation-hydrodynamic methods can alter predictions for these dust-driven outflows \citep[see e.g.][]{krumholz:2012.rad.pressure.rt.instab,kuiper:2012.rad.pressure.outflow.vs.rt.method,wise:2012.rad.pressure.effects,tsang:monte.carlo.rhd.dusty.wind,rosen:massive.sf.rhd}. However, these studies have generally treated the dust in a highly simplified manner, assuming that it is perfectly-coupled to the gas, or moves smoothly as a ``fluid'' (as compared to e.g.\ following individual gyro-orbits of grains), or that it always moves at the local ``terminal'' (homogeneous equilibrium) velocity. 

Since  dust acts as a primary driving mechanism in such outflows, explicitly modeling the dust dynamics is central to understanding whether or not they can  occur. In an extreme limiting case, if the dust-gas coupling were sufficiently weak, radiation pressure would simply expel the grains, entraining little or no gas and resulting in a ``failed'' wind. But even in the opposite limiting case (the ``tight coupling'' regime) where dust grains have small mean-free paths, the recent discovery of dynamical instabilities generic to coupled dust-gas systems (even on scales arbitrarily large compared to the dust mean-free path) provides a motivation to re-visit these dust-driven outflows.

\citet{squire.hopkins:RDI} showed that dust-gas mixtures are unstable to a broad class of instabilities, which they referred to as  ``Resonant Drag Instabilities'' (RDIs). These instabilities manifest whenever dust streams through fluid, gas, or plasma, where there is a difference between the forces acting on the dust or gas. We note that this is always true in ``dust-driven outflows''. In the RDIs, each pair of dust and gas modes (representing modes of the equations for ``dust alone,'' such as drift or gyro motion, and for ``gas alone,'' such as \Alf\ or magnetosonic waves) interact to produce an distinct RDI sub-family, which grows unstably with growth rates maximized around the ``resonance'' where the two modes ``in isolation'' would have similar natural frequencies. These instabilities could {\em qualitatively} change the outcome of dust-driven outflows. For example, it is generally assumed that magnetic fields  ``anchor'' charged dust grains to gas in winds  \citep{hartquist:bfield.dust.coupling.cool.star.winds,yan.2004:lorentz.forces.drag.dust.ism.analytic}. Since  the gyro radii can be smaller than the dust collisional mean-free-path,  a ``tight coupling'' or single-fluid ``dust-plus-gas'' approximation is often invoked, akin to ions in ideal magnetohydrodynamics. However \citet{hopkins:2018.mhd.rdi} and \citet{seligman:2018.mhd.rdi.sims} showed that magnetic forces on dust are in fact violently {\em de-stabilizing} on small-scales, introducing RDIs which can actually act to separate or {\em de-couple} the dust and gas.

\begin{figure}
    \centering
    \includegraphics[width=0.8\columnwidth]{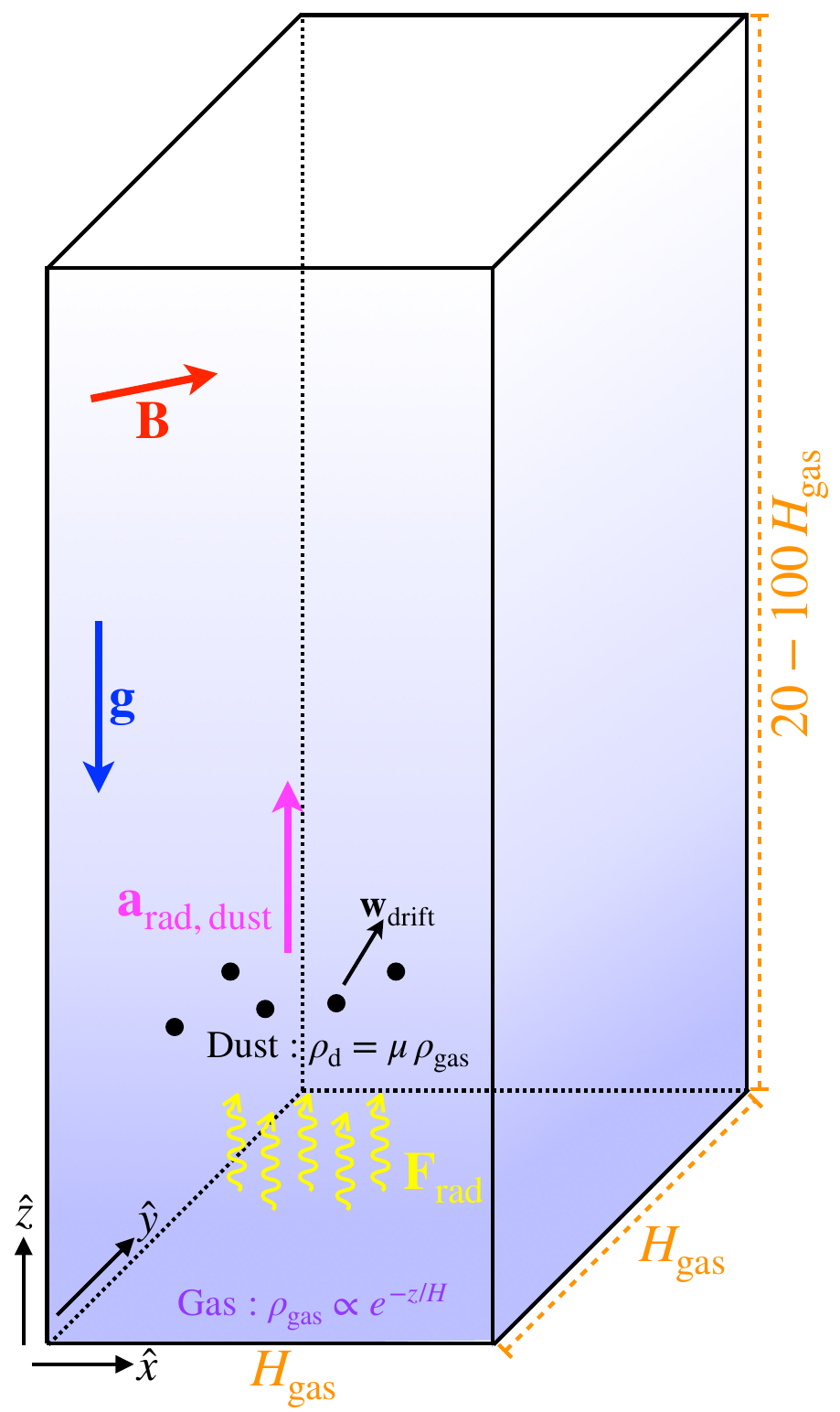}
    \caption{Cartoon illustrating our simulation setup. We simulate 3D boxes with an outflow upper \&\ reflecting lower boundary, and periodic sides. Gas and dust are initially stratified in $\hat{z}$, with $\gasden \propto \exp{(-z/\Lscale)}$, an isothermal ($\gamma=1$) gas EOS with sound speed $\cs$, and magnetic field ${\bf B}_{0} = |{\bf B}|\,(\sin{\Bangle^{0}}\,\hat{x} + \cos{\Bangle^{0}}\,\hat{z})$ in the $\hat{x}-\hat{z}$ plane, uniform dust-to-gas ratio $\dustgas \equiv \dustden/\gasden$, and gravitational acceleration ${\bf g} = -g\,\hat{z}$. The dust grains are modeled with super-particles each representing a given size of grains drawn from a standard MRN spectrum with factor $=100$ range of sizes (with grain charge depending appropriately on grain size). An initial upward radiation flux ${\bf F}_{0} = +F_{0}\,\hat{z}$ is absorbed and scattered by grains giving rise to a (size-dependent) grain acceleration ${\bf a}_{\rm rad,\,dust}  \propto {\bf F}$, which produces a dust drift velocity $\driftvel$. The dust interacts with the gas via collisional+Coulomb drag ($\propto (\dustvel-\gasvel)/\ts$) and Lorentz+electrodynamic forces ($\propto (\dustvel-\gasvel)\times \hat{\bf B} / t_{L})$. Our fidicial boxes have size $\Lscale \times \Lscale \times (20-100)\,\Lscale$ with $\sim 10^{8}$ resolution elements (see \S~\ref{sec:scale} for a description of how $\Lscale$ relates to physical sizes).
    \label{fig:cartoon.ics}}
\end{figure}

Moreover, the RDIs could be important for a wide range of other phenomenology. In HII regions, dust is critical for chemistry and cooling physics as well as depletion of metals, {and} therefore {also affects} observational abundance estimators from emission lines  \citep{shields:1995.dust.HII.region.chem.rad.fx}. Accounting for dust ``drift'' in the fluid limit, or ignoring it, substantially alters HII region expansion rates, densities, and shell structure \citep{akimkin:2017.dustgasHII.expansion.cavity.density.metallicity}.  Dust clumping, which may be induced by RDIs or other dynamics, could enhance leakage of ionizing photons (by creating channels with relatively low opacity) by orders of magnitude \citep{anderson:2010.dust.HII.leakage.clumps.induced.sf.filaments,ma:2015.fire.escape.fractions,ma.2016:binary.star.escape.fraction.effects,ma:2020.no.missing.photons.for.reion.supershells}. Long-wavelength RDIs might drive dust into  fine structures such as filaments, lanes, or ``whiskers,''  which are seen ubiquitously in spatially-resolved HII regions and planetary nebulae \citep{odell:2002.pne.knots.review,apai:2005.HII.region.filamentary.dust.structures}. \citet{padoan:dust.fluct.taurus.vs.sims} argued that dust and gas must become dynamically de-coupled on sufficiently small scales and argued  that this has already been seen in many nearby molecular clouds using a  cross-correlation analysis \citep{thoraval:1997.sub.0pt04pc.no.cloud.extinction.fluct.but.are.on.larger.scales,thoraval:1999.small.scale.dust.to.gas.density.fluctuations,abergel:2002.size.segregation.effects.seen.in.orion.small.dust.abundances,miville-deschenes:2002.large.fluct.in.small.grain.abundances,pineda:2010.taurus.large.extinction.variations,pellegrini:2013.ngc.1266.shocked.molecules,nyland:2013.radio.core.ngc1266}. Fluctuations in the local dust-to-gas ratio on small scales could even play a role in seeding star formation  at high redshifts, where dust grains are rare but play a crucial role in cooling \citep{hopkins.conroy.2015:metal.poor.star.abundances.dust}. This would result in  unique dust abundance signatures.

In this paper, we  explore the role of magnetized dust dynamics in such outflows. We focus  on  outflows in  HII regions and GMCs where the dominant RDIs, dust properties, and relevant radiation-hydrodynamics limits are broadly similar. In a series of previous papers, we analytically identified and studied the properties of the linearized RDIs \citep{hopkins:2017.acoustic.RDI,squire:rdi.ppd,hopkins:2018.mhd.rdi}. In follow-up work, we performed  highly idealized simulations of periodic homogeneous free-falling gas exposed to a uniform radiation field with dust grains with uniform size and charge, obeying an ideal gas law with MHD \citep{moseley:2018.acoustic.rdi.sims,seligman:2018.mhd.rdi.sims,hopkins:2018.mhd.rdi}. These calculations demonstrated that the RDIs in conditions broadly similar to those studied here could (i) have rapid growth rates, (ii) reach large non-linear amplitudes, and (iii) produce potentially ``interesting'' macroscopic effects such as driving strong clumping of dust. However, the  idealized nature of the previous studies means that we could not make meaningful predictions for the questions considered in this paper. Here we perform global, stratified simulations, with a realistic spectrum of grain size and charge that allow for more realistic gas physics, variations in the optical properties of grains, and explicit radiation dynamics. 

{This paper is organized as follows. In} \S~\ref{sec:methods} {we} describe the numerical methods and initial conditions of the simulations presented in this paper. In \S~\ref{sec:results}, we describe the non-linear evolution of the simulations. We explore  the dust morphologies and dynamics,  present observables such as extinction and reddening curves, and investigate the overall evolution of the outflows. {Finally, in} \S~\ref{sec:discussion} {we} discuss our results and summarize our conclusions.


%
%
\begin{figure}
\begin{center}
\includegraphics[width=1.01\columnwidth]{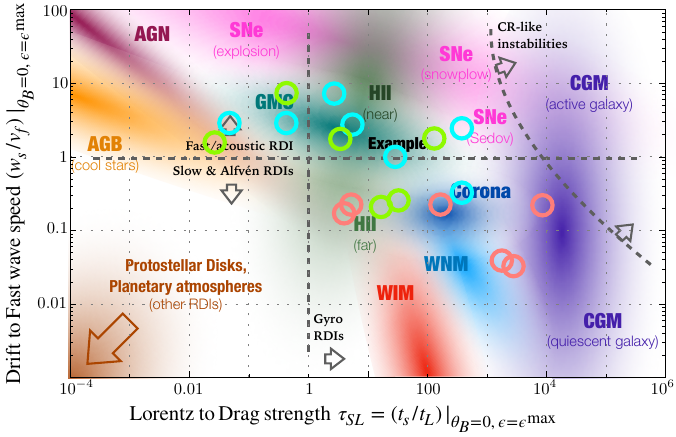}\vspace{-0.35cm}
\caption{Simulations studied in this work are shown with circles (see \tref{table:sims} \& \ref{table:sims.all}), in an illustration of two important parameters of the MHD RDIs (adapted from \citealt{hopkins:2018.mhd.rdi}). Axes show the approximate maximum ``parallel'' dust drift speed $\driftvelmag$ (drift speed of the largest grains, $\grainsize \sim 0.1\,\micron$, assuming magnetic fields are exactly drift-aligned) normalized by the fastest MHD wavespeed $v_{f}^{2} \equiv \cs^{2} + \vA^{2}$, and the maximum parallel ratio of Lorentz force to drag force $\tauparam\equiv \langle t_s \rangle/\langle t_L \rangle$ (again for the largest grains, assuming drift at the maximum parallel speed). Note this is simply a convenient parameterization: the actual drift speeds and $\tauparam$ span a wide range with grain size and are not necessarily parallel. Shaded regions crudely represent typical parameters of different astrophysical environments, including the warm ionized and warm neutral medium (WIM/WNM), giant molecular clouds (GMCs) and near/far vicinity of O-stars in HII regions (HII), supernovae in various phases of evolution (SNe), stellar coronal dust (Corona), cool/giant/AGB star photospheres and outflows (AGB), dusty ``torii'' around active galactic nuclei (AGN), the circum and/or inter-galactic medium around AGN/starburst systems or quiescent galaxies (CGM), and proto-stellar/planetary disks and planetary atmospheres (which extend off the plotted range). Lines/arrows illustrate where different forms of the RDIs should appear: the fast (acoustic) RDI is unstable for $\driftvelmag/v_{f,0}\gtrsim 1$, gyro-resonant RDIs can be dominant at $\tauparam \gtrsim 1$, and cosmic ray-like RDIs can dominate at very large $\tauparam$.
\label{fig:parameter.space}}
\end{center}
\end{figure}

\begin{table*}
\begin{center}
 \begin{tabular}{| l c c c c c c c r |}
 \hline
Name & $\accparammax$ $\left( \frac{|\initvalupper{ \driftvel }|}{\initvalupper{ \cs }} \right)$ & $\sizeparammax$ $\left( \frac{\initvalupper{ \cs}\, \initvalupper{\ts }}{\Lscale} \right)$ & $\chargeparammax$ ($\tauparam$) & $\beta_{0}$ & $\cos{\Bangle^{0}}$ & $\gravparam$ ($\lambda_{\rm Edd}$)  & $\accsizedep$ & Notes \\
 \hline\hline
{\bf GMC} & 65 (3) & 1e-3 (4e-6 - 4e-4) & 400 (1e3 - 1e5) & 0.01 & 0.1 & 110 (6) &  1 & GMC-like region \\ 
{\bf GMC-Q} & 65 (4e-2 - 3) & 1e-3 (6e-6 - 4e-4) & 400 (1e3 - 2e5) & 0.01 & 0.1 & 110 (6) &  0 & -- \\ 
\hline
{\bf HII-N} & 4.5 (0.3) & 3e-3 (2e-5 - 2e-3) & 44 (14 - 1e3) & 4 & 0.1 & 0.01 (1500) &  1 & ``near'' HII-region  \\ 
{\bf HII-N-Q} & 4.5 (3e-3 - 0.3) & 3e-3 (2e-5 - 2e-3) & 44 (14 - 1e3) & 4 & 0.1 & 0.01 (1500) &  0 & -- \\ 
\hline
{\bf HII-F} & 4.8 (0.3) & 3e-2 (2e-4 - 2e-2) & 440 (140 - 1e4) & 4 & 0.1 & 0.001 (1600) & 1 &  ``far'' HII-region  \\ 
{\bf HII-F-Q} & 4.8 (3e-3 - 0.3) & 3e-2 (2e-4 - 2e-2) & 440 (140 - 1e4) & 4 & 0.1 & 0.001 (1600) &  0 & -- \\ 
\hline
\end{tabular}
\end{center}\vspace{-0.25cm}
\caption{Initial conditions for our ``fiducial'' high-resolution simulations (see Appendix~\ref{sec:appendix:simslist} for a full list). By default these adopt an isothermal gas equation-of-state, Epstein+Coulomb drag, Lorentz forces on grains, an MRN spectrum of grain sizes with $\grainsizemin=0.01\,\grainsizemax$, uniform initial dust-to-gas ratio $\dustgas \equiv \initvalupper{\dustden}/\initvalupper{\gasden}=0.01$, stratified $\initvalupper{\gasden}=\rhobase\,\exp{(-z/\Lscale)}$, and $5\times 256^{3} \sim 10^{8}$ resolution elements (4 times as many dust as gas).
Columns show: 
{\bf (1)} Simulation name. 
{\bf (2)} Radiative flux/dust acceleration parameter: $\accparammax \equiv (3/4)\,(F_{0}\,Q_{\rm ext,\,0}/c)/(\rhobase\,\cs^{2})$. In parentheses, we give the range (over all grain sizes) of ${|\initvalupper{ \driftvel }|}/{\initvalupper{ \cs }}$, the initial equilibrium drift velocity in units of the sound speed.
{\bf (3)} Grain size parameter $\sizeparammax \equiv (\internaldensity\,\grainsizemax) / (\rhobase\,\Lscale)$ of the largest grains. Parentheses give range (over grain sizes) of initial stopping time (at initial equilibrium drift) $\initvalupper{\ts}$ in code units ($\Lscale/\cs$). 
{\bf (4)} Grain charge parameter $\chargeparammax \equiv 3\,\initvalupper{ \grainchargeZ }[\grainsizemax] \,e / (4\pi\,c\,(\grainsizemax)^{2}\,\rhobase^{1/2})$ of the largest grains. Parentheses give range of $\tauparam \equiv \initvalupper{\ts}/\initvalupper{\tL}$ (ratio of Lorentz-to-drag force on grains).
{\bf (5)} Initial plasma $\beta_{0} \equiv (\cs/\vA^{0})^{2}$, approximate ratio of thermal-to-magnetic pressure.
{\bf (6)} Angle between initial magnetic field and radiation/gravity direction: $\cos{\Bangle^{0}} \equiv |\hat{\bf B}^{0} \cdot \hat{\bf g}|$.
{\bf (7)} Gravity parameter $\gravparam \equiv |{\bf g}|\,\Lscale/\cs^{2}$. Parentheses give the approximate single-scattering Eddington parameter $\lambda_{\rm Edd} \equiv \dustgas\,\accparammax/\sizeparammax\,\gravparam$.
{\bf (8)} Dust absorption/scattering efficiency scaling $\accsizedep$: $Q_{\rm ext}(\grainsize) \propto \grainsize^{1-\accsizedep}$ (so the optically-thin radiative acceleration of a grain scales $\sim {\bf F}_{\rm rad}\,(Q_{\rm ext}\,\pi\,\grainsize^{2})/(c\,\grainmass) \propto \grainsize^{-\accsizedep}$). Runs denoted ``{\bf -Q}'' have $Q\propto \grainsize$ ($\accsizedep=0$), those without have $Q \sim \,$constant ($\accsizedep=1$).
{\bf (9)} Notes and motivation for each (see \S~\ref{sec:params}).
\label{table:sims}\vspace{-0.25cm}}
\end{table*}


\section{Methods \&\ Parameters}
\label{sec:methods}

\subsection{Dust-Magnetohydrodynamics}
\label{sec:equations}

Most of the numerical methods adopted here have been described in detail in \citet{moseley:2018.acoustic.rdi.sims} and \citet{seligman:2018.mhd.rdi.sims}, so we briefly summarize them here. A cartoon illustrating our setup is shown in Fig.~\ref{fig:cartoon.ics}. Our simulations were run with the code {\small GIZMO} \citep{hopkins:gizmo},\footnote{A public version of the code, including all methods used in this paper, is available at \href{http://www.tapir.caltech.edu/~phopkins/Site/GIZMO.html}{\url{http://www.tapir.caltech.edu/~phopkins/Site/GIZMO.html}}} using the Lagrangian ``meshless finite mass'' (MFM) method for {solving the equations of magneto-hydrodynamics (}MHD{)}, which has been extensively tested on problems involving multi-fluid MHD instabilities, the magneto-rotational instability (MRI), shock-capturing, and more \citep{hopkins:mhd.gizmo,hopkins:cg.mhd.gizmo,hopkins:gizmo.diffusion,su:2016.weak.mhd.cond.visc.turbdiff.fx}. Grains are integrated using the ``super-particle'' method \citep[see, e.g.][]{carballido:2008.grain.streaming.instab.sims,johansen:2009.particle.clumping.metallicity.dependence,bai:2010.grain.streaming.vs.diskparams,pan:2011.grain.clustering.midstokes.sims,2018MNRAS.478.2851M}, whereby the motion of each dust ``super-particle'' in the simulation follows Eq.~\eqref{eq:eom} below, but each represents an ensemble of dust grains with similar size ($\grainsize$), mass ($\grainmass$), and charge ($\graincharge$). The numerical methods for grain integration are tested in \citet{hopkins.2016:dust.gas.molecular.cloud.dynamics.sims,lee:dynamics.charged.dust.gmcs}, with back-reaction accounted for as in \citet{moseley:2018.acoustic.rdi.sims} (see App.~B therein) and the Lorentz force is evolved using a Boris integrator.

Each individual grain (or dust super-particle) in the code obeys:
\begin{align}\label{eq:eom}
\Dt{\dustvel} &=  -\frac{\driftvel}{\ts} - \frac{\driftvel\times\Bhat}{\tL} + \acc_{\rm ext,\, dust}\, ,
\end{align}
where $\acc_{\rm ext,\, dust}$ is an external acceleration, $\ts$ is the drag coefficient or ``stopping time,'' $\tL$ the gyro or Larmor time,\footnote{For convenience in Eq.~\ref{eq:eom} we define $t_{\rm L}$ to be positive definite and assume the grain charge is negative, but our simulations are manifestly invariant to swapping the sign of the grain charge.} ${\bf B}$ is the magnetic field vector ($\hat{\bf B} \equiv {\bf B}/|{\bf B}|$ its direction), and $\driftvel \equiv \dustvel - \gasvel$ is the drift velocity defined as the difference between the grain velocity $\dustvel$ and gas velocity $\gasvel$ at the same position ${\bf x}$. The gas obeys the ideal MHD equations in an external gravitational field ${\bf g}$, with the addition of a back-reaction force from the grains in the momentum equation. In particular, whenever drag or Lorentz forces exert a force $\grainmass\,\mathrm{d}\dustvel/\mathrm{dt}$ on a grain within a given gas cell, an equal-but-opposite force is applied to the gas (guaranteeing exact force balance and momentum conservation). In our default simulations, gas obeys an exactly polytropic equation of state with thermal pressure $P=\initvallower{\gaspressure}\,( {\gasden}/{\initvalupper{\gasden}} )^{\gamma}$ and sound speed $\cs^{2} \equiv \partial P / \partial \gasden$ (with $\gasden$ the gas density), though we have tested a model with a simply dynamical cooling/heating prescription instead and find this makes little difference to our results.

In our default simulations we assume Epstein drag (with Stokes and/or Coulomb drag contributing negligible corrections for our purposes here; see Appendix~\ref{sec:appendix:drag}),  which can be approximated to very high accuracy with the expression (valid for both sub and super-sonic drift)
\begin{align} \label{eq:ts}
\ts &\equiv \sqrt[]{\frac{\pi \gamma}{8}}\frac{\internaldensity \,\grainsize}{\gasden\,\cs}\, \bigg( 1+\frac{9\pi\gamma}{128} \frac{|\driftvel|^{2}}{\cs^{2}} \bigg)^{-1/2},
\end{align}
where $\internaldensity$ and $\grainsize$ are the \textit{internal} grain density and radius, respectively. The Larmor time is:
\begin{align} \label{eq:tL} 
\tL &\equiv \frac{\grainmass\,c}{|\graincharge\,\B |} = \frac{4\pi\,\internaldensity\,\grainsize^{3}\,c}{3\,e\,|\grainchargeZ\,\B |}
\end{align}
where $\grainmass$ and $\graincharge = \grainchargeZ\,e$ are the grain mass and charge. 
We adopt a standard empirical \citet{mathis:1977.grain.sizes}-like grain size spectrum with differential number $d N_{\rm d} / d \grainsize \propto \grainsize^{-3.5}$, from a maximum grain size $\grainsizemax$ to minimum $\grainsizemin \approx \grainsizemax/100$ (representative of the range of grain sizes excluding the smallest PAHs, where an aerodynamic description is not appropriate),\footnote{Each dust super-particle $i$ represents an ensemble of $\Delta N_{i}$ grains of (identical within the super-particle) size $\grainsize = \grainsize^{i}$, with ensemble mass $\Delta m_{d} = \grainmass^{i}\,\Delta N_{i}$ (where $\grainmass^{i}=(4\pi/3)\,\internaldensity\,(\grainsize^{i})^{3}$. We choose $\Delta N_{i} \propto (\grainsize^{i})^{-2.5}$, so that the number of discrete super-particles sampling each logarithmic interval in grain size is uniform, i.e.\ $d N_{\rm superparticles}/ d \ln{\grainsize} \sim (\grainsize^{i}/\Delta N_{i})\,d N_{\rm d} / d \grainsize \sim $\,constant (given our assumption that the box-averaged grain size distribution follow the MRN scaling $dN_{\rm d}  / d \grainsize \propto \grainsize^{-3.5}$). This ensures that we do not under or over-sample the dynamics or interactions of large or small grains.} with $\grainsize$-independent $\internaldensity$, and assume the grain charge-to-mass ratio $|\graincharge|/\grainmass \propto \grainsize^{-2}$ (e.g.\ $\graincharge \propto \grainsize$, appropriate for grains primarily charged by collisional, Coulomb, photo-electric, or electrostatically-limited processes; \citealt{draine:1987.grain.charging,tielens:2005.book}).\footnote{We ignore charge quantization effects, but this is a good approximation for the grains of interest here (large enough for aerodynamic behavior to be valid). The normalization of the grain charge is given by the $\chargeparammax$ in Table~\ref{table:sims}, motivated by the scalings in \S~\ref{sec:params}, for which we adopt the larger of the collisional charge from \citet{draine:1987.grain.charging} or photo-electric charging from \citet{tielens:2005.book}, both with the appropriate maximum/minimum (e.g.\ electrostatically limited) charges defined therein (see \citealt{hopkins:2018.mhd.rdi} for a summary).} Simulation parameters are given in \tref{table:sims} and \fref{fig:parameter.space}.

\subsection{Radiation-Dust-Magnetohydrodynamics}
\label{sec:rad}

We are generally interested in situations where radiation absorbed by dust or gas pushes against gravity. This means $\acc_{\rm ext,\,gas} = {\bf g} + \acc_{\rm rad,\,gas}({\bf x},\,\rho,\,...)$, and 
\begin{align}
\acc_{\rm ext,\,dust} = {\bf g} + \acc_{\rm rad,\,dust}({\bf x},\,\grainsize,\,...).
\end{align}
We will consider the case where absorption and scattering are dominated by dust, i.e.\ $\acc_{\rm rad,\,gas} \rightarrow \mathbf{0}$. 

Assuming isotropic scattering and re-emission in the rest frame and keeping terms up to $\mathcal{O}(v^{2}/c^{2})$ in the radiation-dust-hydrodynamics equations \citep{mihalas:1984oup..book.....M,lowrie:1999.radiation.hydro.coupling}, given an incident flux ${\bf F}_{\nu}$ the acceleration induced by absorption and scattering is: 
\begin{align}
\acc_{\rm rad,\,dust} \approx \frac{1}{\grainmass\,c}\,\int Q_{\rm ext,\,\nu}\,\pi\,\grainsize^{2}\,{\bf G}_{\nu}\,d\nu \approx \frac{\pi\,\grainsize^{2}}{\grainmass\,c}\,\langle Q \rangle_{\rm ext}\,{\bf G}_{r}
\end{align} 
where ${\bf G}_{r} \equiv {\bf F}-\dustvel\cdot (e_{\rm rad}+\mathbb{P}_{\rm rad})$. Here $Q_{\nu}$ and $\langle Q \rangle$ are the frequency-dependent and averaged extinction efficiencies, $e_{\rm rad}$ and $\mathbb{P}_{\rm rad}$ {are} the radiation energy density and pressure tensor. 

\subsubsection{Optically-Thin Simulations}

In the optically-thin limit, $|{\bf F}| \gg |\dustvel\cdot (e_{\rm rad}+\mathbb{P}_{\rm rad})|$ by $\mathcal{O}(c/v)$, and (because we adopt a plane-parallel geometry) ${\bf F}({\bf x},\,...) \rightarrow {\bf F}_{0} = F_{0}\,\hat{z}$ is constant, and $\langle Q({\bf x},\,\grainsize,\,...) \rangle \rightarrow \langle Q(\grainsize,\,...) \rangle$ is a function only of grain properties. Then $\acc_{\rm rad,\,dust} \rightarrow a_{\rm rad,\,dust}\,\hat{z}$ with $a_{\rm rad,\,dust} \approx F_{0}\,f_{r}(\grainsize,\,...)$ where $f_{r} \equiv \langle Q(\grainsize,\,...) \rangle\,\pi\,\grainsize^{2}/\grainmass\,c$. In general, for an incident spectrum peaked at some wavelength $\langle \lambda_{\rm rad} \rangle$, $Q$ depends primarily on grain size, with two relevant limits: 
\begin{align}
Q \sim \left( \frac{\grainsize}{\langle \lambda_{\rm rad} \rangle} \right)^{1-\accsizedep} \sim &
\begin{cases}
1\  &\ (\accsizedep=1;\ \ \grainsize \gg \langle \lambda_{\rm rad} \rangle) \\
\frac{\grainsize}{\langle \lambda_{\rm rad} \rangle} \ &\ (\accsizedep=0; \ \ \grainsize \ll \langle \lambda_{\rm rad} \rangle) \\
\end{cases}
\end{align}
Using this and $\grainmass \propto \grainsize^{3}$, we will conveniently parameterize the radiative acceleration as: 
\begin{align}
a_{\rm rad,\,dust} \equiv \accabsmax\,\left( \frac{\grainsize^{\rm max}}{\grainsize} \right)^{\accsizedep}
\end{align}
with $\accsizedep = 0$ corresponding to the ``long wavelength'' incident radiation case with $\langle \lambda_{\rm rad} \rangle \gg \grainsize$, and $\accsizedep=1$ corresponding to the ``short wavelength'' case with $\langle \lambda_{\rm rad} \rangle \ll \grainsize$. Since ${F}$ is constant, no explicit ``on the fly'' radiation transport is needed, and we can simply add this term directly to $\acc_{\rm ext}$. 

Briefly, note that if the grains are drifting at the equilibrium drift velocities in a homogeneous background, this dependence of $a_{\rm rad,\,dust}$ on $\grainsize$ translates to a drift velocity which is independent of grain size for $\accsizedep=1$, or increases with grain size (as $\grainsize^{1}$ or $\grainsize^{1/2}$ depending on if the drift is in the sub-sonic or super-sonic limit) for $\accsizedep=0$.

\subsubsection{Semi-Opaque, RDMHD Simulations}

In this paper we only consider modest optical depths $\lesssim 1$ and $\sim 1$. Nonetheless, at the larger {optical} depths we consider the ``optically thin'' approximation above might break down, especially locally in dense dust clumps that could become self-shielding to an external radiation field. We therefore consider an additional set of explicit radiation-dust-magnetohydrodynamics (RDMHD) simulations. We solve the radiation transport equations in {\small GIZMO} using the M1 moments method \citep{levermore:1984.FLD.M1}, as detailed and explicitly tested in a number of other applications \citep{lupi:2017.gizmo.galaxy.form.methods,lupi:2018.h2.sfr.rhd.gizmo.methods,hopkins:2019.grudic.photon.momentum.rad.pressure.coupling,hopkins:radiation.methods,grudic:starforge.methods}, with a single broad-band frequency interval. Neglecting terms $\mathcal{O}(v^{2}/c^{2})$, emission and re-emission by dust (since we do not follow these bands), relativistic beaming, and thermal/internal physics of grains, the M1 transport equations we solve are: 
\begin{align}
\label{eqn:rad.egy} \frac{1}{\tilde{c}}\frac{\partial e_{r}}{\partial t} + \nabla \cdot \left( \frac{{\bf F}_{r}}{c} \right) &= -R_{a}\,e_{r} + ({R_{a}-R_{s}})\,\frac{\dustvel \cdot {\bf G}_{r}}{{c^{2}}} \\ 
\label{eqn:flux} \frac{1}{\tilde{c}}\frac{\partial}{\partial t}\left( \frac{{\bf F}_{r}}{c} \right) + \nabla \cdot \mathbb{P}_{r} &= -({R_{a}+R_{s}})\, \frac{{\bf G}_{r}}{c}
\end{align}
where $\mathbb{P}_{r} \equiv e_{r}\,\mathbb{D}_{\rm M1}$ with $\mathbb{D}_{\rm M1}$ the Eddington tensor given by the usual M1 closure, ${\bf G}_{r} \equiv {\bf F}_{r} - {\bf v}_{d}\cdot(e_{r}\,\mathbb{I} + \mathbb{P}_{r})$, $\tilde{c} \le c$ is the reduced speed of light, and $R_{a,\,s} \equiv \rho\,\kappa_{a,\,s}$ are the absorption (``$a$'') and scattering (``$s$'') coefficients. Emission $\dot{e}_{\rm em}=0$ everywhere except the $z=0$ boundary (``base'') of the box, where we initialize a set of boundary cells that inject a constant vertical photon flux such that the flux is exactly equal to the desired flux ${\bf F}=F_{0}\,\hat{z}$ at $z\rightarrow 0$.

In most M1 implementations, including {\small GIZMO}, these transport equations are solved on the mesh defined by the gas cells. But the opacities  $R_{a,\,b}$ in this problem are defined by the dust, which is sampled by point-like super-particles. We therefore interpolate from the dust particles onto the mesh to determine $R_{a,\,b}$ in each cell, and from the mesh back to the dust to determine the flux at each dust particle location ${\bf F}({\bf x}={\bf x}_{\rm grain},\,...)$.  
\begin{align}
R_{a,\,s}^{i,\,{\rm cell}} &\approx \sum^{\rm grains}_{j}\,\Delta m_{j}\,W({\bf x}_{j}-{\bf x}_{i},\,H_{i})\,\left( \frac{\langle Q_{a,\,s} \rangle^{j}\,\pi\,\epsilon_{{\rm grain},\,j}^{2}}{m_{{\rm grain},\,j}} \right) \\
{\bf F}^{j,\,{\rm grain}} &\approx \frac{\sum_{i}^{\rm cells}\,{\bf F}^{i,\,{\rm cell}}\,\Delta m_{i}\,W({\bf x}_{i}-{\bf x}_{j},\,H_{j})}{\sum_{i}^{\rm cells}\,\Delta m_{i}\,W({\bf x}_{i}-{\bf x}_{j},\,H_{j})}
\end{align}
(with an identical interpolation for $e_{r}$, $\mathbb{P}_{r}$ to grains and $\dustvel$ to gas), where $W$ is the normalized kernel function used in the {\small GIZMO} hydrodynamics operations \citep{hopkins:gizmo} with the properties: $\sum_{\rm j}^{\rm cells} W({\bf x}_{j}-{\bf x}_{i},\,H_{i}) \equiv 1/V_{i} \equiv \Delta m_{i}/\rho_{i}$, $H_{i}=2\,h_{i}=2\,V_{i}^{-1/3}$ ($\Delta m_{i}$ is the total mass of a gas cell or grain ``super-particle''). These interpolation functions have the advantages that (1) they interpolate exactly to the correct $\kappa_{a,\,s}$ or ${\bf F}$ in a field with constant $\kappa$ ($Q$ and $\epsilon$) or ${\bf F}$ respectively, and (2) the discretized integral/sum over the interpolated fields exactly conserves total grain mass/area and total photon momentum. 

In the optically-thin limit, $Q_{a}$ and $Q_{s}$ are degenerate with $F$ (only the product $F\,Q_{\rm ext} = F\,(Q_{a}+Q_{s})$ appears), so we need only to specify the parameters $\accabsmax$ and $\accsizedep$. If there is non-negligible optical depth, however, the degeneracy between $F$ and $Q$ and between absorption and scattering is broken (note the different terms in Eq.~\ref{eqn:rad.egy} for $R_{a}$ and $R_{s}$), and we need to specify additional quantities. In our RDMHD simulations, we parameterize the {\em input} or ``base'' flux ${\bf F}(z=0) = F_{0}\,\hat{z}$ via $\accabsmax \equiv F_{0}\,Q_{\rm ext,\,0}\,\pi\,(\grainsize^{\rm max})^{2}/\grainmass^{\rm max}\,c$, which is simply the value we would have in the optically-thin case with $Q=1$. We then parameterize $\langle Q_{a,\,s} \rangle \equiv Q_{a,\,s}^{\rm max}\,(\grainsize/\grainsize^{\rm max})^{(1-\accsizedep)}$, and define the albedo $A_{0} \equiv Q_{s}^{\rm max}/Q_{\rm ext,\,0}$ where $Q_{\rm ext,\,0} \equiv Q_{s}^{\rm max} + Q_{a}^{\rm max} = \langle Q_{\rm ext}(\grainsize=\grainsize^{\rm max})\rangle$. So we must specify $Q_{\rm ext,\,0}$ and $A_{0}$ in addition to $\accabsmax$ and $\accsizedep$.

Since the radiation field is now evolved explicitly, we have a Courant-type timestep condition: $\Delta t_{\rm rad} < C\,\Delta x_{i}/\tilde{c}$. To make the simulations computationally tractable we follow standard practice adopting a reduced speed of light (RSOL), $\tilde{c} < c$. However, we must still choose $\tilde{c}$ much larger than any other signal speed or global velocity in the problem: in particular, if $\tilde{c}$ is not larger than the speed of e.g.\ the fastest outflowing dust or gas, then the radiation can unphysically ``lag behind'' the outflow, causing it to artificially stall. We find converged solutions here require $\tilde{c} \gtrsim 300\,\cs^{0}$, so for safety our default RDMHD simulations adopt $\tilde{c}\sim1000\,\cs^{0}$ ($\sim 500\,{\rm km\,s^{-1}}$ in GMCs, $\sim 10^{4}\,{\rm km\,s^{-1}}$ in HII regions). Thus, this still means our timesteps must be $\sim 10$ times smaller (hence simulations 10 times more expensive) in RDMHD compared to the optically-thin simulations above. We are therefore restricted to lower resolution for RDMHD.

\subsection{Initial \&\ Boundary Conditions}
\label{sec:ics}

We initialize a 3D box as illustrated in \fref{fig:cartoon.ics}, with $L_{\rm xy}$ in the $\hat{x}-\hat{y}$ plane, and long-axis length\footnote{Because of the exponential decrease in density with $z$, it makes essentially no difference how long we make the long-axis of the box, once $L_{\rm z} \gtrsim 10\,L_{\rm xy}$, and our testing with $L_{\rm z}=(10,\,20,\,100,\,500)\,L_{\rm xy}$ confirms this. However the accuracy of the vertically stratified box approximation breaks down once $z\gg L_{\rm xy}$, so we focus our analysis on material at less than $\sim 20$ scale-heights.} $L_{\rm z} = 20-100\,L_{\rm xy} \gg L_{\rm xy}$ in the $\hat{z}$ direction with $L_{\rm xy}=\Lbox$. The boundary conditions are periodic in $\hat{x}$ and $\hat{y}$: the ``base'' ($z=0$) $\hat{z}$ boundary is reflecting, while the ``upper'' ($z=+L_{\rm z}$) $\hat{z}$ boundary allows gas and dust to escape (outflow). Gas is initialized with a vertically stratified density 
\begin{align}
\initvalupper{\gasden} \equiv \rho(t=0) =\rhobase\,\exp{(-z/\Lbox)}
\end{align} 
(so $\rhobase \approx M_{\rm gas,\,box}/L_{\rm xy}^{3}$), velocity $\initvalupper{\gasvel} = 0$, and uniform magnetic field\footnote{Instead initializing constant plasma $\beta$ so $|{\bf B}(t=0)| \propto (\gasden\,\cs^{2})^{1/2}$ does not substantively change our results.} $\initvallower{\B} \equiv \initvallower{B}\,\initvallower{\Bhat}$ (with $\initvallower{\Bhat} \equiv \sin{\Bangle^{0}}\,\hat{x} + \cos{\Bangle^{0}}\,\hat{z}$ in the $\hat{x}-\hat{z}$ plane), and initially-uniform dust-to-gas ratio 
\begin{align}
\dustgas \equiv \frac{\initvalupper{\dustden}}{\initvalupper{\gasden}} \ .
\end{align} Dust velocities $\dustvel$ are initialized with with the local homogeneous steady-state equilibrium values\footnote{$\initvalupper{\driftvel} = |\acc|\, \initvalupper{\ts}\,(1+\dustgas)^{-1}\,(1+\tauparam^{2})^{-1}\,[\hat{\acc} - \tauparam\,(\hat{\acc} \times  \initvallower{\Bhat} ) + \tauparam^{2}\,(\hat{\acc}\cdot\initvallower{\Bhat})\,\initvallower{\Bhat}]$, which is $\sim \acc\,\ts^{0}$ for $\tauparam \ll 1$ and $\sim |\acc|\,{\ts^{0}}\,\cos{(\Bangle^{0})}\,\Bhat_{0}$ for $\tauparam \gg 1$.} (see \paperone, \S~3.1), but it makes no difference (outside of eliminating a brief initial transient) if we initialize $\dustvel=0$. All elements feel a uniform ``downward'' gravitational acceleration ${\bf g} = -g\,\hat{z}$, and there is an initial radiation flux ${\bf F}(z=0) = F_{0}\,\hat{z}$ in the ``upward'' direction which gives rise to the radiative acceleration of dust grains.

We can then fully-specify the initial conditions with a number of dimensionless parameters, given in \tref{table:sims}: 
(1) initial magnetic strength  $B_{0}$ given by the plasma $\beta_{0} \equiv (\cs^{0} / v_{A,\,{\rm base}}^{0})^{2} = 4\pi\,\rhobase\,(\cs^{0}/B_{0})^{2}$, and direction $\Bangle^{0}$; 
(2) gas polytropic index $\gamma$; 
(3) strength of gravity relative to pressure forces $\gravparam \equiv g / (\cs^{2} / \Lbox)$; 
(4) dust-to-gas ratio $\dustgas$; 
(5) grain ``size parameter'' (normalization of the drag force scaling Eq.~\ref{eq:ts}), $\sizeparammax \equiv \internaldensity\,\grainsizemax / \initvalupper{ \gasden }\, \Lbox$, evaluated at $\grainsize = \grainsizemax$; 
(6) grain ``charge parameter'' (normalization of the Lorentz force scaling in Eq.~\ref{eq:tL}), $\chargeparammax \equiv -3\,\initvalupper{ \grainchargeZ } \,e / (4\pi\,c\,(\grainsizemax)^{2}\,(\initvalupper{ \gasden })^{1/2})$; 
(7) scaling of the flux and dust opacities, which we parameterize by $\accabsmax \equiv  F_{0}\,Q_{\rm ext,\,0}\,\pi\,(\grainsize^{\rm max})^{2}/\grainmass^{\rm max}\,c$ and $\accsizedep=0$ or $=1$; 
(8) for our RDMHD simulations we also specify albedo $A_{0}$ and absolute value of the absorption efficiency $Q_{\rm ext,\,0}$.

\begin{figure}
    \centering
    \includegraphics[width=0.9\columnwidth]{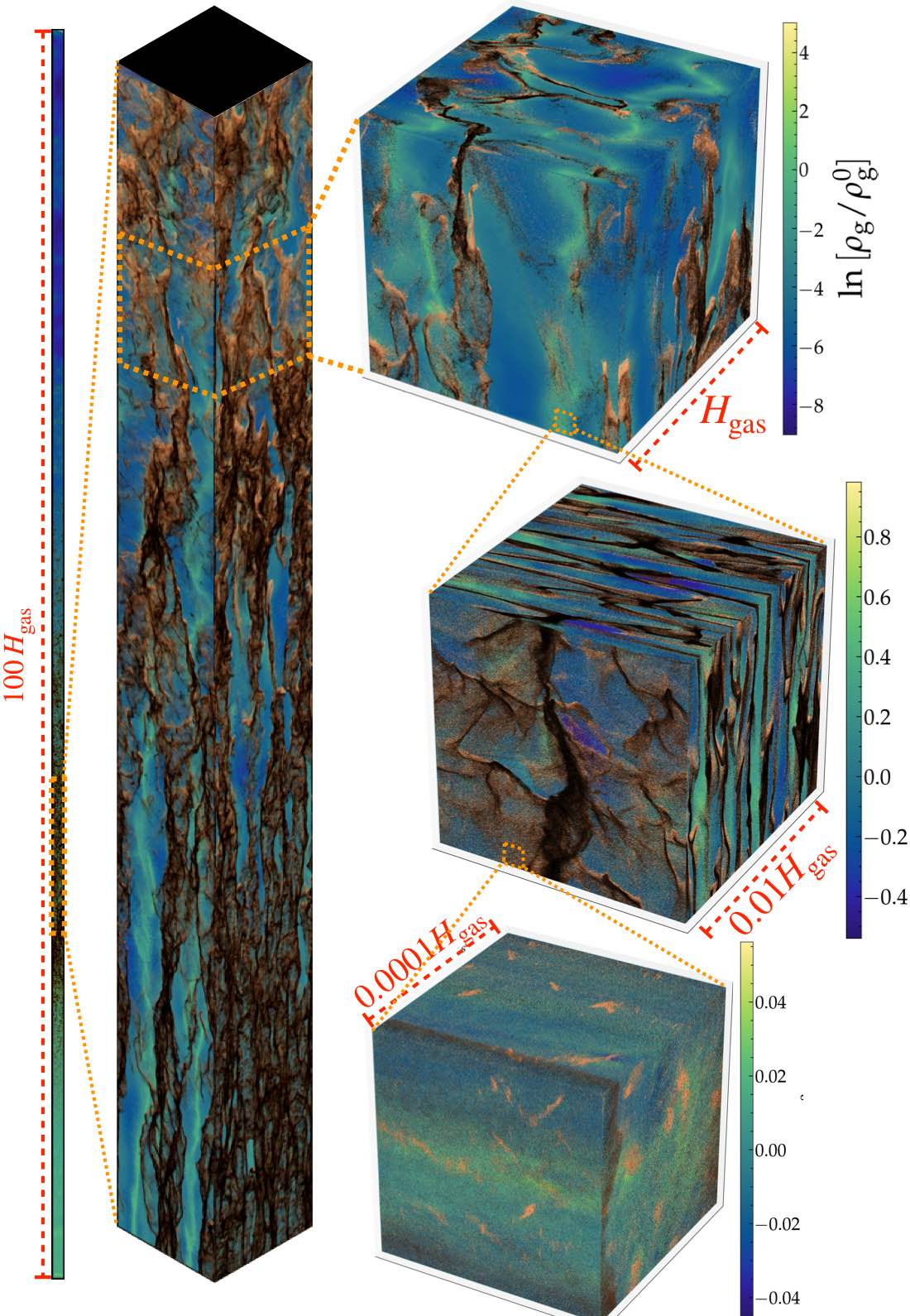}
    \caption{Illustration of the dynamic range of scales probed by our fiducial stratified boxes (the two vertically extended columns and cube with side-length $\sim \Lscale$ show sub-volumes of a single box), as well as our successive ``zoom-in'' unstratified periodic boxes (separate simulations from Table~\ref{table:sims.all} shown as the two cubes with side-lengths $10^{-4}-10^{-2}\,\Lscale$, with identical physical parameters as the ICs of the base of the stratified box, but re-scaled box size). The smooth blue-green-yellow colorscale shows the gas density, projected onto the plotted surfaces (as labeled). The copper-brown-black colorscale plots (as individual pixels) individual dust grains on each surface, colored by size (lightest are smallest grains, darkest/black are largest grains). The dynamic range of the simulations and physical structure of dusty outflows spans from the linear size of our stratified boxes (up to $\sim 100\,\Lscale$, where we see the large-scale outflow structure) through $\sim 1-10\,\Lscale$ (where we see filamentary structure and where, at any time, most of the dust and gas mass in the outflow is contained). ``Horsehead'' type structures are obvious on intermediate ($\sim \Lscale$) scales. In the ``zoom in'' simulation boxes boxes we see that strong structure in the dust persists on all spatial scales these simulations can explore (well below observable scales); although the gas becomes smoother and less compressible on smaller scales.
    \label{fig:boxtower.dynamic.range}}
\end{figure}

Our fiducial simulations adopt fixed mass resolution, with gas resolution $\Delta m_{g} \approx 10^{-7}\,\rhobase\,H^{3}$ and 4 times as many dust elements with mean resolution $\langle \Delta m_{d} \rangle \approx 2.5\times10^{-10}\,\rhobase\,H^{3}$ (giving $N=0.5\times10^{8}$ resolution elements). Our low-resolution parameter-survey simulations use 8 times fewer gas+dust elements. Note that because the simulations are Lagrangian (both gas cells and dust super-particles), the mass resolution is fixed, but the effective spatial resolution can be much higher in dense regions.

We also consider a number of small-scale unstratified, periodic boxes (see \tref{table:sims.all}), meant to represent a ``zoom in'' onto roughly a single resolution element in our stratified boxes. We adopt identical {\em physical} parameters to the stratified boxes at the ``base'' (e.g.\ $\gasden=\rhobase$), in a uniform periodic, cubic box, {which} effectively scales to size $\Lbox \sim 10^{-4} - 10^{-2}\,\Lscale$ (see \fref{fig:boxtower.dynamic.range}). As noted in \citet{hopkins:2017.acoustic.RDI,moseley:2018.acoustic.rdi.sims}, the results in these unstratified boxes are analytically and numerically invariant to any value of a uniform acceleration (like $g$).

\begin{figure}
    \centering
    \includegraphics[width=0.85\columnwidth]{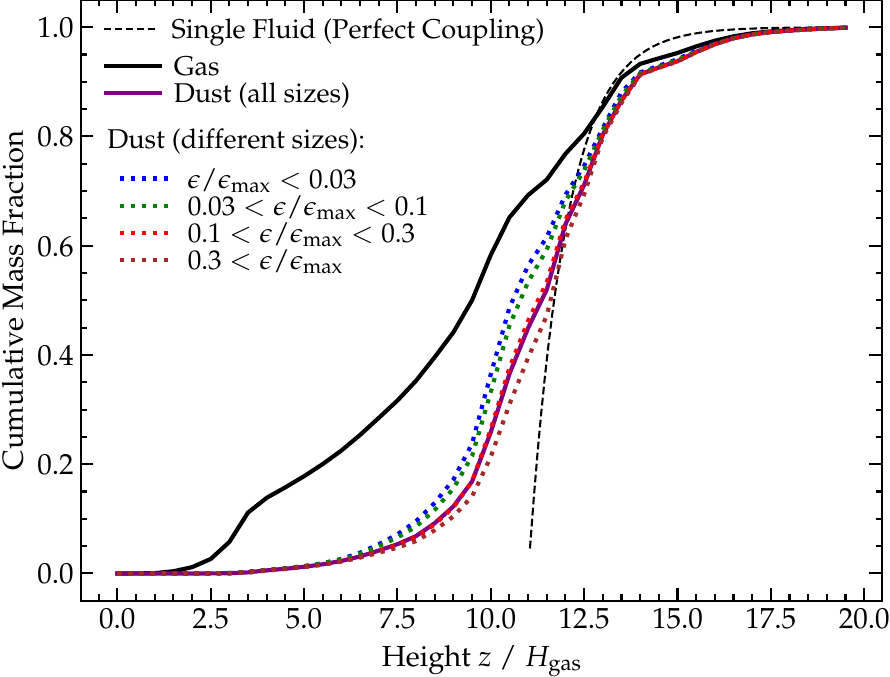} \\
    \hspace{-0.2cm}\includegraphics[width=0.88\columnwidth]{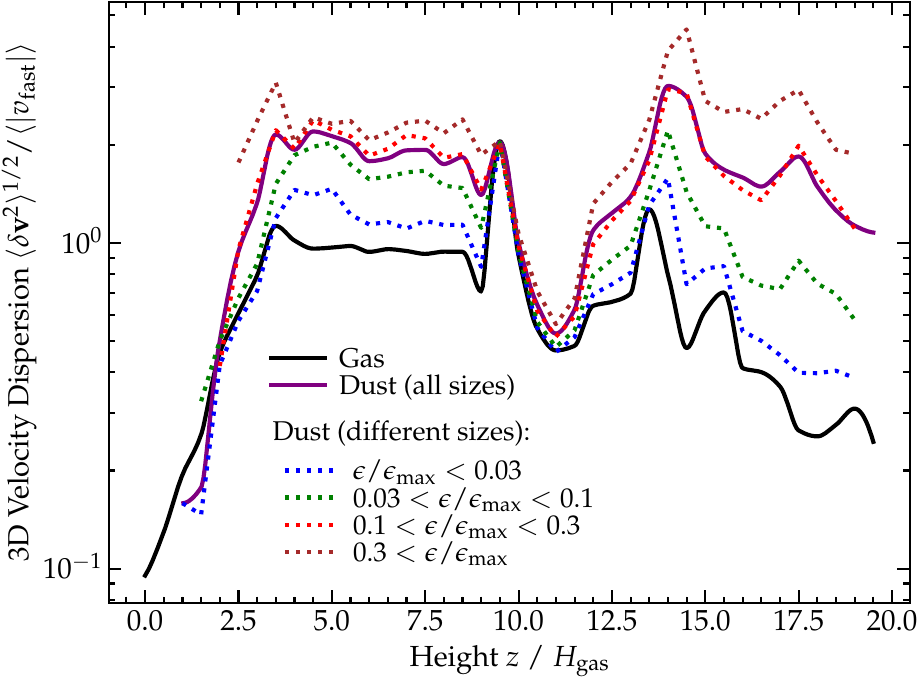} \\
    \includegraphics[width=0.85\columnwidth]{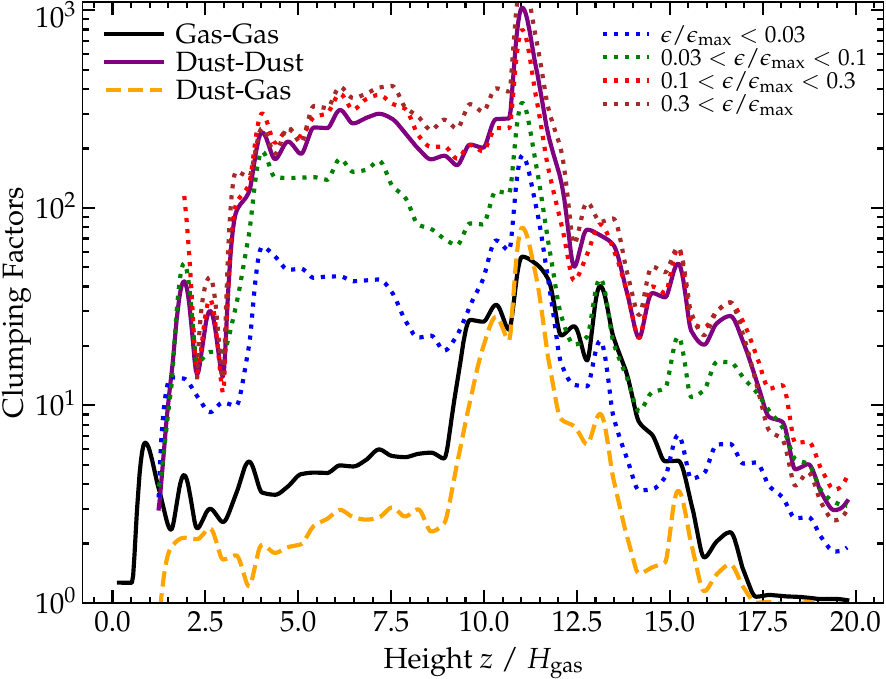} \\
    \vspace{-0.1cm}
    \caption{Properties of dust-driven outflows at a given time for {\bf GMC-Q}, chosen at $t\sim 3\,t_{\rm acc}$ (where $t_{\rm acc}$ is the characteristic acceleration timescale defined by the ICs; $t_{\rm acc} \equiv (2\,\Lscale/\langle a_{\rm eff} \rangle)^{1/2}$ with $\langle a_{\rm eff} \rangle \equiv \langle M_{\rm dust}\,\langle a_{\rm dust,\,rad} \rangle / M_{\rm total} \rangle$), so a perfectly-uniform outflow would have reached a height $z\sim 10\,\Lscale$. 
    {\em Top:} Cumulative mass profile of gas and dust (and grains in different bins of size $\epsilon$). We compare the profile that would be obtained for a single dust+gas fluid with a spatially-uniform acceleration (the ``perfect-coupling'' limit). The dust is at roughly the same position, without much dependence on grain size. The gas has mostly been entrained to similar height, though a non-negligible fraction $\sim 10-30\%$ has been ``left behind'' at the base of the wind. 
    {\em Middle:} 3D rms random velocity dispersion (subtracting the bulk flow) within narrow bins of $z$, relative to the fast magnetosonic speed (since $\beta \lesssim 1$ here, this is approximately the \Alf\ speed). The gas and dust both reach qualitatively similar trans-\Alf{ic} random velocities (while the bulk outflow speed is substantially super-\Alf{ic}). Dust has larger random motion, with dispersions larger for larger (less-strongly-coupled) grains.
    {\em Bottom:} Clumping factors for gas-gas ($\langle \gasden^{2} \rangle / \langle \gasden \rangle^{2}$), dust-dust ($\langle \dustden^{2} \rangle/\langle \dustden \rangle^{2}$), and gas-dust ($\langle \gasden\,\dustden \rangle / \langle \gasden \rangle\,\langle \dustden \rangle$). Gas-gas and gas-dust clumping is significant, especially around $\sim 10\,\Lscale$ where most of the mass resides. Dust-dust clumping is extremely strong, and stronger for larger grains.
    \label{fig:bulk.props.vs.z.gmc.Q}}
\end{figure}

\begin{figure}
    \centering
    \includegraphics[width=0.85\columnwidth]{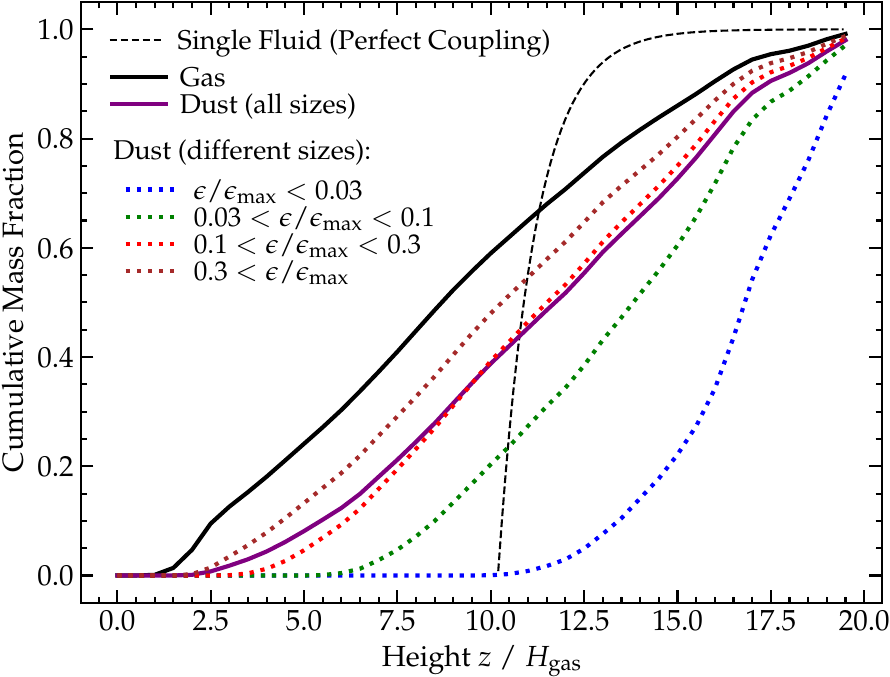}\\
    \hspace{-0.14cm}\includegraphics[width=0.855\columnwidth]{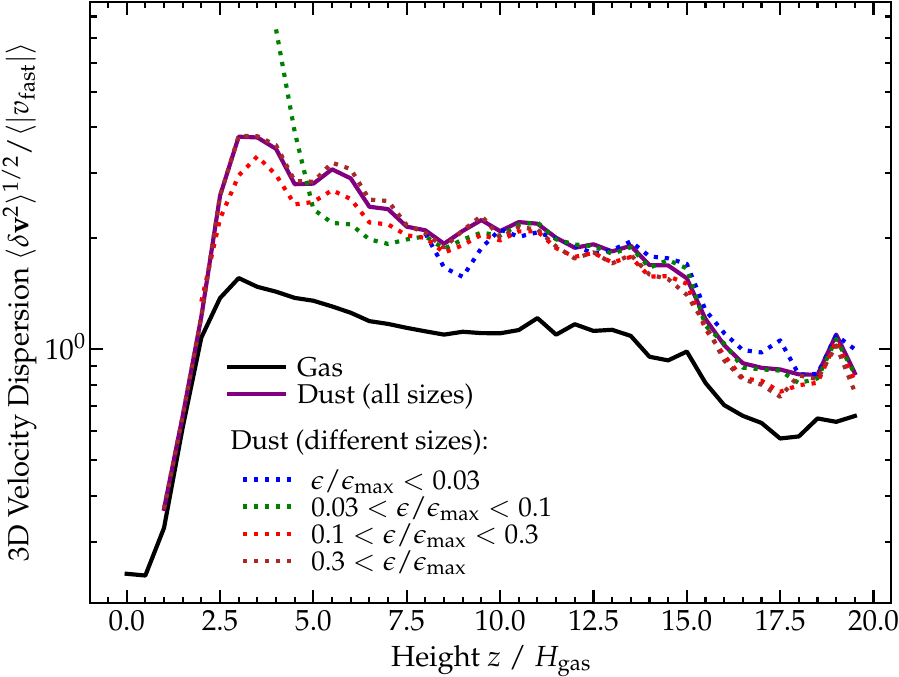}\\
    \includegraphics[width=0.86\columnwidth]{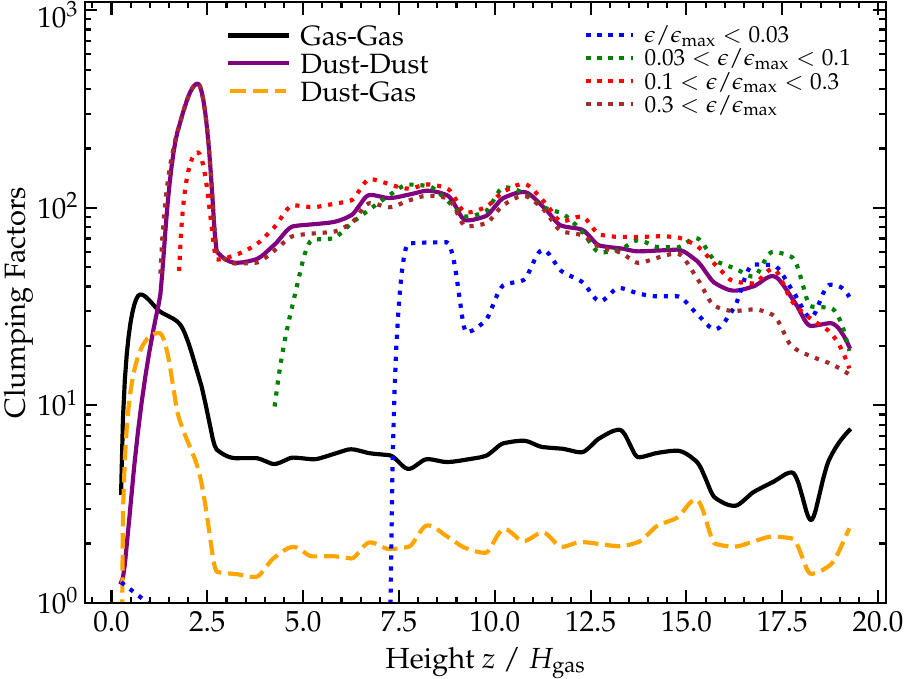}\\
    \caption{Same as Fig.~\ref{fig:bulk.props.vs.z.gmc.Q} for {\bf GMC}. Here the dust acceleration scales $\propto \grainsize^{-1}$, so the homogeneous equilibrium drift speed is independent of $\grainsize$, yet the small grains non-linearly end up moving faster and accelerate somewhat past the gas and large grains. The dust clumping/clustering is now more similar across $\grainsize$, and the dust again drives trans-magnetosonic turbulence in the gas. 
    \label{fig:bulk.props.vs.z.gmc}}
\end{figure}

\subsection{Parameter Space Explored}
\label{sec:params}

We now discuss some scalings that motivate the parameters of our study. These are outlined in greater detail in \paperone\ and \citet{hopkins:2018.mhd.rdi}, so we briefly summarize them here. In most GMCs and HII regions, we expect maximum grain sizes $\grainsize^{\rm max} \sim 0.1\,\micron$, and minimum $\grainsize^{\rm min} \sim {\rm nm} \sim 0.01\,\grainsize^{\rm max}$, with an MRN-like size spectrum as we adopt and $\internaldensity \sim 1.5\,{\rm g\,cm^{-3}}$, and dust-to-gas-ratio $\dustgas \sim 0.01\,(Z/Z_{\odot})$. \tref{table:sims} and \fref{fig:parameter.space} give a complete list of parameters adopted and illustrate their values relative to other astrophysical systems.

%
{\bf GMC:} Consider dust with $\grainsize^{\rm max} \sim \epsilon_{0.1}\,0.1\,\micron$ in a GMC with a typical gas surface density $\Sigma_{\rm gas} \sim M_{\rm GMC}/\pi\,R_{\rm cl}^{2} \sim \Sigma_{100}\,100\,M_{\odot}\,{\rm pc}^{-2}$ (where $M_{\rm GMC}$ and $R_{\rm cl}$ are the cloud mass and radius), gas mass $M_{\rm GMC} \sim M_{6}\,10^{6}\,M_{\odot}$ and a fraction $\epsilon^{\ast} \equiv M_{\ast}/M_{\rm GMC} \sim \epsilon^{\ast}_{0.1}\,0.1$ of its mass in young stars, with a near-isothermal $\gamma\sim1$ (owing to rapid cooling) at $T\sim T_{100}\,100\,$K, and $\langle \Bmag^{2}\rangle^{1/2} \sim B_{5}\,5\,\microGauss$, all similar to values observed in the massive complexes that dominate Milky Way star formation \citep{crutcher:cloud.b.fields,rice:2016.gmc.mw.catalogue,grudic:sfe.gmcs.vs.obs,guszejnov:2019.imf.variation.vs.galaxy.props.not.variable,guszejnov:fire.gmc.props.vs.z,benincasa:2020.gmc.lifetimes.fire,lee:2020.hopkins.stars.planets.born.intense.rad.fields}. Taking this (with $\Lscale\sim R_{\rm cl}$), with a flux given by $M_{\ast}$ and the light-to-mass ratio and SED for a young stellar population ($\sim 1200\,L_{\odot}/M_{\odot}$) combined with typical grain properties above (with $Q_{\rm ext,\,0}\sim 0.2\,Q_{0.2}$ appropriate for observed GMC grains at optical/NUV wavelengths; \citealt{weingartner:2001.dust.size.distrib,weingertner.draine:photo.forces.on.dust.hard.ism.rad}), with collisional charging dominated by interactions with the WNM as grains move through multi-phase gas so (since our default simulations adopt a simple EOS) we take the collisional WNM scaling from \citet{weingartner:2001.grain.charging.photoelectric}, and we obtain $\sizeparammax \sim 0.001\,\epsilon_{0.1}/\Sigma_{100}$; $\accparammax \sim 70\,M_{6}^{1/2}\,Q_{0.2}\,\epsilon^{\ast}_{0.1}/\Sigma_{100}^{1/2}\,T_{100}$; $\gravparam \sim 110\,(M_{6}\,\Sigma_{100})^{1/2}/T_{100}$, $\chargeparammax \sim 300\,M_{6}^{3/4}\,T_{100}^{1/2}\,\epsilon_{0.1}^{\ast}/(\epsilon_{0.1}\,\Sigma_{100}^{5/4})$, $\beta \sim 0.02\,\Sigma_{100}^{3/2}\,T_{100}\,B_{5}^{-2}\,M_{6}^{-1/2}$. This motivates the parameters of our ``GMC-like'' simulations.

%
{\bf HII:} Consider dust at a distance $r \sim 0.1-1\,$pc around an HII region near e.g.\ an O5 star with $L\sim L_{6}\,10^{6}\,L_{\odot}$ ($M_{\ast} \sim 20\,M_{\odot}$), at $T\sim 10^{4}\,$K ($\gamma\approx1$ regulated by photo-heating) with an isothermal sphere-like density profile $n/{\rm cm^{-3}} \sim 100\,(r/{\rm pc})^{-2} \sim n_{x}\,10^{x}$. We then have 
$\sizeparammax \sim 0.003\,\epsilon_{0.1}/(n_{4}\,r_{0.1}) \sim 0.03\,\epsilon_{0.1}/(n_{2}\,r_{1})$; 
$\accparammax \sim 4.5\,L_{6}\,Q_{\rm ext,\,0}/(n_{4}\,r_{0.1}^{2}) \sim 4.8\,L_{6}\,Q_{\rm ext,\,0}/(n_{2}\,r_{1}^{2})$; 
$\gravparam \sim (0.001-0.01)\,n_{4}\,r_{0.1}^{2} \sim 0.001\,n_{2}\,r_{1}^{2}$ (the first $\gravparam \rightarrow 0.01$ if we include gravity of the star itself). Because of the strong UV radiation field photo-electric charging likely dominates, which for the scalings in \citet{tielens:silicate.dust.composition,tielens:1998.dust.is.amorphous.iron.poor.silicates,tielens:2005.book} gives 
$\chargeparammax \sim 44/(\epsilon_{0.1}\,n_{4}^{1/2}) \sim 440/(\epsilon_{0.1}\,n_{2}^{1/2})$. The two radii chosen here, corresponding to our ``near'' and ``far'' setups, are qualitatively motivated roughly to lie on either side of the Stromgren radius in \citet{hopkins:2018.mhd.rdi}, representing gas in the WIM inside the HII region and WNM just outside, but this should not be taken too literally.

Note that $\sizeparam \ll 1$ for all conditions here: $\sizeparam$ is approximately the ratio of the dust drag/collisional mean-free path ({the} scale over which dust momentum is redistributed to gas; $L_{\rm drag} \sim |\driftvel|\,\ts$) to gas scale-length $\Lscale=\Lbox$, so the grains are ``well-coupled.''
As discussed in \citet{hopkins:2017.acoustic.RDI,hopkins:2018.mhd.rdi}, the characteristic wavelength dividing the ``long-wavelength'' or ``pressure-free'' RDIs and the ``intermediate wavelength'' or ``mid-$k$'' magnetosonic RDIs is $\lambda_{\rm crit} \sim \driftvelmag\,\ts/\dustgas \sim (\sizeparam/\dustgas)\,\Lbox$ -- so the largest-wavelength modes of interest here ($\lambda \gtrsim \Lbox$) are in the mid-$k$ regime for the largest grains and long-wavelength regime for the smallest grains. 

Another closely-related parameter is the extinction optical depth integrated to infinity: 
\begin{align}
\opticaldepth \equiv \int_{0}^{\infty}\int_{\epsilon^{\rm min}}^{\epsilon^{\rm max}}\,\left(\frac{Q_{\rm ext}\,\pi\,\epsilon^{2}}{\grainmass}\right)\,\frac{d\dustgas}{d\epsilon}\,\rho(z)\,d\epsilon\,d z\ ,
\end{align} 
giving for our initial conditions (with an MRN size spectrum and exponentially-stratified density)
\begin{align}
\label{eqn:taugeo}. \langle \opticaldepth \rangle = \frac{3\,f_{\psi}\,Q_{\rm ext,\,0}\,\dustgas}{4\,\sizeparam_{m}} \sim 0.4\,\frac{f_{\psi}\,L_{\rm pc}\,n_{100}\,\dustgas_{0.01}\,Q_{\rm ext,\,0}}{\epsilon_{0.1}\,\bar{\rho}^{\,i}_{\rm grain,\,cgs}}\ .
\end{align} 
where $f_{\psi} = 1$ for $\accsizedep=0$ and $f_{\psi}=10$ for $\accsizedep=1$. From this and \tref{table:sims}, we see that GMCs and HII regions, as expected, correspond to $\opticaldepth \sim 0.1-$\,a few, i.e.\ relatively small, but not completely negligible optical depths for $Q\sim 1$ (corresponding to optical/near-IR wavelengths -- e.g. similar to $A_{V} \sim 1$). 

Note that our simulations are defined entirely by these dimensionless parameters {and therefore} do not necessarily represent one specific set of physical conditions -- any system which results in the same dimensionless parameters will give identical results (in the idealized setups here). But also there is a large range expected for plausible ISM conditions, and large uncertainties on some parameters (like grain charge). We therefore take these scalings only as {an} order-of-magnitude motivation, and systematically vary some of the relevant parameters in lower-resolution tests, to identify the most robust behaviors.

\begin{figure}
    \centering
    \includegraphics[width=0.51\columnwidth]{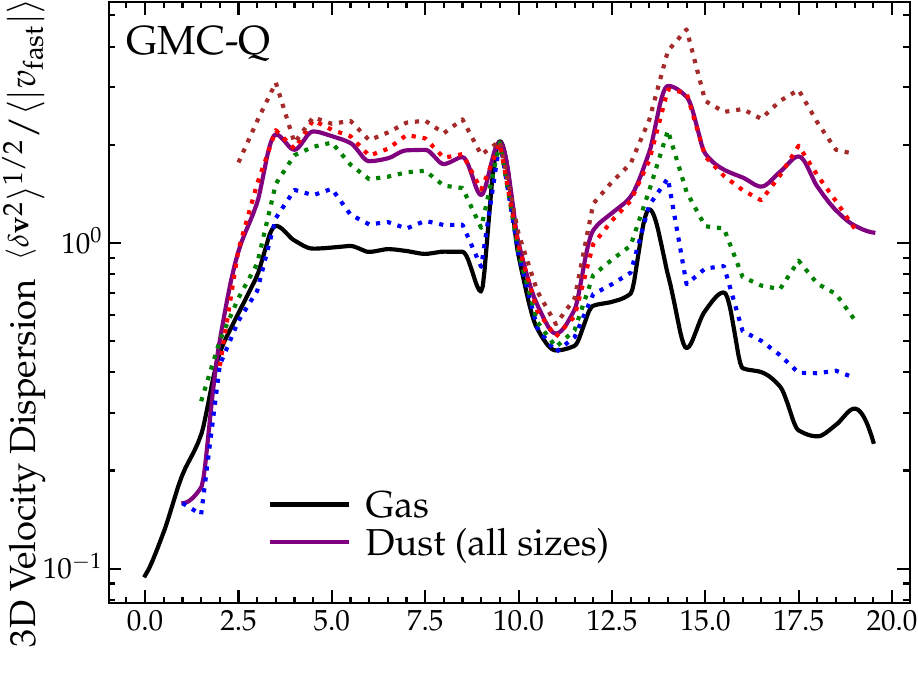}
    \hspace{-0.2cm}\includegraphics[width=0.50\columnwidth]{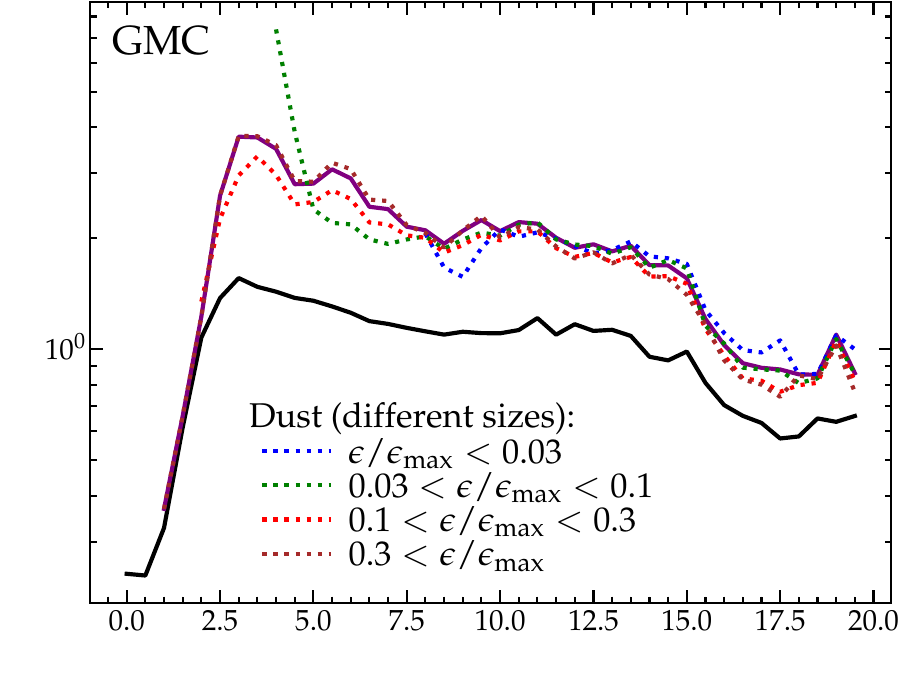}\\
    \includegraphics[width=0.50\columnwidth]{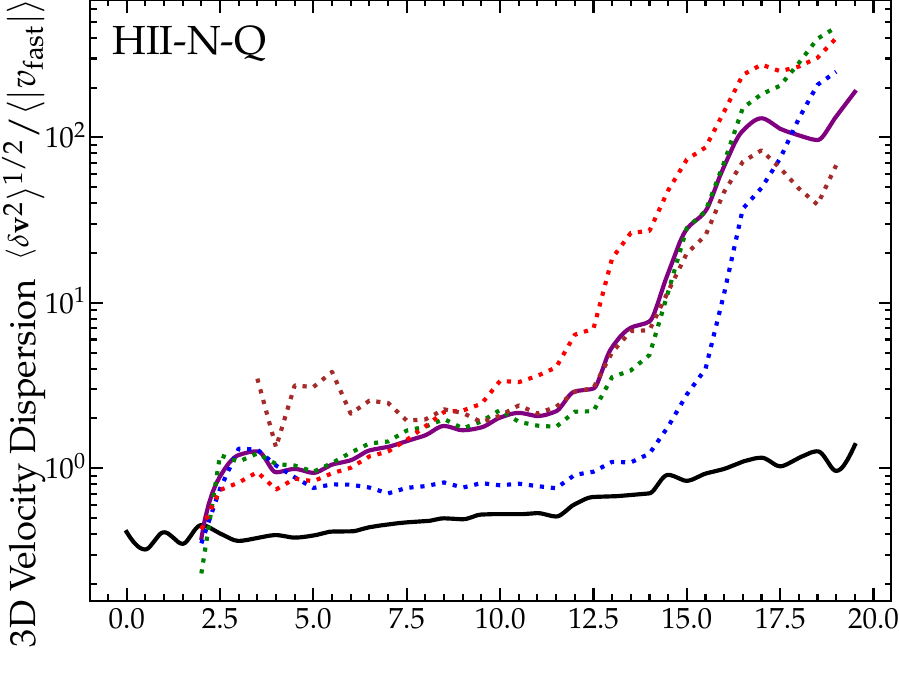}
    \hspace{-0.2cm}\includegraphics[width=0.50\columnwidth]{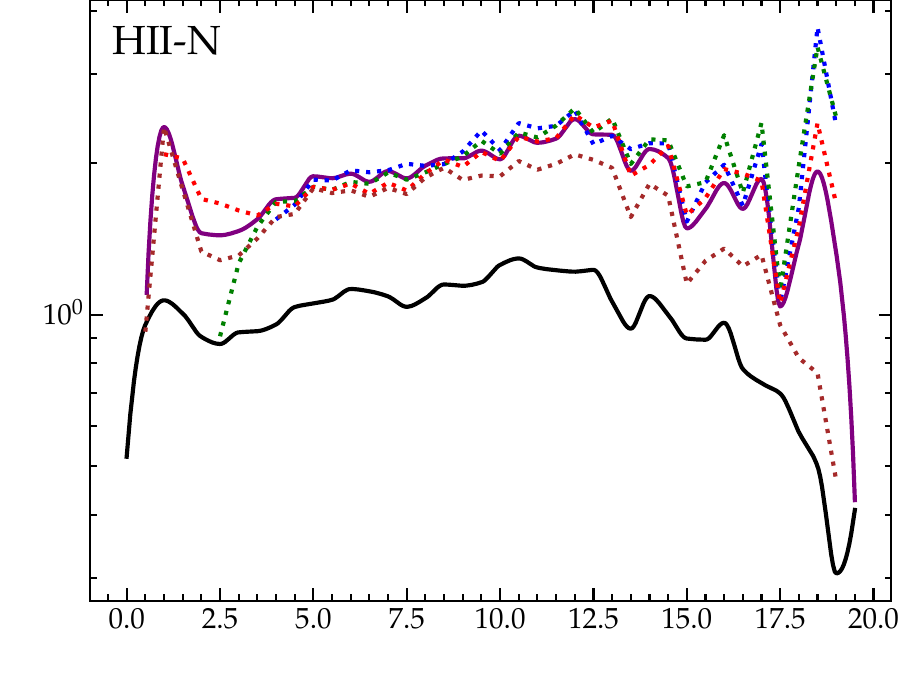}\\
    \includegraphics[width=0.51\columnwidth]{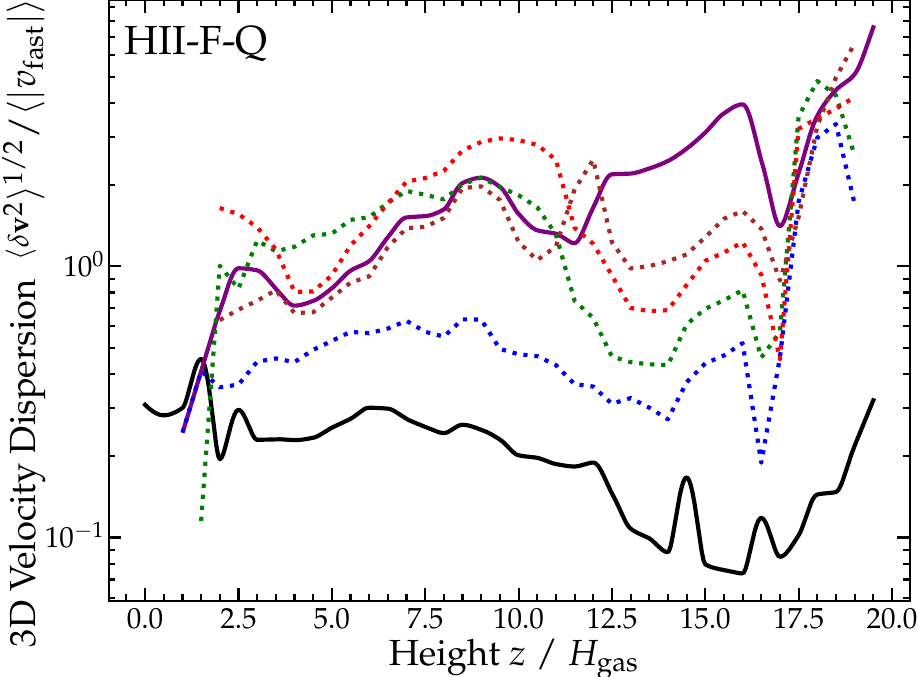}
    \hspace{-0.2cm}\includegraphics[width=0.50\columnwidth]{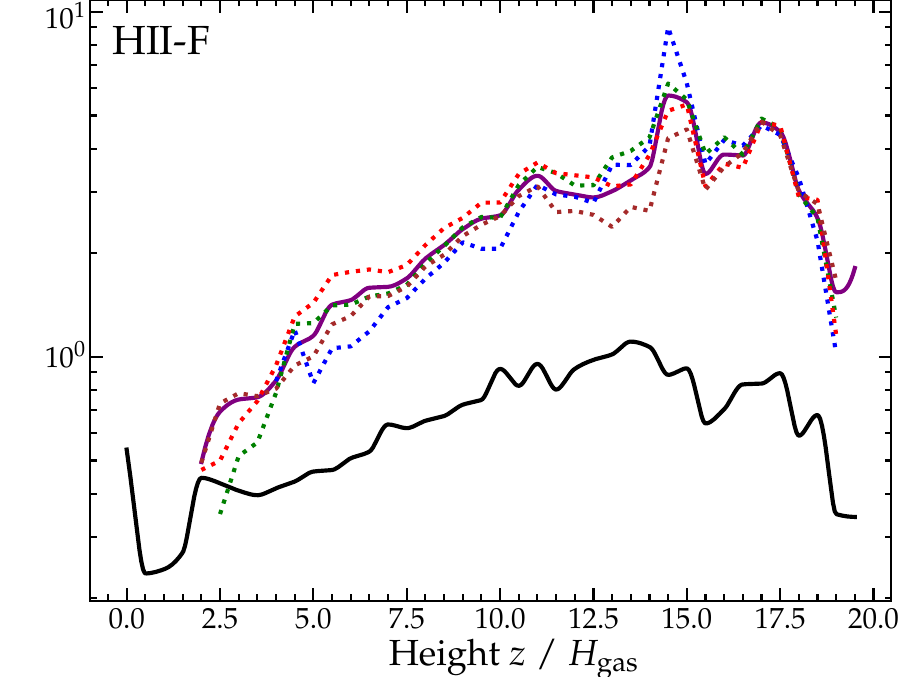}\\
    \vspace{-0.15cm}
    \caption{Velocity fluctuation profiles (as Fig.~\ref{fig:bulk.props.vs.z.gmc.Q}). 
        {\em Left:} Q runs, {\rm Right:} no-Q runs. {\em Top:} GMC, {\em Middle:} HII-N, {\em Bottom:} HII-F. The qualitative trends with height and magnitude of the gas+dust turbulence are similar to those in Figs.~\ref{fig:bulk.props.vs.z.gmc.Q}-\ref{fig:bulk.props.vs.z.gmc}; runs with $Q\propto \grainsize$ ($a_{\rm rad} \sim \,$constant) show nearly grainsize-independent fluctuations; runs with $Q\sim $\,constant ($a_{\rm rad} \propto \grainsize^{-1}$) show stronger velocity dispersion in larger grains. HII-F, with more sub-sonic grain acceleration, produces slightly weaker dispersions.
    \label{fig:turbulence}}
\end{figure}

\subsection{Translating the Simulations to Physical Scales}
\label{sec:scale}

As noted above, the simulations are entirely defined by dimensionless parameters, motivated by the scalings in \S~\ref{sec:params}. This means that the units of length (e.g.\ $\Lscale$), time ($t_{\rm acc}$, defined below), mass, etc., can be arbitrarily scaled to any physical system which has the same dimensionless parameters. Nonetheless, it is useful to consider how this scaling behaves in the context of real systems. Obviously, most of the parameters in Table~\ref{table:sims} like plasma $\beta_{0}$, dust to gas ratio $\dustgas$, grain albedo/absorption efficiencies, or magnetic field direction $\Bangle^{0}$, contain no information about the absolute units/scale of the problem. Similarly the acceleration parameter $\accparammax$ is essentially the ratio of radiation to thermal energy density in the optically-thin limit, which does not set a scale, and the charge parameter is defined by the ratio of Lorentz to drag forces, which depends only weakly and indirectly on scale through the details of the assumed scalings of the dust charge law (and does not strongly influence our conclusions). 

The one parameter which does define a scale in a meaningful sense is the ``size parameter'' $\sizeparammax$, which we show above (Eq.~\ref{eqn:taugeo}) relates (when combined with the dust-to-gas-ratio) directly to the average geometric optical depth of the system. Given some assumed properties of the dust, this defines the column density of the system. So for each set of ICs, the column density is relatively well-specified. For the GMC simulations ($\sizeparammax=10^{-3}$), our choices correspond to column densities of 
\begin{align}
\Sigma_{\rm gas}^{\rm GMC} \sim 100\,{\rm M_{\odot}\,pc^{-2}}\,\left(\frac{\internaldensity}{2\,{\rm g\,cm^{-3}}}\right)\,\left( \frac{\grainsizemax}{0.1\,{\rm \mu m}}\right)
\end{align}
 (as we assumed for our scalings in \S~\ref{sec:params}, motivated by typical observed GMC surface densities), or $N_{\rm H} \sim 10^{22}\,{\rm cm^{-2}}\,(\internaldensity/2\,{\rm g\,cm^{-3}})\,(\grainsizemax/0.1\,{\rm \mu m})$. But the cloud size/mass/density scale can be freely re-scaled, so long as the column matches this above, and because the observed molecular cloud population in the Galaxy all exhibit similar surface densities, this means the simulation can be rescaled to more or less any ``typical'' cloud size, with $\Lscale$ representing the characteristic gradient scale length of the cloud, i.e.:
 \begin{align}
 \Lscale^{\rm GMC} \sim 5\,{\rm pc}\,\left( \frac{{\rm M}_{\rm GMC}}{10^{4\,{\rm M_{\odot}}}}\right)^{1/2}\,\left(\frac{100\,{\rm M_{\odot}\,pc^{-2}}}{\Sigma_{\rm gas}^{\rm GMC}} \right)^{1/2}
 \end{align}
Similarly, we can estimate the characteristic timescale $t_{\rm acc} \equiv (2\,\Lscale/\langle {a}_{\rm eff} \rangle)^{1/2}$, where $a_{\rm eff}$ is the net acceleration of the initial homogeneous dust+gas mixture (i.e.\ the time to accelerate the gas past its initial scale length) in more physical units. Combining the scalings above, we have for the GMC runs that $t_{\rm acc} \sim 0.5\,t_{\rm ff,\,GMC} \sim 0.5\,\sqrt{3\pi/32\,G\,\langle \rho \rangle_{\rm GMC}} \sim 1\,{\rm Myr}\,({\rm M_{\rm GMC}}/10^{4}\,{\rm M_{\odot}})^{1/4}\,(\Sigma_{\rm gas}^{\rm GMC}/100\,{\rm M_{\odot}\,pc^{-2}})^{-3/4}$, where $t\_{\rm ff}$ is the cloud free-fall time. Note that this is not an accident, as the observed typical star formation efficiency used to estimate the radiation forces ($\accparammax$) in \S~\ref{sec:params} has been shown by many to around the critical value that should unbind the cloud in about a free-fall time \citep[see][]{grudic:sfe.cluster.form.surface.density,2018ApJ...859...68K,2019Natur.569..519K}.
 
Repeating this exercise for the HII simulation ICs, we have
\begin{align}
N_{\rm H}^{\rm HII-N} &\sim 3\times10^{21}\,{\rm cm^{-2}}\,\left(\frac{\internaldensity}{2\,{\rm g\,cm^{-3}}}\right)\,\left( \frac{\grainsizemax}{0.1\,{\rm \mu m}}\right) \\ 
N_{\rm H}^{\rm HII-F} &\sim 3\times10^{20}\,{\rm cm^{-2}}\,\left(\frac{\internaldensity}{2\,{\rm g\,cm^{-3}}}\right)\,\left( \frac{\grainsizemax}{0.1\,{\rm \mu m}}\right) 
\end{align}
and corresponding physical scales of 
\begin{align}
\Lscale^{\rm HII-N} &\sim 0.1\,{\rm pc}\,\left( \frac{N_{\rm H}^{\rm HII-N}}{3\times10^{21}\,{\rm cm^{-2}}} \right)\,\left( \frac{10^{4}\,{\rm cm^{-3}}}{\langle n_{\rm gas} \rangle} \right) \\ 
\Lscale^{\rm HII-F} &\sim 1\,{\rm pc}\,\left( \frac{N_{\rm H}^{\rm HII-F}}{3\times10^{20}\,{\rm cm^{-2}}} \right)\,\left( \frac{10^{2}\,{\rm cm^{-3}}}{\langle n_{\rm gas} \rangle} \right) 
\end{align}
where $\langle n_{\rm gas} \rangle$ is the mean gas density at a radius $r \sim \Lscale$ from the O-star powering the HII region. Finally for these systems the characteristic outflow acceleration time translates to $t_{\rm acc} \sim 10^{5}\,{\rm yr}\,(\Lscale/{\rm pc})^{3/2}$. 

Because it is reasonably well-specific in our setup, we will below occasionally present values of column densities and extinction from the simulations in physical units. However, because of the rescaling freedom above, we will present units of length and time in units of $\Lscale$ and $t_{\rm acc}$.

\section{Results}
\label{sec:results}

\subsection{General Behaviors}
\label{sec:results:general}

Fig.~\ref{fig:boxtower.dynamic.range} illustrates some of {the} key results generic to our ``full physics'' simulations (showing run {\bf GMC}): the RDIs grow rapidly, as expected according to linear theory, though the very longest-wavelength modes have growth timescales comparable to the flow time.\footnote{For details and comparisons of linear growth rates, see \citet{moseley:2018.acoustic.rdi.sims,seligman:2018.mhd.rdi.sims,hopkins:2019.mhd.rdi.periodic.box.sims}. Crudely, the {\em slowest}-growing modes of interest here, with wavelengths $\sim \Lscale$, have linear growth timescale $t_{\rm grow}/t_{\rm acc}\sim 0.2\,(\dustgas/\accparammax\,\sizeparammax)^{1/6}$ (assuming these correspond to the long-wavelength ``pressure free'' modes). Essentially all magnetosonic, gyro and/or shorter-wavelength RDIs have faster growth times. At different wavelengths $\lambda$ for different grain sizes $\grainsize$, their growth rates can be very approximately order-of-magnitude estimated (depending on the mode and wavelength regime, see \citealt{hopkins:2018.mhd.rdi}) by $t_{\rm grow}/t_{\rm acc} \sim 0.1\,(\lambda/\Lscale)^{0.5-0.7}\,(\grainsizemax/\grainsize)^{0.25-0.5}$, so the fastest-growing resolved modes in the box ($\lambda \sim 0.01\,\Lscale$ for smaller grains) often have $t_{\rm grow} \sim 0.001\,t_{\rm acc}$.} These produce non-linear fluctuations in the dust and gas density and filamentary structures (often with ``horsehead'' or ``cap'' morphologies at their endpoints) elongated along the vertical direction. 

Using our ``zoom-in'' boxes, we confirm as expected that all spatial scales are unstable to RDIs. The gross qualitative behavior and dominant RDIs are broadly similar over scales from $\sim 0.001-10\,\Lscale$; they change in form below $\lesssim 10^{-4}\,\Lscale$ in Fig.~\ref{fig:boxtower.dynamic.range} -- this transition corresponds to scales smaller than the collisional mean-free-path of the large dust grains, so the large grains form more diffuse non-linear structures. On small scales, the intuition from our previous studies of idealized periodic boxes in \citet{hopkins:2019.mhd.rdi.periodic.box.sims} applies: RDI growth rates become more rapid (growth rates scaling as $\propto k^{0.3-0.7}$, on average, depending on the mode), and dust continues to cluster, but the {\em gas} becomes less compressible.

\begin{figure}
    \centering
    \includegraphics[width=0.495\columnwidth]{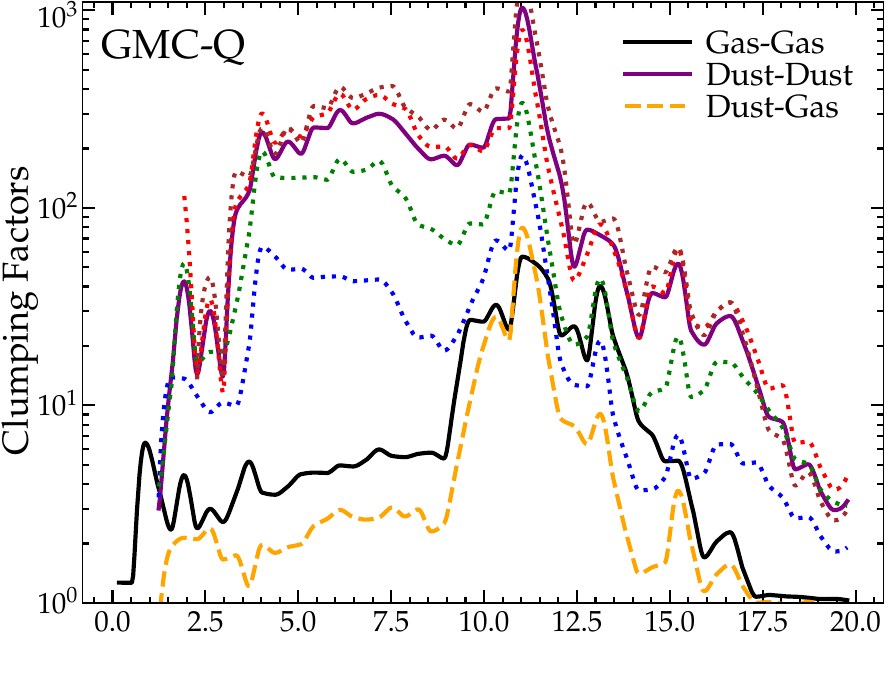}
    \hspace{-0.2cm}\includegraphics[width=0.50\columnwidth]{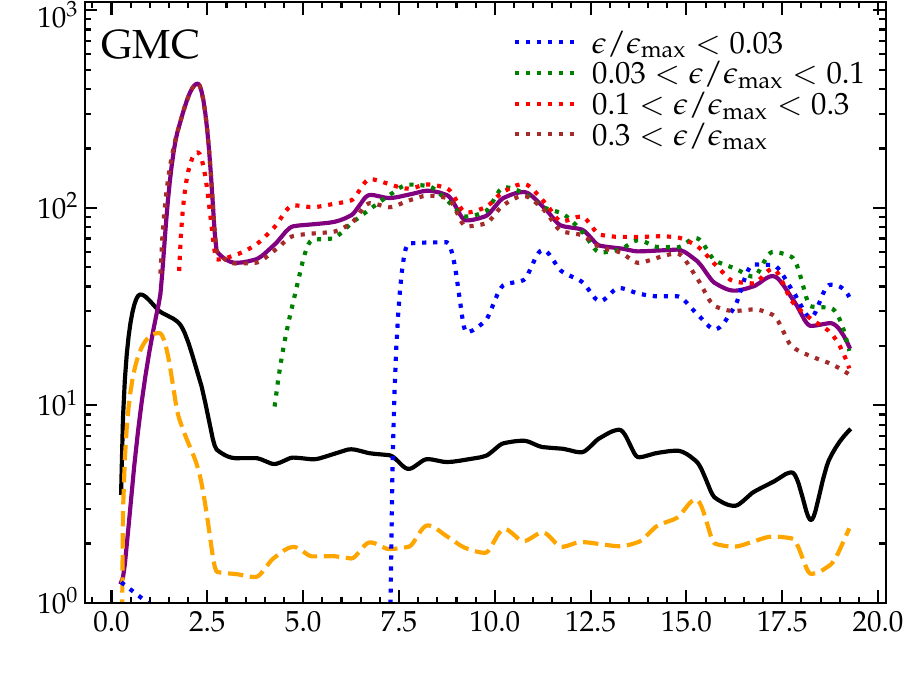}\\
    \includegraphics[width=0.50\columnwidth]{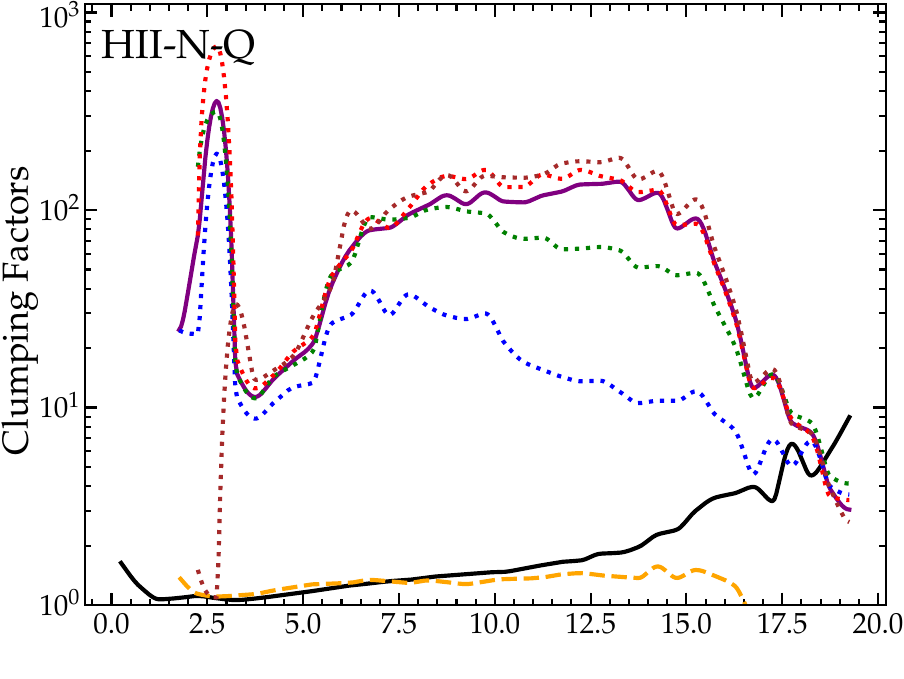}
    \hspace{-0.2cm}\includegraphics[width=0.50\columnwidth]{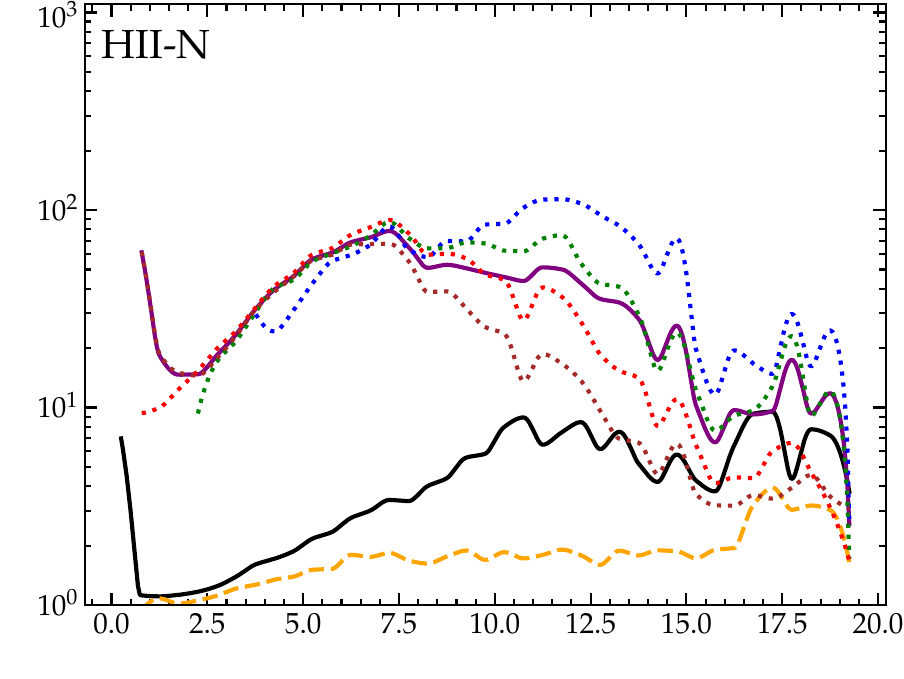}\\
    \includegraphics[width=0.50\columnwidth]{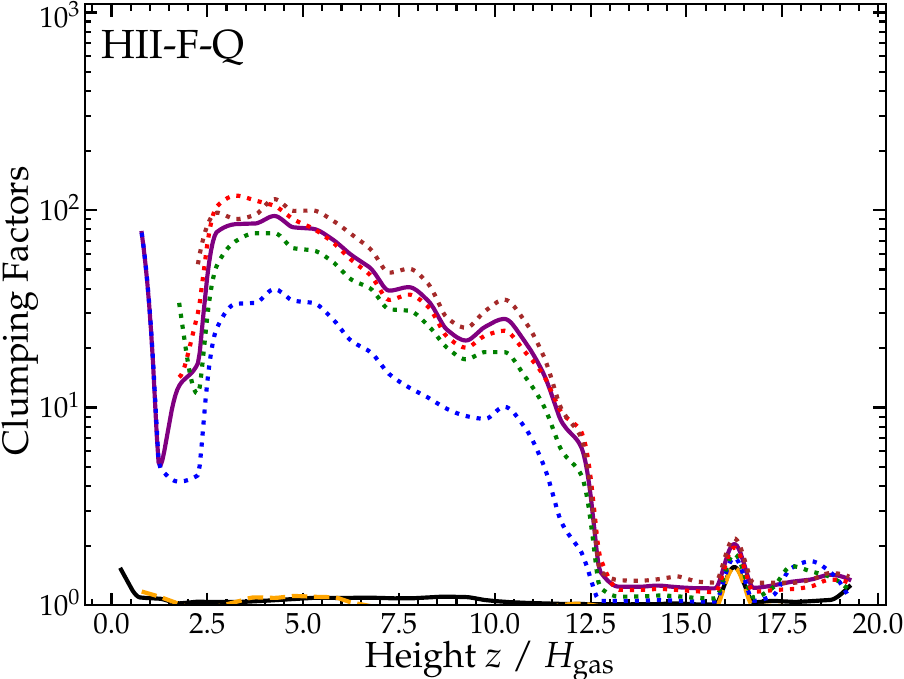}
    \hspace{-0.2cm}\includegraphics[width=0.50\columnwidth]{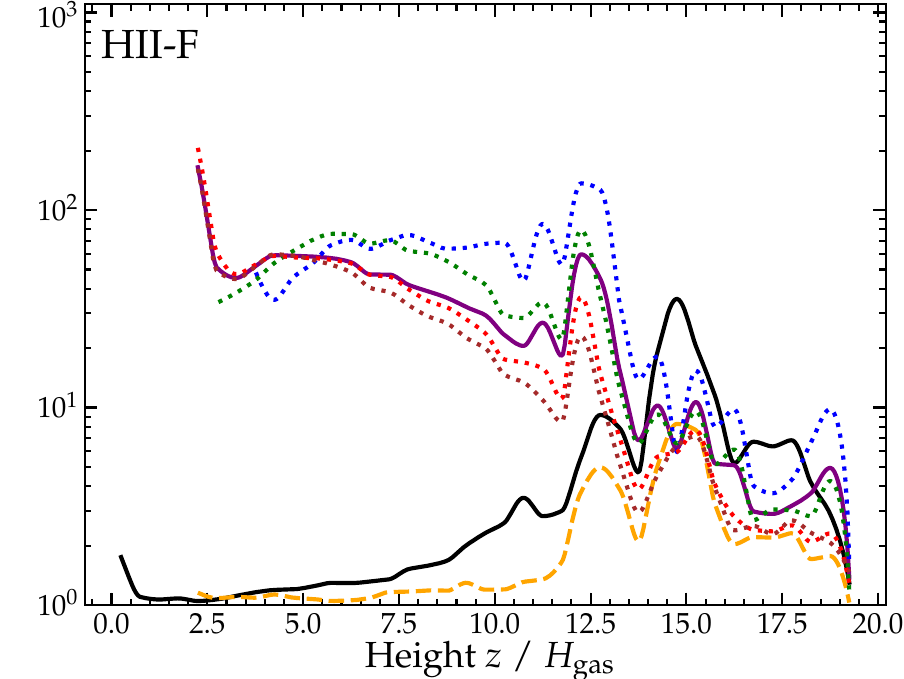}\\
    \vspace{-0.15cm}
    \caption{Clumping factor profiles (as Fig.~\ref{fig:bulk.props.vs.z.gmc.Q}). 
        {\em Left:} Q runs, {\rm Right:} no-Q runs. {\em Top:} GMC, {\em Middle:} HII-N, {\em Bottom:} HII-F. The qualitative trends with height and magnitude of the gas+dust turbulence are similar to those in Figs.~\ref{fig:bulk.props.vs.z.gmc.Q}-\ref{fig:bulk.props.vs.z.gmc}. Grains that dominate the opacity (large grains for $Q\propto \grainsize$, small for $Q\sim\,$constant) exhibit the strongest clumping. Dust-dust clumping is much stronger than gas-gas or gas-dust.
    \label{fig:clumping.factors}}
\end{figure}

\subsubsection{Outflow Launching \&\ Stratification}
\label{sec:results:general:outflow}

Fig.~\ref{fig:bulk.props.vs.z.gmc.Q} shows some key properties of the outflow in {\bf GMC-Q} at a few times the characteristic bulk acceleration timescale $t_{\rm acc} \equiv (2\,\Lscale/\langle {a}_{\rm eff} \rangle)^{1/2}$ where $a_{\rm eff} = F_{\rm total}/M_{\rm total}$ is the effective acceleration defined by the {\em total} upward force on all dust $F_{\rm total}$ and total dust+gas mass. We see the outflows are accelerated, and broadly reach the same height they would if we ignored the RDIs (treated dust and gas as perfectly-coupled). However some dust, and a significant fraction of the gas (tens of percent) lags behind, while a small fraction of the dust+gas move ``ahead'' of the expectation for a perfectly-coupled mixture, as the RDIs make the outflow highly inhomogeneous. Fig.~\ref{fig:bulk.props.vs.z.gmc} shows the same for {\bf GMC}: here $Q\sim\,$constant, so $\accsizedep = 1$ ($a_{\rm rad,\,dust} \propto \grainsize^{-1}$). The qualitative behaviors are similar, except in the behaviors of differently-sized grains.

If the dust and gas remained in {a} totally homogeneous steady-state, the dust would move relative to the gas at the equilibrium drift speed $\driftvel$ (exact expressions {are given} in \S~\ref{sec:ics}), crudely $|\driftvel| \sim |a_{\rm rad,\,dust}|\,\ts \propto \grainsize^{1-\accsizedep}$, for $a_{\rm gas,\,dust} \propto \grainsize^{-\accsizedep}$. So we would naively expect that in runs with $\accsizedep=0$ ($Q \propto \grainsize$), the drift velocity is smaller for smaller grains and so large grains will ``lead,'' while for runs with $\accsizedep=1$ ($Q\sim\,$constant), the drift velocity is $\grainsize$-independent, so dust will move in unison. But we see that with $\accsizedep=0$ (e.g.\ {\bf GMC-Q}), the grains are close to in-unison: small grains do ``lag,'' but by a very small amount. With $\accsizedep=1$ (e.g.\ {\bf GMC}), on the other hand, the small grains push noticeably ``ahead'' of the large grains. It appears that {\em non-linearly}, the {\em acceleration} of grains ($\grainsize$-independent in {\bf GMC-Q}, larger for smaller grains in {\bf GMC}) matters more for its motion as compared to its drift velocity. This arises naturally if the RDIs segregate grain sizes on micro-scales, so each obeys its own quasi-independent equilibrium solution (the local dust+gas mix accelerates at $\sim \dustgas(\grainsize)\,a_{\rm rad,\,dust}(\grainsize)$).

\begin{figure}
    \centering
    \includegraphics[width=0.50\columnwidth]{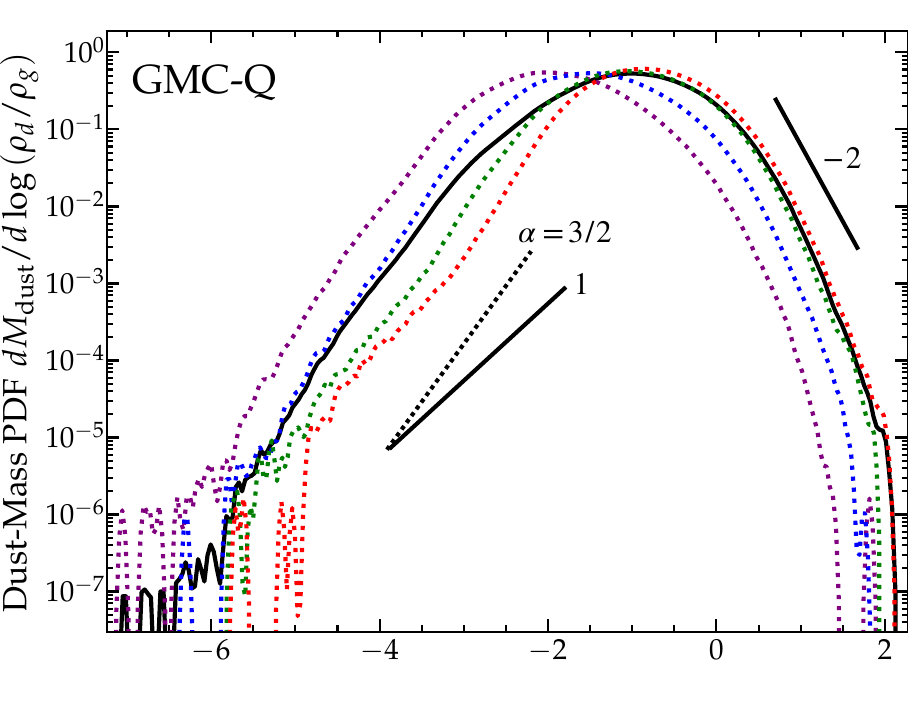}
    \hspace{-0.2cm}\includegraphics[width=0.50\columnwidth]{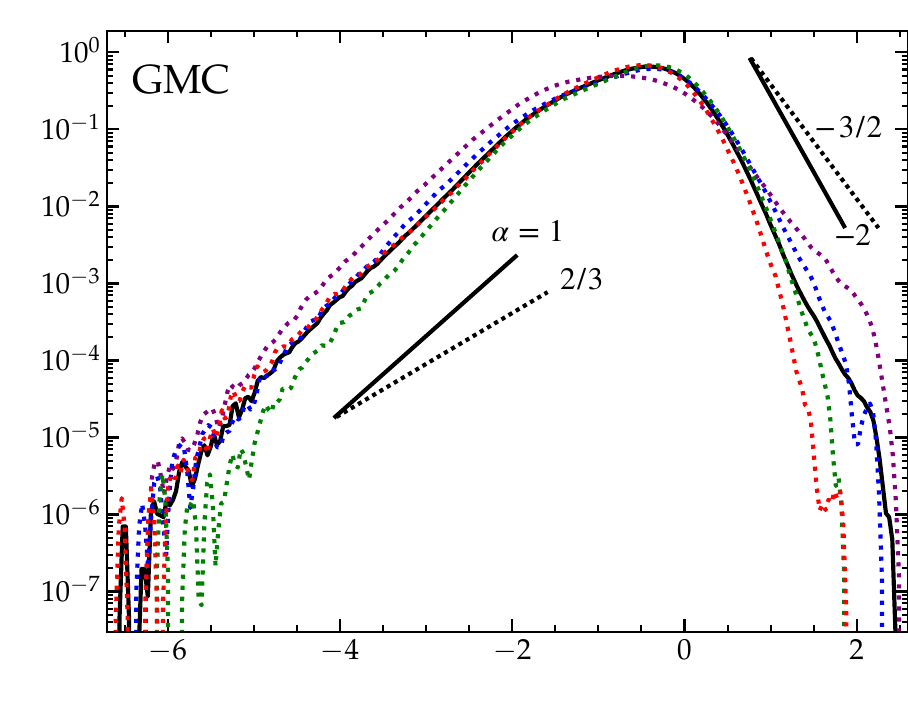}\\
     \vspace{-0.3cm}
    \includegraphics[width=0.50\columnwidth]{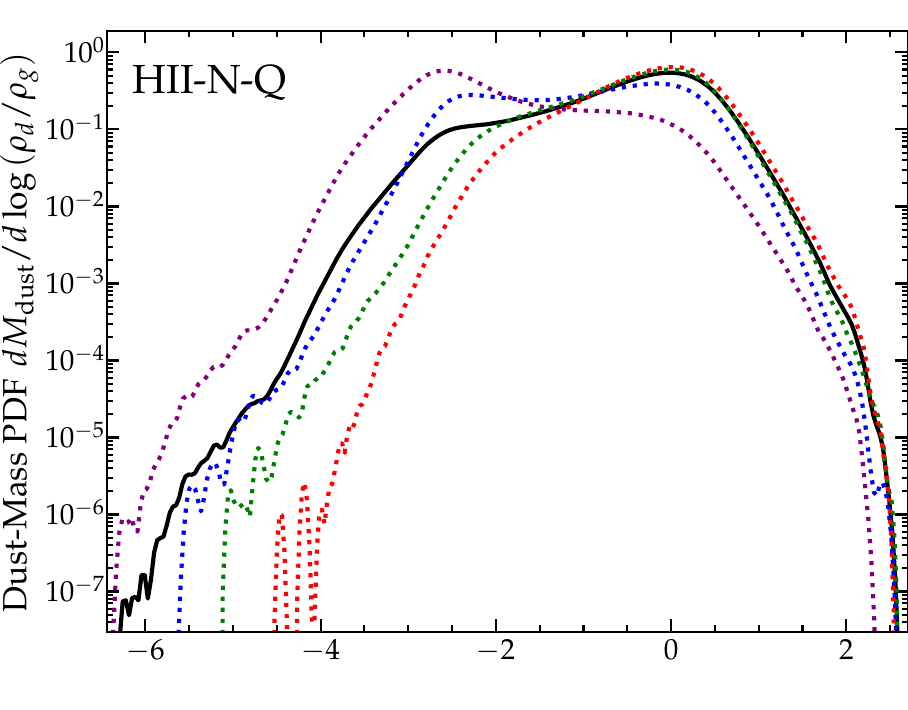}
     \hspace{-0.2cm}\includegraphics[width=0.50\columnwidth]{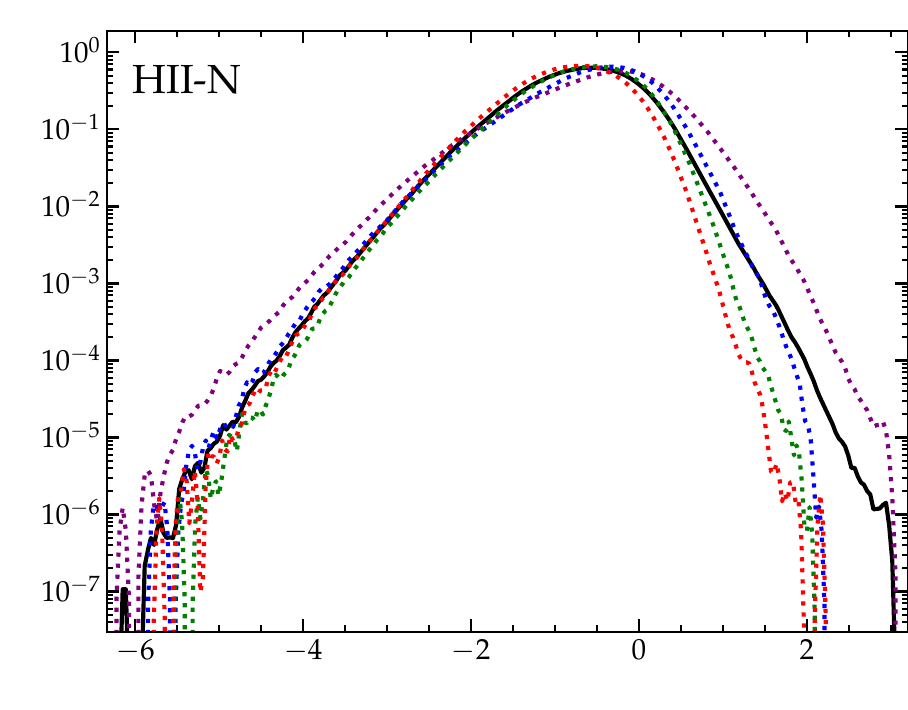}\\
     \vspace{-0.3cm}
    \includegraphics[width=0.50\columnwidth]{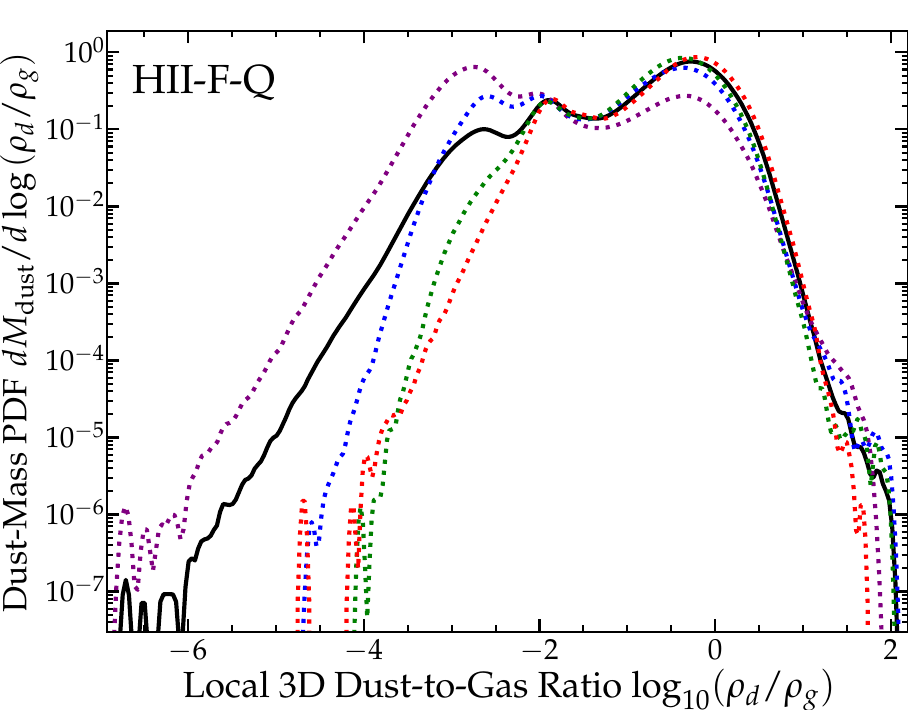}
    \hspace{-0.2cm}\includegraphics[width=0.50\columnwidth]{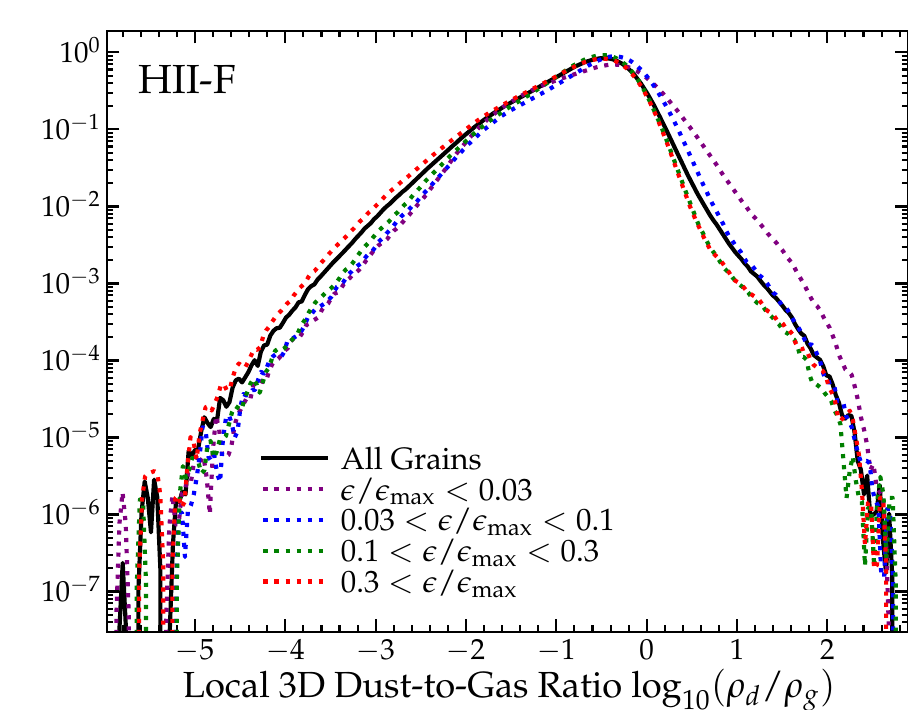}\\
    \vspace{-0.15cm}
    \caption{Dust-mass-weighted PDF of the local (resolution-scale) dust-to-gas ratio $(\dustden/\gasden)$ (around $3\,t_{\rm acc}$, in the range of $z$ containing $\sim 95\%$ of the dust mass). 
    {\em Left:} Q runs, {\rm Right:} no-Q runs. {\em Top:} GMC, {\em Middle:} HII-N, {\em Bottom:} HII-F. 
    The no-Q runs follow very robust power-law profiles, $dM_{\rm dust}/d\log{(\dustden/\gasden)} \propto (\dustden/\gasden)^{\alpha}$ with $\alpha\sim 1$ ($\alpha\sim-1.5$) at low (high) $(\dustden/\gasden)$. The Q runs are more lognormal with more curvature, and steeper, so over range here given steeper $\alpha \sim 1.5$ ($\alpha\sim-2$). Although the clumping factor of small grains is smaller in the Q runs (with $Q\propto \grainsize$), their width in $\log{(\dustden/\gasden)}$ is similar, but the fluctuations are skewed to lower absolute $\dustden/\gasden$. The PDF peaks at $\sim 1$, i.e.\ most grains locally ``see'' $\dustden \sim \gasden$, while the {\em volume}-weighted PDF peaks at $\sim \dustgas \sim 0.01$ (i.e.\ most random points in space have $\dustden/\gasden \sim 0.01$). In the tails, fluctuations can span a range of $\sim 10^{8}$.
    \label{fig:dust.gas.pdf}}
\end{figure}

\subsubsection{Turbulence \&\ Random Motions in The Outflows}
\label{sec:results:general:turbulence}

\begin{figure}
    \centering
    \includegraphics[width=0.88\columnwidth]{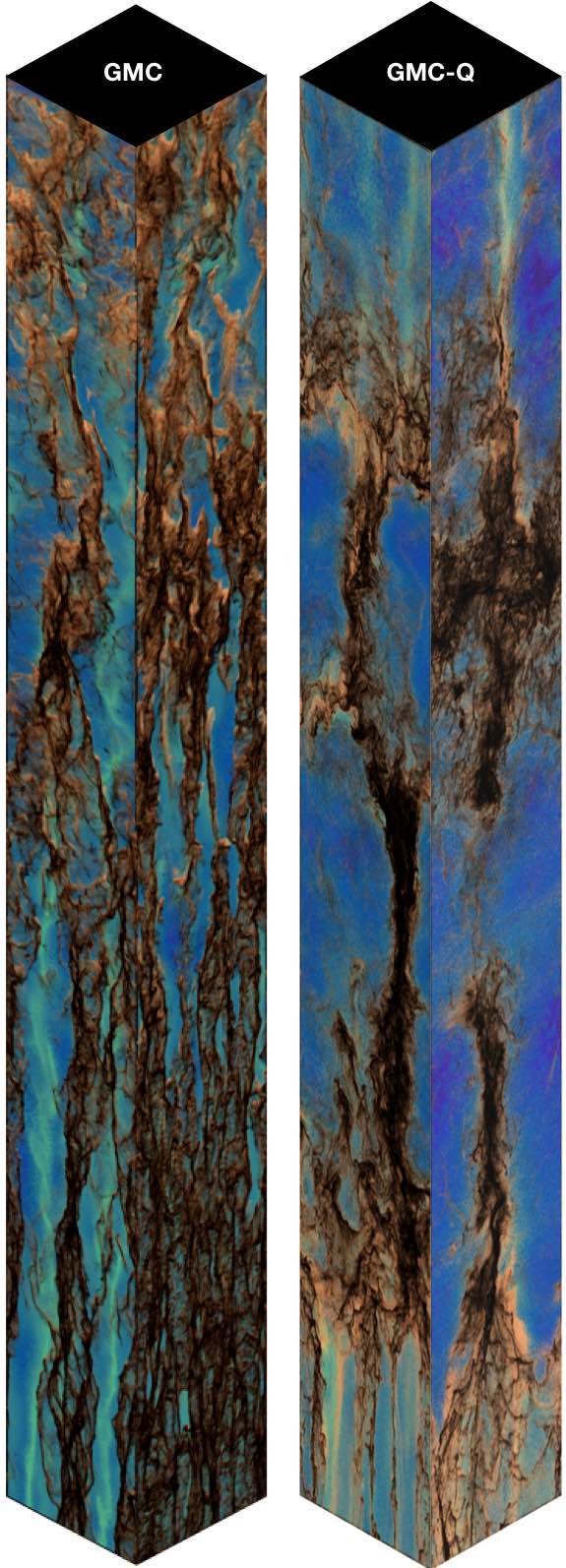}
    \caption{Isometric images of runs {\bf GMC} \&\ {\bf GMC-Q} (same style as Fig.~\ref{fig:boxtower.dynamic.range}), at the same time as Figs.~\ref{fig:bulk.props.vs.z.gmc.Q}-\ref{fig:bulk.props.vs.z.gmc}, surrounding the median dust+gas position, to show the visual morphology. The ``base'' of each is the box size $\Lscale\times\Lscale$. The optical properties of the grains, here whether $Q\sim\,$constant ({\bf GMC}) or $Q\propto \grainsize$ ({\bf GMC-Q}), change the dependence of dust acceleration on grain size and ultimate morphology of the system. The filamentary structures share many morphological features with observed dust in GMCs (see \S~\ref{sec:results:general:morphology}).
    \label{fig:isometric.view.GMC}}
\end{figure}
\begin{figure}
    \centering
    \includegraphics[width=0.85\columnwidth]{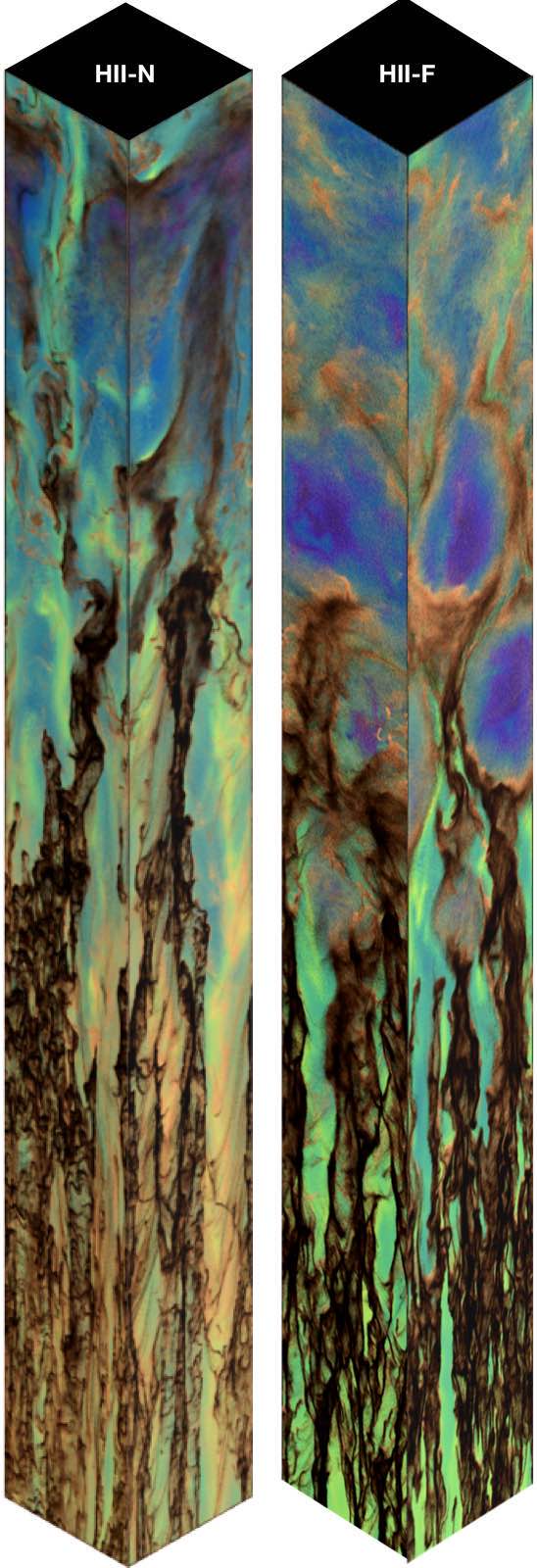}
    \caption{Isometric images of runs {\bf HII-N} and {\bf HII-F} as Fig.~\ref{fig:isometric.view.GMC} at the same time as Figs.~\ref{fig:turbulence}-\ref{fig:dust.gas.pdf} (and same scale $\sim \Lscale \times \Lscale \times 10\,\Lscale$). The visual morphologies share some common features with the {\bf GMC} cases, like the ubiquitous filamentary structure, but also feature more narrow filaments at the base and more cloud or cirrus-like structures at larger heights (\S~\ref{sec:results:general:morphology}).
    \label{fig:isometric.view.HII}}
\end{figure}

{The middle panels of} Fig.~\ref{fig:bulk.props.vs.z.gmc.Q}  \&\ Fig.~\ref{fig:bulk.props.vs.z.gmc} next compare the 3D velocity dispersion in dust and gas within slabs at a given height. While this is somewhat anisotropic (with modestly-larger dispersion in the $\hat{z}$ direction of outflow), the anisotropy is only an order-unity effect. This is because the magnetic RDIs involve a quite complicated spectrum of resonant angles in $\hat{\bf k}$ even at a single wavelength (see \citealt{hopkins:2018.mhd.rdi}, \S~5.2, Figs.~4-5), and those angles shift as a function of wavelength, isotropizing the injected power. As we show below, this does not occur if we neglect magnetic fields and have only the acoustic RDI, which features only a single resonance angle (across all wavelengths). The turbulence in {\em gas} saturates at trans-magnetosonic speeds $\langle| \delta \gasvel |^{2} \rangle^{1/2} \sim \langle |v_{\rm fast}| \rangle$ (in {\bf GMC}, $v_{\rm fast} \sim \vA$, but in the {\bf HII} runs $v_{\rm fast} \sim \cs$). Dust, being pressure-free, can easily reach higher global velocity dispersions compared to gas (i.e.\ $\langle |\delta \dustvel|^{2} \rangle^{1/2} \gtrsim \langle| \delta \gasvel |^{2} \rangle^{1/2}$). But we stress that this is the grain velocity dispersion on {\em large} scales (slabs of size $\sim \Lscale$), which involves coherent modes/structures and is much larger than the {\em local} micro-scale grain-grain approach velocities relevant for e.g.\ grain collisions or coagulation (which will be studied in more detail in Squire et al., in prep.). Note that the bulk outflow velocity, from the previous panel, is significantly larger $\sim \langle {a}_{\rm eff} \rangle\,t \sim 4\,\langle v_{\rm fast} \rangle$. Fig.~\ref{fig:turbulence} shows this is generically true across our simulations.

In all cases, runs with $\accsizedep=0$ ($Q\propto \grainsize$) exhibit stronger dust velocity fluctuations for larger grains, while runs with $\accsizedep=1$ ($Q\sim\,$constant) show $|\delta \dustvel|$ nearly-independent of $\grainsize$. This matches our simple expectation for characteristic local {\em relative} velocities to scale as $|\delta \dustvel | \propto  |{\bf a}_{\rm ext,\,dust} - {\bf a}_{\rm ext,\,gas}|\,\ts \propto  \grainsize^{1-\accsizedep}$. 

Examining our un-stratified ``zoom-in'' boxes allows us to confirm the behavior seen in \paperone, wherein the turbulent gas motions become smaller on progressively smaller scales and the gas becomes less compressible, while the dust dispersion drops more slowly. For e.g.\ {\bf GMC-U-M} and {\bf GMC-U-Q-M}, with box size $\sim 10^{-2}\,\Lscale$ of {\bf GMC}, $\langle| \delta \gasvel |^{2} \rangle^{1/2} \sim 0.05\,v_{\rm fast}$ (with $\langle| \delta \dustvel |^{2} \rangle^{1/2} \sim 0.5\,v_{\rm fast}$) while for {\bf GMC-U-S} ($\sim 10^{-4}\,\Lscale$), $\langle| \delta \gasvel |^{2} \rangle^{1/2} \sim 0.001\,v_{\rm fast}$; for ({\bf HII-N-U-Q-L}, {\bf HII-N-U-Q-M}, {\bf HII-N-U-Q-S}), with box sizes $\sim (0.5, 10^{-3}, 10^{-5})\,\Lscale$, $\langle| \delta \gasvel |^{2} \rangle^{1/2} \sim (1.4, 0.4, 0.15)\,v_{\rm fast}$ and $\langle| \delta \dustvel |^{2} \rangle^{1/2} \sim (2.4, 2.5, 1.6)\,v_{\rm fast}$. 
In some of these unstratified boxes (e.g.\ {\bf GMC-U-S}), after the initial rapid exponential RDI growth, the dust velocity dispersion continues to rise roughly linearly in time instead of saturating -- this owes to coherent dust filamentary structures accelerating at different speeds, as seen in e.g.\ the periodic-box simulations of \citet{moseley:2018.acoustic.rdi.sims}. This appears to be an artifact of the periodic box setup: in our global stratified boxes such structures rapidly separate/stratify (per \S~\ref{sec:results:general:outflow}) rather than remaining artificially ``adjacent'' to one another. Partially as a consequence, the ``zoom-in'' boxes do not appear to exhibit as clear a separation in the behavior of $\langle| \delta \dustvel |^{2} \rangle^{1/2}$ versus grain size between runs with $\accsizedep=0$ or $=1$.

\subsubsection{Dust Clustering \&\ Gas-Dust Clumping Factors}
\label{sec:results:general:clumping}

\begin{figure*}
    \centering
    \includegraphics[width=0.96\textwidth]{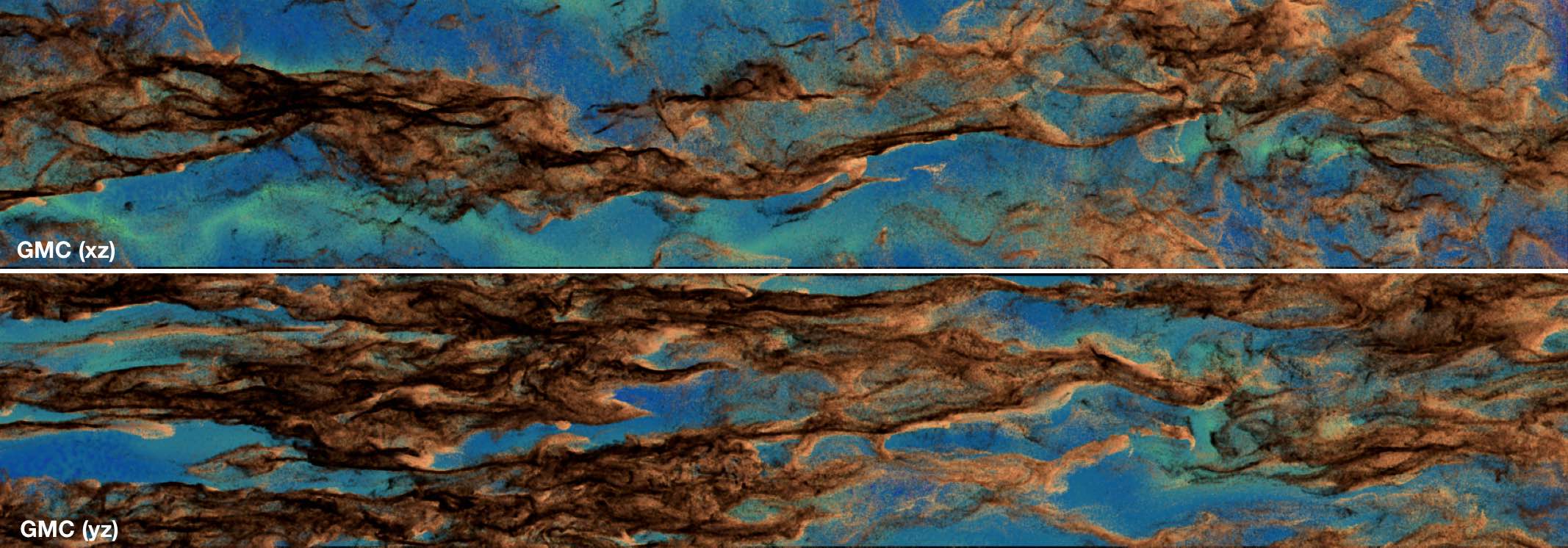}\\
    \includegraphics[width=0.96\textwidth]{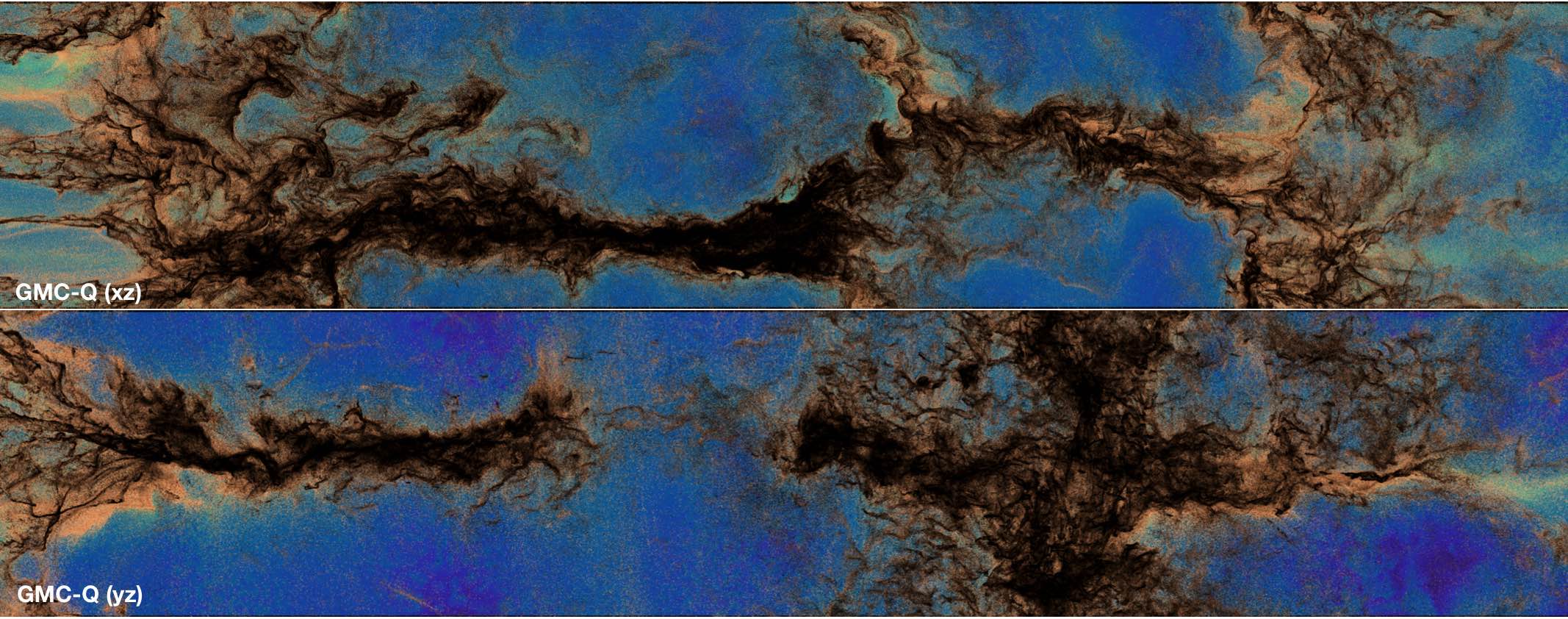}\\
    \includegraphics[width=0.96\textwidth]{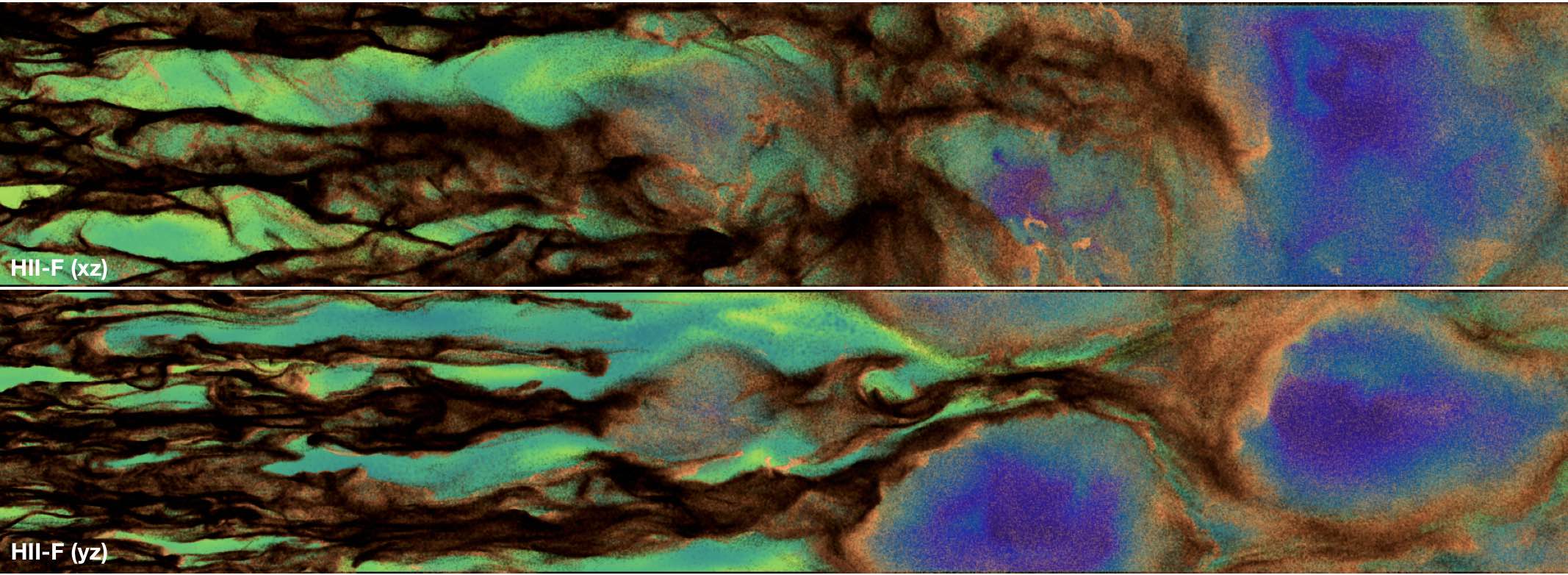}\\
    \includegraphics[width=0.96\textwidth]{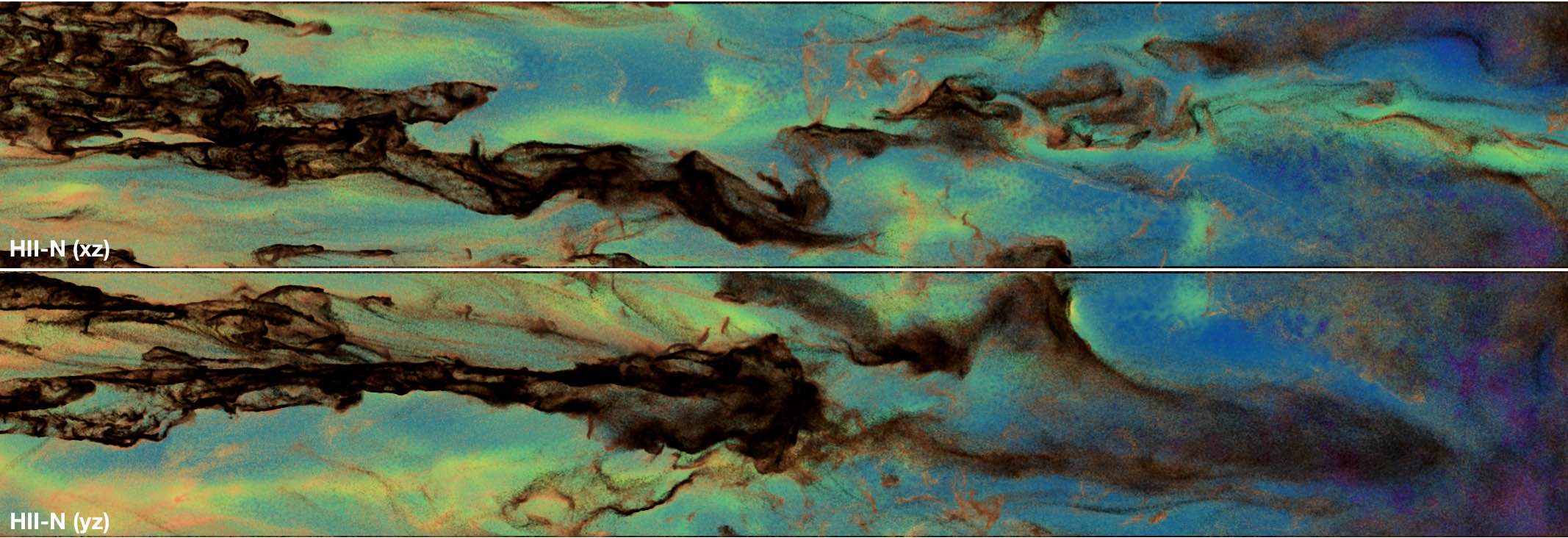}\\
    \vspace{-0.3cm}
    \caption{2D projection, zoomed-in to show more small-scale structure, of the images from Figs.~\ref{fig:isometric.view.GMC}-\ref{fig:isometric.view.HII}. The short axis of each image has scale $\Lscale$.
    \label{fig:side.view.multi}}
\end{figure*}

{The bottom panels of} Fig.~\ref{fig:bulk.props.vs.z.gmc.Q} \&\ Fig.~\ref{fig:bulk.props.vs.z.gmc} show the clumping factors -- a crucial quantity for understanding grain growth and chemistry -- of dust and gas versus height {for runs GMC-Q and GMC}. Fig.~\ref{fig:clumping.factors} extends this to our fiducial simulations. 
The clumping factor $C_{nm}$ for species $n$ and $m$ is the integral of the auto- (if $m=n$) or cross-correlation function of the local density, i.e.\ 
\begin{align}
C_{nm} &\equiv \frac{\langle \rho_{n}\,\rho_{m} \rangle_{V}}{\langle \rho_{n} \rangle_{V}\,\langle \rho_{m}\rangle_{V}} = \frac{V\,\int_{V} \rho_{n}({\bf x})\,\rho_{m}({\bf x})\,d^{3}{\bf x}}{\left[ \int_{V} \rho_{n}({\bf x})\, d^{3}{\bf x} \right] \, \left[ \int_{V} \rho_{m}({\bf x})\, d^{3}{\bf x} \right]} = 
\frac{\langle \rho_{n} \rangle_{M_{m}}}{\langle \rho_{n} \rangle_{V}}
\end{align}
where $V\equiv \int d^{3}{\bf x}$ is some volume, $\langle ... \rangle_{V}$ denotes the {\em volume-weighted} mean, and $\langle ... \rangle_{M_{m}}$ denotes the mean weighted by mass of species $m$.

The gas-gas clumping factor $C_{\rm gg}$ can be broadly understood as arising from the continuity equation (gas velocity fluctuations) akin to trans- or super-sonic turbulence with sonic Mach number $\mathcal{M}_{s}$ \citep{vazquez-semadeni:1994.turb.density.pdf,scalo:1998.turb.density.pdf,hopkins:2012.intermittent.turb.density.pdfs}. 
If $\gasden$ follows a lognormal PDF with variance $S_{\ln\rhogas} = \ln[1+(b\,\mathcal{M}_{s})^{2}]$, then $C_{\rm gg} \approx 1 + (b\,\mathcal{M}_{s})^{2}$, giving $C_{\rm gg} \sim 1-10$ for the values of $\mathcal{M}_{s}$ {here} (assuming $b \sim (1/5-1/3)$, plausible values for MHD turbulence; \citealt{konstantin:mach.compressive.relation,squire.hopkins:turb.density.pdf}). Thus, pressure effects mean that the gas clumping is somewhat restricted, and would not exceed what we would generically expect in supersonic turbulence.

The gas-dust cross-correlation factor $C_{\rm dg}$ is almost always $>1$, which indicates that dust and gas indeed remain {\em positively} correlated (as opposed to anti-correlated, which would give $C_{\rm dg} < 1$). That immediately distinguishes the scenario here from some classic ``turbulent concentration'' scenarios for incompressible turbulence with negligible dust back-reaction \citep[see][]{cuzzi:2001.grain.concentration.chondrules,yoshimoto:2007.grain.clustering.selfsimilar.inertial.range,bec:2009.caustics.intermittency.key.to.largegrain.clustering,pan:2011.grain.clustering.midstokes.sims,monchaux:2012.grain.concentration.experiment.review,hopkins:2013.grain.clustering} and some radiative Rayleigh-Taylor instabilities (discussed below), which would predict a strong anti-correlation. However, we generally have $C_{\rm dg}<C_{\rm gg}$ by a small amount, and $C_{\rm dg} \ll C_{\rm dd}$. This is consistent with the visual picture in Fig.~\ref{fig:boxtower.dynamic.range} and \citet{hopkins:2019.mhd.rdi.periodic.box.sims}: on large scales (which contain most of the power for {\em gas} turbulence and therefore density fluctuations), gas and dust broadly trace one another (with some weak de-coupling giving $C_{\rm dg}<C_{\rm gg}$ as some dust can drift through some dense gas structures). On small scales, dust continues to cluster, and is still positively correlated with gas density, but the gas behaves increasingly incompressibly (so the dust fluctuations become progressively larger {\em relative to gas} at high-$k$). This is particularly important for questions of e.g.\ dust growth and interactions with the ambient gas (e.g.\ grain growth via accretion), which will be enhanced in dense regions owing to the positive cross-correlation, but not (on average) to the degree the dust-dust or gas-gas clumping factors might imply.

The dust-dust clumping factor $C_{\rm dd}$ can be quite large, $\sim 10-1000$. Again note the same trend as with $\delta \dustvel$: for $\accsizedep=0$, large-grains exhibit larger clumping, while for $\accsizedep=1$, the results are weakly-$\grainsize$-dependent or even reversed. Since $C_{\rm dd}$ is just an integral of the dust-density PDF, we examine those PDFs directly in Fig.~\ref{fig:dust.gas.pdf}.\footnote{Note these are measured {\em at the resolution scale}: given the large dynamic range of the RDIs, the PDF width might continue to grow if we went to infinite resolution \citep[although see][]{hopkins:2013.grain.clustering}, and of course the fluctuations must become smaller if we average over larger spatial scales \citep[for quantitative examples, see][]{hopkins.2016:dust.gas.molecular.cloud.dynamics.sims,lee:dynamics.charged.dust.gmcs}.} The PDF shape and behavior in the ``tails'' can be especially important for some rare phenomena (as opposed to the peak or dispersion, which dominate $C_{\rm dd}$; \citealt{hopkins:totally.metal.stars,hopkins.conroy.2015:metal.poor.star.abundances.dust}), so it is significant that the PDFs are notably non-Gaussian. At low $(\dustden/\gasden)$, the PDFs (especially for runs with $\accsizedep=1$, $Q\sim\,$constant) approximately follow $d M_{\rm dust}/ d\ln{(\dustden/\gasden)} \propto (\dustden/\gasden)^{\alpha}$ with $\alpha \approx +1$, i.e.\ $d M_{\rm dust} / d (\dustden/\gasden) \sim \,$constant. At high $(\dustden/\gasden)$, $\alpha \approx -1.5$. That, in turn, means that with probability $\gtrsim 10^{-7}$ we see events with $(\dustden/\gasden) \sim 10^{-6} - 10^{3}$ ($\sim 10-30$ times the nominal ``1$\sigma$'' core PDF width). We also see that even where the clumping factor is smaller for small grains (e.g.\ $\accsizedep=0$ runs), the PDFs of $(\dustden/\gasden)$ for small grains are comparably broad or even more broad in log-space as those of large grains: they are simply biased towards lower $(\rhodust/\rhogas)$ (which gives lower $C_{\rm dd}$). This can follow if grains are ``expelled'' from certain regions by e.g.\ vorticity \citep{yakhot:1997.scalar.field.turb.pdfs,monchaux:2010.grain.concentration.experiments.voronoi,hopkins:2013.grain.clustering,colbrook:passive.scalar.scalings} as grains with large mean-free path are less-efficiently expelled.

One striking feature of the PDFs of $(\dustden/\gasden)$ is that they peak order-of-magnitude around $(\dustden/\gasden) \sim 1$, i.e.\ $\dustden \sim \gasden$ (100 times the mean dust-to-gas ratio $\dustgas \sim 0.01$). As shown in idealized experiments in \citet{hopkins:2019.mhd.rdi.periodic.box.sims}, and confirmed by our own experiments with varied $\dustgas$ here, this remains robust regardless of the initial $\dustgas$. In other words, dust in lower-$\dustgas$ runs clumps more strongly relative to its initial conditions, while in higher-$\dustgas$ cases it clumps less strongly, giving peak $(\dustden/\gasden) \sim 1$. These are the {\em dust-mass-weighted} PDFs, so that means a substantial fraction of the dust mass ends up clumping until the local $\dustden \sim \gasden$ {\em in the local vicinity of the dust grains}, beyond which point it is harder to significantly increase the dust density (the PDF turns over).\footnote{The {\em volume} weighted PDF $P_{V}$ of $(\dustden/\gasden)$ is skewed to lower $(\dustden/\gasden)$ by one power of $\dustden^{-1}$, by construction. $P_{V}$ peaks around (or somewhat below) $\dustden \sim \dustgas\,\gasden \sim 0.01\,\gasden$, as expected, since $\dustgas \equiv \langle \dustden \rangle_{V} / \langle \gasden \rangle_{V}$ by definition.} This is plausible from simple linear and non-linear considerations. In linear theory the growth rates of the RDIs depend on ${\dustgashat} \equiv \dustgas/(1+\dustgas)$ to some positive power \citep{squire.hopkins:RDI,hopkins:2017.acoustic.RDI} -- so RDI growth rates increase as the local $(\dustden/\gasden)$ increases until they saturate when $\dustden\sim\gasden$. Moreover, non-linearly, once the dust dust dominates the local density, the ``confining'' force from gas pressure which helps to retain coherent small-scale structures becomes weaker \citep{hopkins:2013.grain.clustering}.

Examination of our unstratified ``zoom-in'' simulations allows us to immediately verify that the gas is increasingly incompressible on smaller scales: e.g.\ ({\bf GMC}, {\bf GMC-U-M}, {\bf GMC-U-S}), with box sizes $(\gg 1,\,10^{-2},\,10^{-4})\,\Lscale$, have $C_{\rm gg}-1 \sim (1,\,0.02,\,10^{-5})$. This corresponds roughly to our analytic expectation in \paperone\ for saturation of the gas turbulence when the decay rates become comparable to RDI driving rates (giving a dispersion in gas density $\propto \lambda / (\cs\,t^{\rm rdi}_{\rm grow}[\lambda]) \propto \lambda^{0.5-0.66}$ where $\lambda$ is the scale and $t_{\rm grow}^{\rm rdi}[\lambda]$ the RDI growth time on that scale). The dust clumping (or equivalently, width of the dust density or dust-to-gas ratio PDF) in these zoom-in runs is a weaker function of scale: for e.g.\ ({\bf GMC}, {\bf GMC-U-M}, {\bf GMC-U-S}) we have $C_{\rm dd}-1 \sim (200,\,10,\,0.2)$ and ({\bf HII-N-U-Q-L}, {\bf HII-N-U-Q-M}, {\bf HII-N-U-Q-S}) with sizes $(0.5,\,10^{-3},\,10^{-5})\,\Lscale$ have $C_{\rm dd}-1 \sim (240,\,1.9,\,0.08)$ (with a dispersion in $\log_{10}(\dustden/\gasden)$ of $\sim (1.3,\,0.5,\,0.13)\,$dex, close to $\propto \lambda^{0.2}$). Interestingly, in most of the unstratified ``zoom-in'' runs, including the ``-Q'' ($\accsizedep=0$) variations, small grains exhibit larger clumping factors on these much smaller scales compared to large grains, unlike the behavior for $\accsizedep=0$ in our stratified global boxes in Fig.~\ref{fig:clumping.factors}. This owes in part to the fact that the ``zoom-in'' box sizes $\lesssim 10^{-2}\,\Lscale$ become comparable to or smaller than the collisional mean-free path or ``stopping length'' of the largest grains,\footnote{We will refer to the dust ``mean free path'' to collisions or ``stopping length, defined as the distance over which the dust must travel relative to the gas before being significantly decelerated by drag forces, $\lambda_{\rm mfp}^{\rm grain} \sim \driftvelmag\,\ts$.} so they cannot fully capture the clustering.


Note that, because of the continued clustering seen in ever-smaller boxes, we hesitate to define any specific threshold for dust ``clumps'' or ``clusters'' as objects (hence focusing on robust statistics such as the clumping factor and/or auto/cross-correlation functions, which do not depend on a specific physical or observational definition of a ``clump''). However in more detailed studies of dust-dust interactions, which as we noted could be dramatically enhanced by the clustering above, it might be important to consider this (see \citealt{squire:2022.acoustic.rdi.size.spectrum} for additional discussion). In future work it will be particularly interesting to examine in more detail the effects of this grain-grain clustering on grain collisions and subsequent coagulation (and/or bouncing or shattering); not only will the clustering enhance these interaction rates, but the conditions and relative velocities are quite radically distinct from those assumed in classic studies of the grain collision/coagulation kernel usually modeling passive grains in a subsonic turbulent flow without any radiative forcing (compare e.g.\ \citealt{pan:2010.grain.velocity.sims,pan:2011.grain.clustering.midstokes.sims,pan:2013.grain.relative.velocity.calc}). As shown in e.g.\  \citet{squire:2022.acoustic.rdi.size.spectrum}, radiative accelerations alone could significantly alter some of these historical conclusions. It will also be interesting to include other physics that may influence the dynamics of grains indirectly via their evolution over the timescales here, such as growth from accretion from the ISM itself, or sputtering (although we find the local dust-gas relative velocities here in the dust clumps are quite modest and so do not expect sputtering to be significant in the scenarios simulated here). However we stress that the actual spatial scales for coagulation of grains are vastly smaller than those resolved here (and of course the ``clumps'' in our simulations represent regions of locally-enhanced grain and/or gas density, not regions where grains have necessarily coagulated), and many uncertainties remain in growth models which depend on grain chemistry in a way we are not yet explicitly modeling in our simulations.

\subsection{Morphological Structure}
\label{sec:results:general:morphology}

\begin{figure*}
    \centering
    \includegraphics[width=0.97\textwidth]{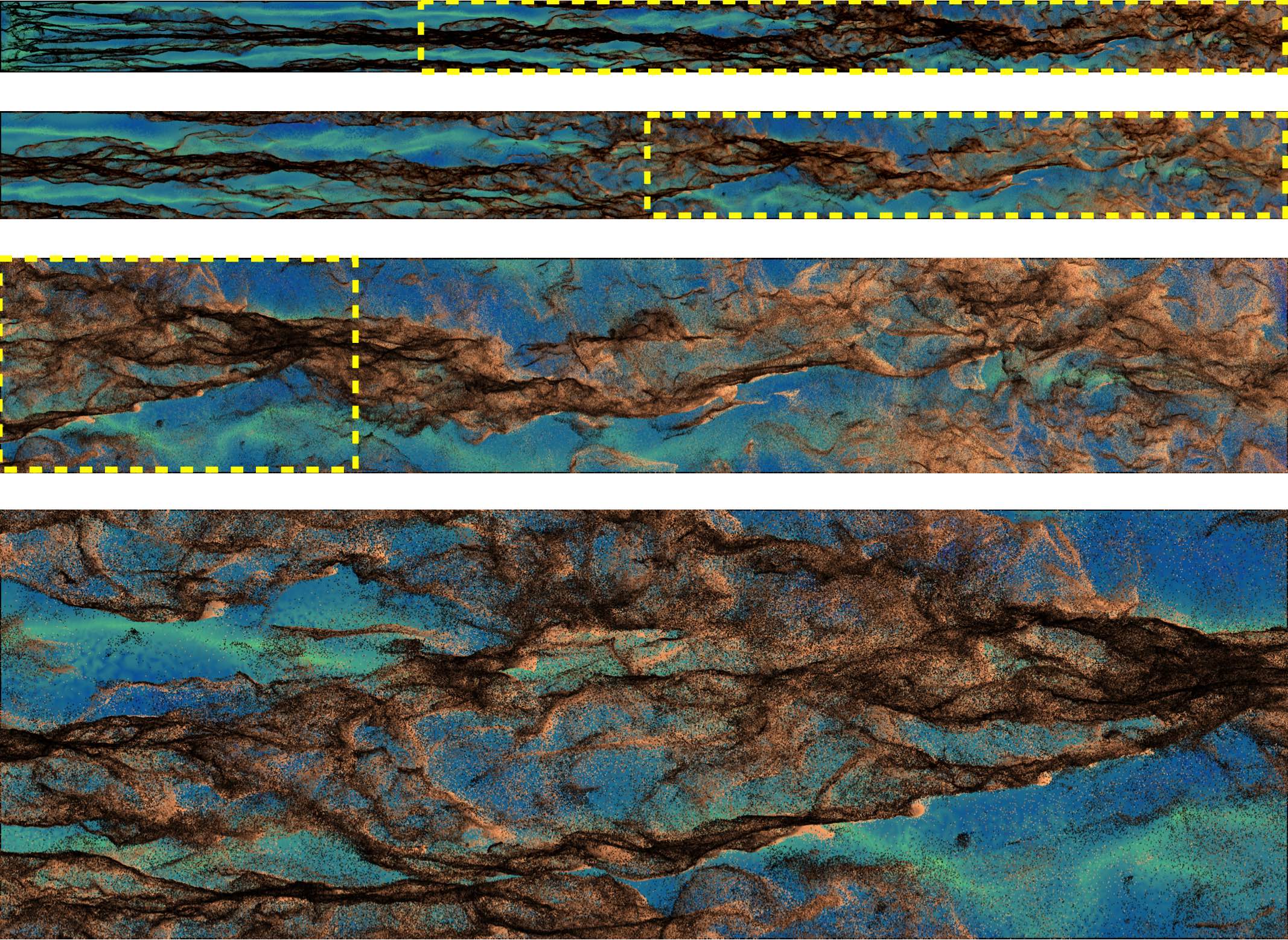}
    \caption{Continued zoom-in of the image of {\bf GMC} in Fig.~\ref{fig:side.view.multi} to show further small-scale structure below that shown in the previous Figure, resolved in the same simulation. From top to bottom, each dashed rectangle corresponds to the area of the image immediate below (short axis in each has scale $\Lscale$). The structure is almost self-similar, with sub-filaments embedded within larger filamentary structures, reflecting the structure of the RDIs across different scales.
    \label{fig:resolution.zoomin}}
\end{figure*}

Figs.~\ref{fig:isometric.view.GMC} \&\ \ref{fig:isometric.view.HII} show isometric projections of {the GMC and HII} fiducial high-resolution boxes, and Fig.~\ref{fig:side.view.multi} zooms in on these in separate face-on projections (with a further ``zoom-in'' on substructure in Fig.~\ref{fig:resolution.zoomin}). We discuss these in relation to observations below, but here {we} describe some robust physical effects. 

Obviously, the dust and gas form characteristic structures elongated in the vertical (outflow or $\hat{z}$) direction. This is seen generically even in idealized periodic cubic boxes \citep{moseley:2018.acoustic.rdi.sims,hopkins:2019.mhd.rdi.periodic.box.sims}, so owes not to the stratified or elongated nature of the simulation boxes here, but to a very simple phenomenon. The vertical (radiation pressure) force scales with the dust column/opacity, while the total mass is gas-dominated, so fluctuations in $\dustden/\gasden$ along different ``columns'' translate to variations in the mean vertical acceleration, which quickly shear any structures along the outflow direction $\hat{z}$. 

As the outflow launches, the morphology develops in the generic \citet{zeldovich:1970.pancakes}-style manner: initially, the fastest-growing resonance leads to collapse of the dust from uniform 3D to 2D sheet-like structures (on large scales here, this is often the aligned ``quasi-sound'' mode, which has fastest growth rates when $\hat{\bf k} \approx \bhat$ for the trans/super-sonic drift conditions here with $\tauparam \gtrsim 1$, so the sheets form in the $\hat{y}-\hat{z}$ plane);\footnote{Because the RDIs are generically unstable at all wavenumbers $k$ with growth rates that increase with $k$, short-wavelength (high-$k$) modes grow first (generating e.g.\ multiple parallel sheets/filaments on small scales), then merge into larger structures as longer-wavelength modes grow. The same occurs in more idealized simulations (see \citet{hopkins:2019.mhd.rdi.periodic.box.sims}, Figs.~4 \&\ 11).} secondary modes with nearly-perpendicular resonant angles generate ``corrugation'' in the sheets which break into quasi-1D filaments (these are often the magnetosonic MHD-wave RDIs, which for super-sonic drift have resonant growth rates when $\hat{\bf k}$ lies near the plane perpendicular to $\bhat$); finally tertiary modes (e.g.\ the gyro RDIs or parasitic instabilities) break these up into smaller clumps (point-like 0D structures). Again this is seen even in idealized simulations (without stratification), and the hierarchical 3D$\rightarrow$2D$\rightarrow$1D$\rightarrow$0D process is generic to any anisotropic collapse/condensation process \citep[see e.g.][]{hopkins:frag.theory}, so this is not surprising. However, as we show below, in simulations where we neglect dust charge, the lack of any more complex resonances with different preferred directions means that the process is largely arrested at the ``sheet'' stage.

It is often (though not always) the case that the ``base'' of the outflow features a larger number of thinner, more-vertical dust columns, while larger heights feature a smaller number of thicker structures (although note these have substantial sub-structure; Fig.~\ref{fig:resolution.zoomin}), and the ``uppermost'' end of the outflow features more diffuse cloud-like structures (see e.g.\ Fig.~\ref{fig:isometric.view.HII}). This {\em is} driven by the stratification, as a couple of key properties depend on vertical height $z$: (1) the $\tauparam$ parameter (ratio of Lorentz-to-drag force on dust) increases in our ICs (approximately as $\tauparam \propto \rho^{-1/2} \propto \exp{(z/2\Lscale)}$), (2) the dust collisional mean free path or ``stopping length,'' $\lambda^{\rm grain}_{\rm mfp} \sim \driftvelmag\,\ts \propto \rho^{-(1/2-1)}$ (for trans/super-sonic $\driftvelmag$) also increases. Effect (1) means that in linear stages, the dominant modes change with $z$, from the nearly-acoustic (more weakly-magnetized) mid-$k$ modes that produce nearly-vertical dust ``jets'' as shown in non-magnetized dust simulations in \citet{moseley:2018.acoustic.rdi.sims} at lower $\tauparam$ (low-$z$); to the mix of quasi-sound, magnetosonic \&\ \Alf\ MHD-wave instabilities that merge into larger more ``wavy'' or ``bent'' filaments with more internal structure (see Fig.~\ref{fig:resolution.zoomin}) at larger-but-not-extremely-large $\tauparam$ (intermediate $z$); to the ``cosmic-ray-like'' instabilities which dominate at very high-$\tauparam$ (high-$z$) and, as shown in \citet{hopkins:2019.mhd.rdi.periodic.box.sims}, excite \Alf{ic} fluctuations which scatter dust grains akin to resonant and non-resonant cosmic ray streaming instabilities \citep{1975MNRAS.172..557S,bell.2004.cosmic.rays}, isotropizing the dust velocity distribution function and dispersing the grains. Effect (2) means that even well into non-linear stages (where the dust has a substantial velocity dispersion in the $\hat{x}-\hat{y}$ plane), increasing $\lambda_{\rm mfp}^{\rm grain}$ with $z$ naturally leads to more ``dispersed'' structures at higher-$z$. 

Finally, we also see that the optical properties of grains, specifically whether $\accsizedep=0$ or $=1$ (how $Q$ depends on $\grainsize$) has significant effects on the detailed morphology. We discuss this further below, but it should not be surprising. The dependence of $Q(\grainsize)$ determines how the radiative grain acceleration $|{\bf a}_{\rm rad}| \propto \grainsize^{\accsizedep}$ scales with $\grainsize$, which determines how the drift velocity $\driftvelmag \sim |{\bf a}_{\rm rad}|\,\ts$ scales with $\grainsize$. As a result, runs {\bf GMC} and {\bf GMC-Q} differ by a factor {of} $\sim \grainsize^{-1} \sim 100$ in the drift speed of the smallest grains. That, in turn, directly changes the growth rates of the RDIs, the mode geometry (changing the direction of the resonant mode angles $\hat{\bf k}$, which depend on $\driftvelmag$), and the wavelengths of some resonances (e.g.\ the gyro-RDIs), per \citet{hopkins:2018.mhd.rdi}. This also changes whether certain modes are even present: the difference in drift speed in {\bf GMC} vs.\ {\bf GMC-Q} for the smallest grains translates to super-vs-sub-sonic drift, which changes whether the fast-magnetosonic MHD-wave RDI is unstable. Likewise, $\tauparam$ depends on $\driftvelmag$, and determines whether e.g.\ the cosmic-ray like modes can grow.

\begin{figure}
    \centering
    \includegraphics[width=0.97\columnwidth]{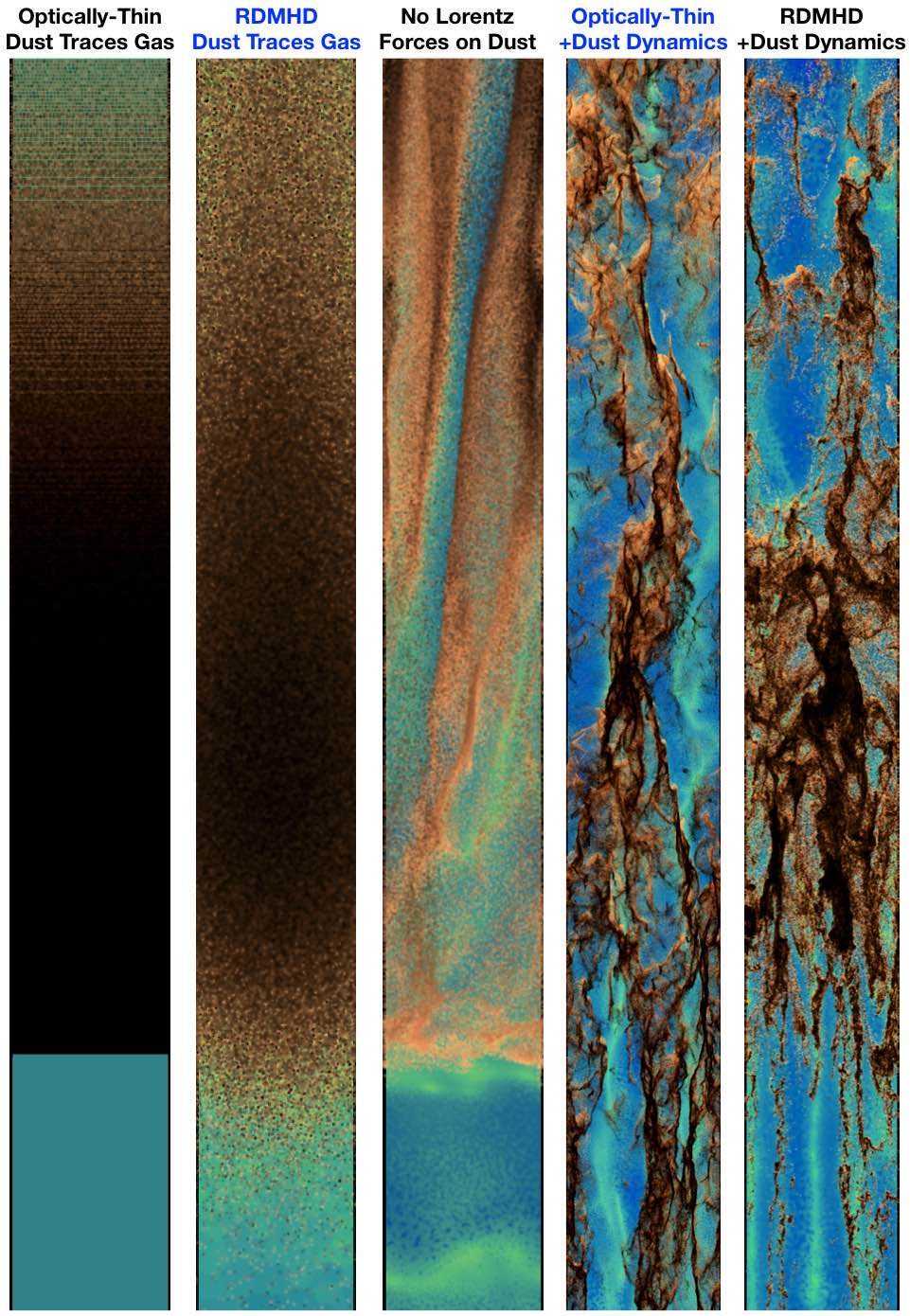}
    \caption{Image of the outflow at similar time $t\sim 3\,t_{\rm acc}$ as Fig.~\ref{fig:resolution.zoomin}, comparing otherwise-identical simulations of run {\bf GMC} with different physics (see \S~\ref{sec:physics}). {\em Left:} An ``optically-thin'' (uniform radiation flux; see \S~\ref{sec:rad}) simulation where we assume dust simply moves exactly with the gas (i.e.\ ignore separate dust+gas dynamics). The system is vertically accelerated perfectly-uniformly. {\em Second:} A ``full radiation-dust-MHD'' simulation (evolving the flux explicitly) simulation where dust moves with exactly with gas. Finite optical-depth effects ``smear out'' the ``base'' of the outflow, but the system accelerates stably and no structure develops. {\em Middle:} A simulation including dust dynamics but removing the Lorentz forces on grains (e.g.\ treating dust as neutral, so it does not see magnetic fields). While RDIs can and do develop, producing some structure, the only available RDI is the acoustic RDI which has a vastly-simpler resonant structure and single resonant angle (corresponging to the common angle of the mode seen here). {\em Second-from-right:} Optically-thin simulation with our full dust dynamics (restoring the grain charge/Lorentz forces). {\em Right:} Full RDMHD simulation with the full dust dynamics (albedo $A=1$). The differences between full-RDMHD and optically-thin cases are clearly second-order compared to the effect of dust dynamics.
    \label{fig:rdi.physics.demo}}
\end{figure}

\begin{figure}
    \centering
    \includegraphics[width=0.473\columnwidth]{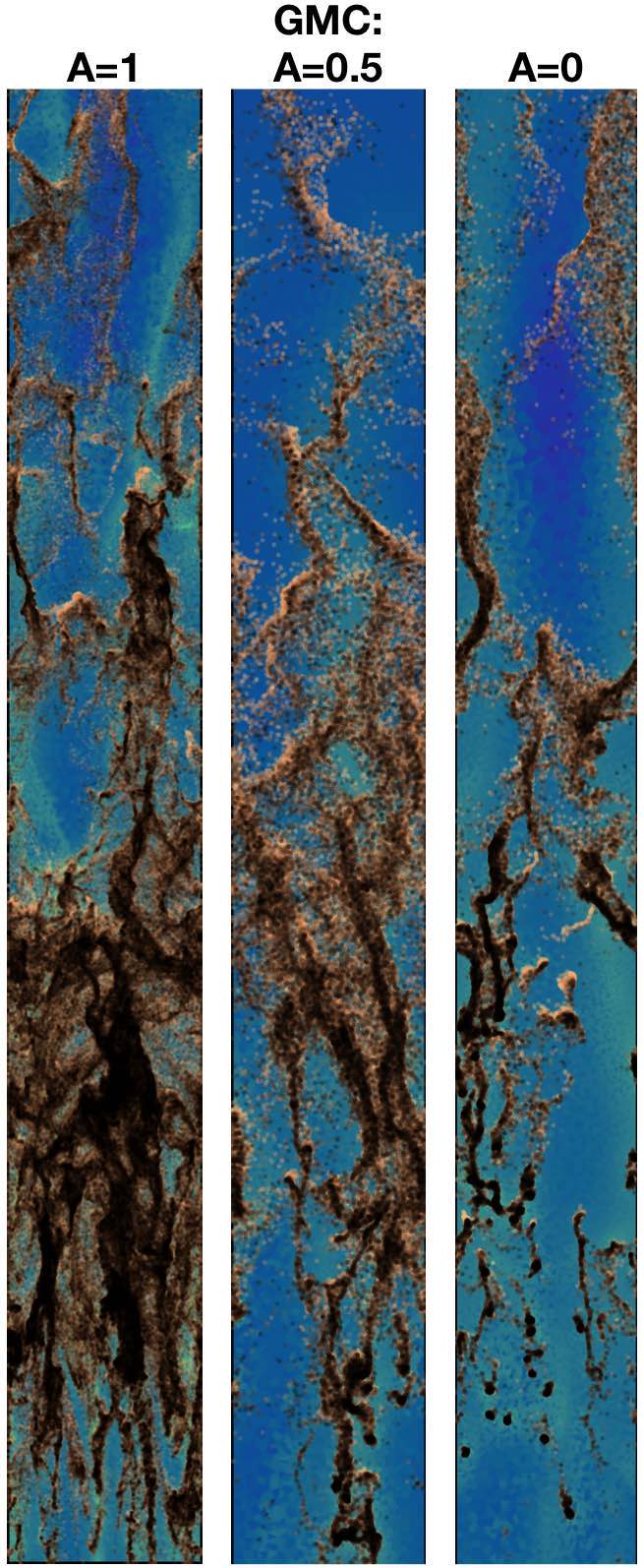}
    \includegraphics[width=0.48\columnwidth]{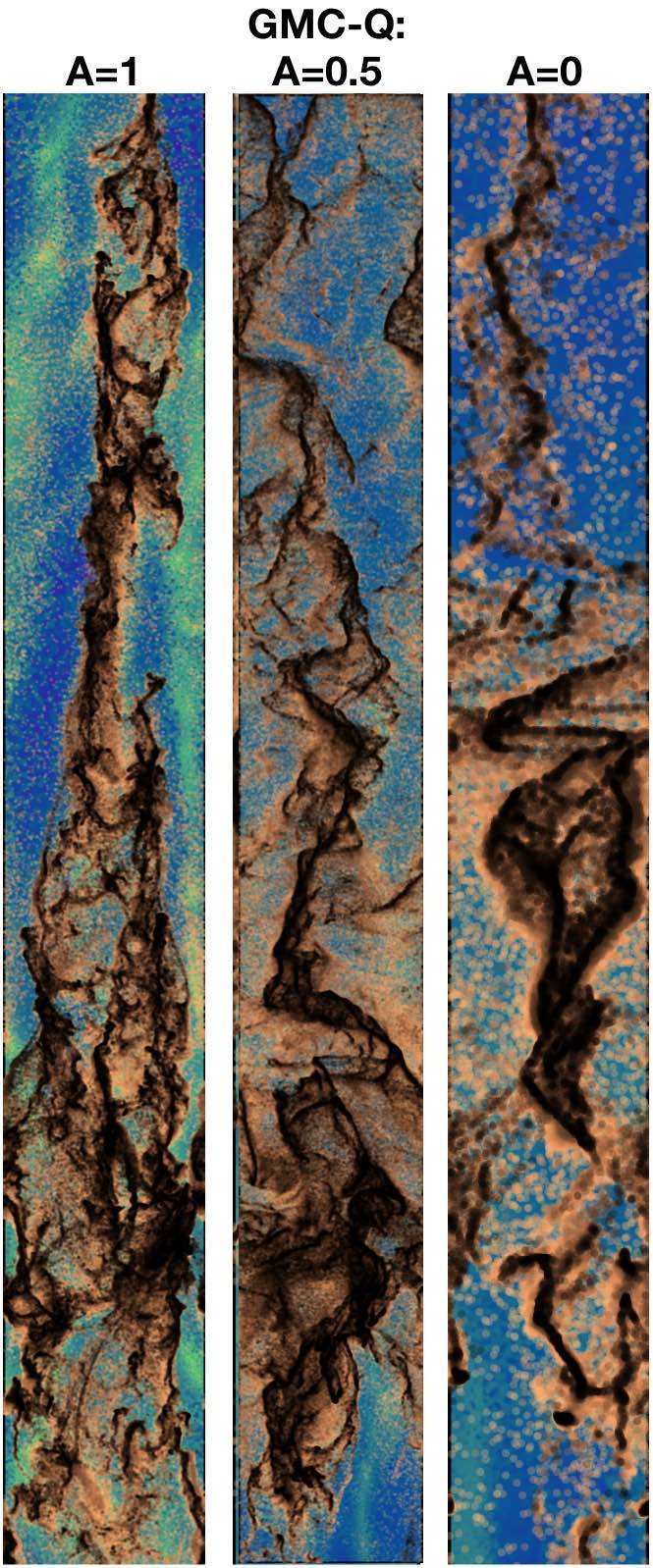}
    \caption{As Fig.~\ref{fig:rdi.physics.demo}, except now we compare otherwise identical {\bf GMC} runs with full RDMHD, with different grain optical properties. {\em Left 3:} Runs with $Q\sim\,$constant, with dust albedo $A=1,\,0.5,\,0$. {\em Right 3:} Runs with $Q\propto \grainsize$, with albedo $A=1,\,0.5,\,0$. All runs have an extinction at the wavelength of the ``driving'' radiation field of $\sim 1.6\,$mag. All produce broadly similar qualitative phenomena, but the detailed morphologies are sensitive to optical properties. $A=0$ is probably the least physically realistic (as we expect $A\sim0.5$ for optical/UV single-scattering and $A\sim1$ to approximate the multiple-scattering IR regime), but it produces significantly denser small-scale dust structures.
    \label{fig:albedo.comparison}}
\end{figure}

\begin{figure}
    \centering
    \includegraphics[width=0.97\columnwidth]{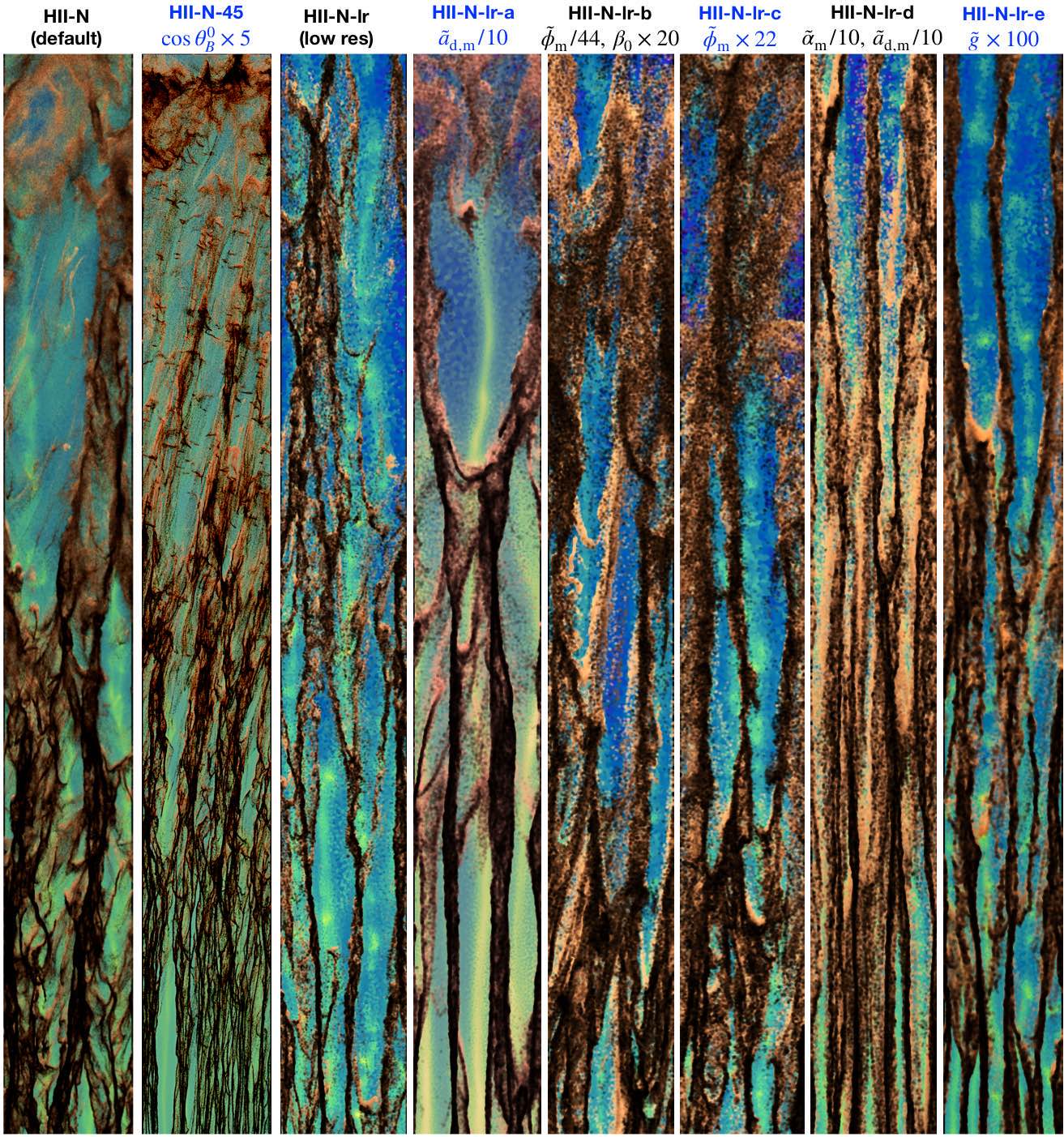}
    \caption{As Figs.~\ref{fig:rdi.physics.demo}-\ref{fig:albedo.comparison}, comparing a number of the physics variations of our {\bf HII-N} runs (see \tref{table:sims.all}). These correspond to changing the alignment and strength of magnetic fields, grain sizes or charge, strength of gravity, incident radiation flux, numerical resolution, and related quantities (see \S~\ref{sec:physics:additional} for details). While these do have non-trivial quantitative effects, they are generally sub-dominant to other variations in physics above.
    \label{fig:parameter.morphology.effects}}
\end{figure}

\begin{figure*}
    \includegraphics[width=0.987\textwidth,left]{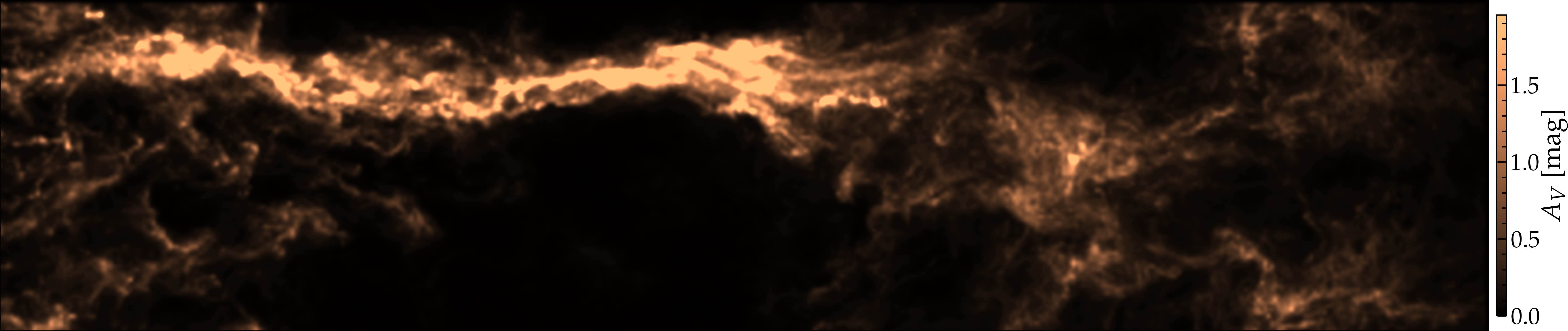}\\
    \includegraphics[width=0.998\textwidth,left]{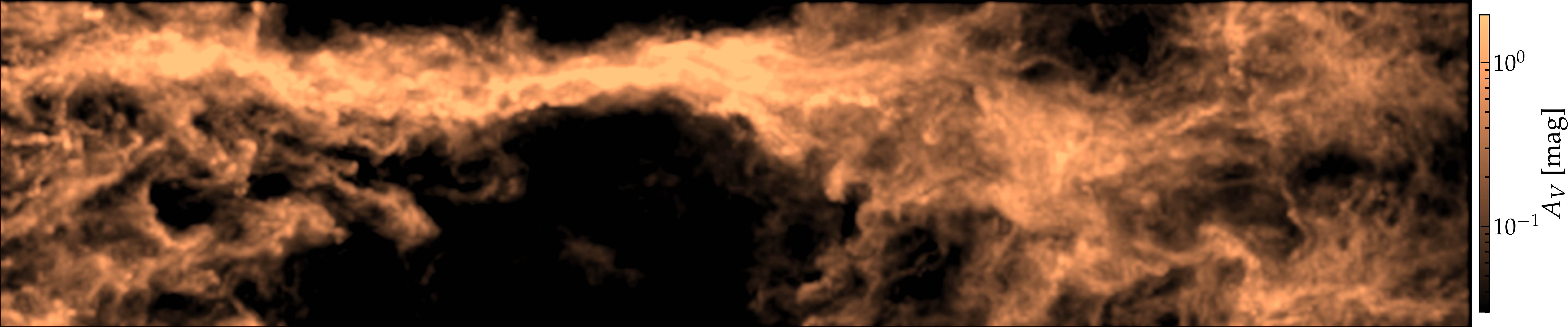}\\
    \includegraphics[width=0.995\textwidth,left]{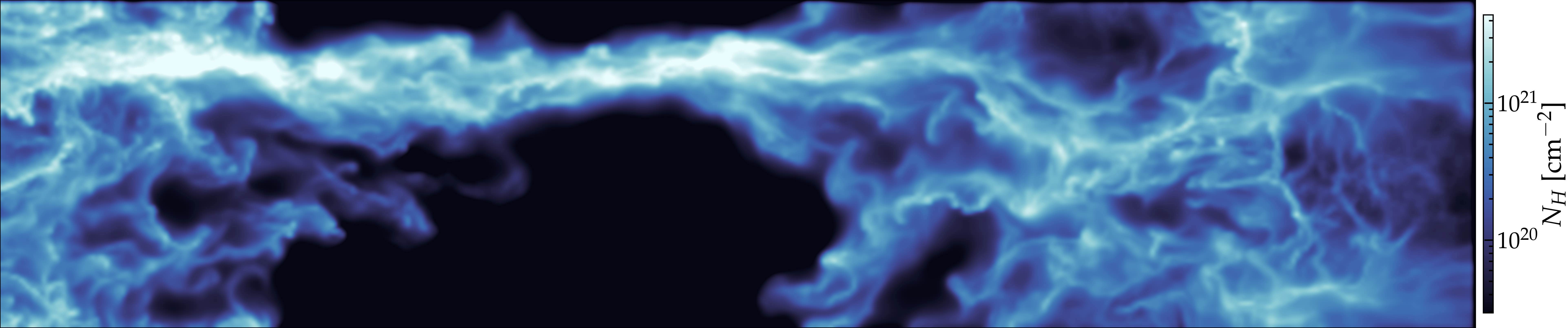}\\
    \includegraphics[width=1.000\textwidth,left]{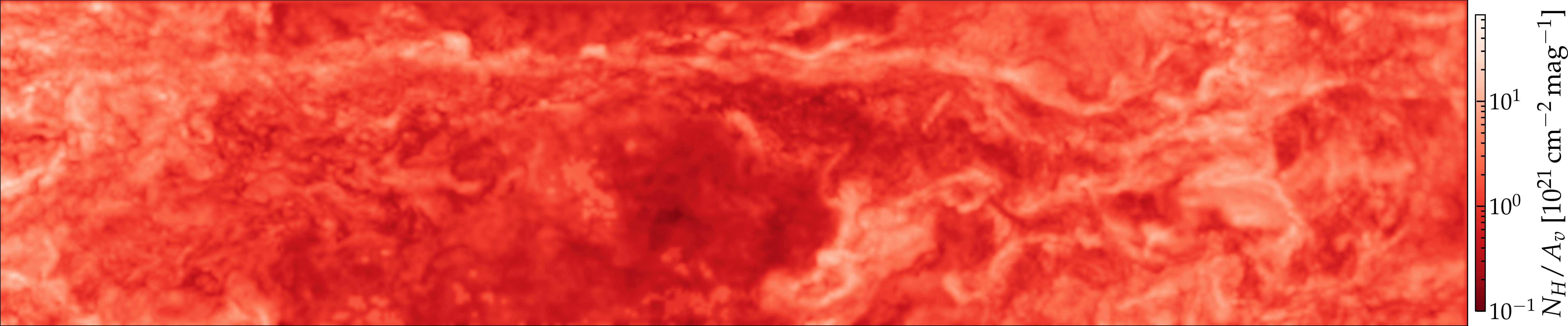}\\
    \includegraphics[width=0.998\textwidth,left]{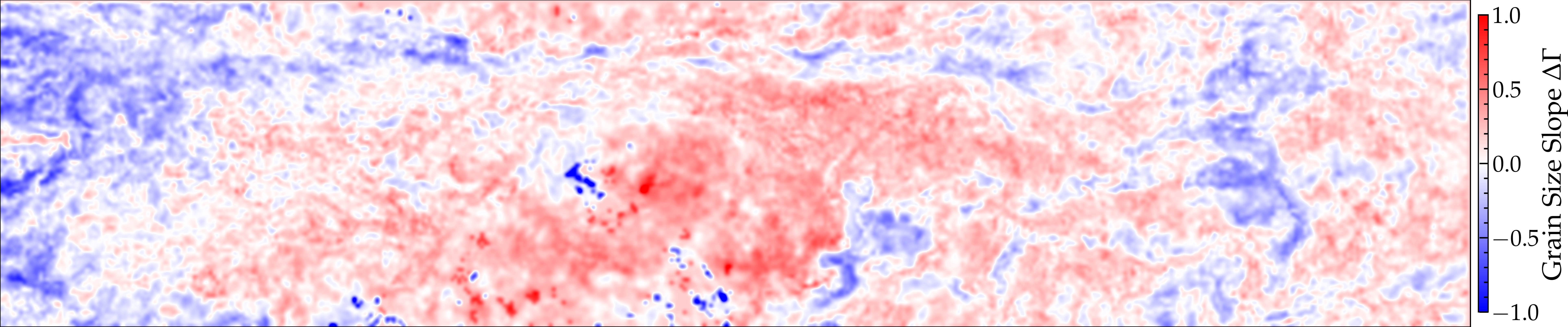}\\
    \caption{Example 2D integrated projection maps of our proto-typical GMC-like simulation ({\bf GMC-Q}), at $t\sim3\,t_{\rm acc}$, along the $y-z$ axis (outflow moving to the right), with the short image axis being size $\Lscale$, long axis length chosen to include $\sim30\%$ of the dust+gas, and pixel size $=0.01\,\Lscale$. {\em First-from-Top:} Extinction $A_{V}$ (linear scale). {\em Second:} $A_{V}$ (log scale, to highlight lower columns $A_{V}\sim 0.1$). {\em Third:} Gas column $N_{H}$ (similar log-stretch). {\em Fourth:} $N_{H}/A_{V}$ ratio. {\em Fifth:} Deviation from the mean (MRN) size spectrum, $\Delta\Gamma$ (positive means more large grains/``greyer'' extinction, negative more small grains/``steeper'' extinction). The gross filament morphology resembles observed filamentary GMCs. On large scales, dust ($A_{V}$) and gas ($N_{H}$) closely trace one another, with relatively small variation in $A_{V}/N_{H}$ (a sightline-integrated quantity) even in very small pixels. The RDIs produce variation in line-of-sight grain size distribution (GSD); the maximum variations here $\Delta\Gamma \sim \pm 1$ correspond to a factor $\sim 1/3-3$ shift in the mean extinction-weighted grain size.
    \label{fig:projection.maps.GMC.Q}}
\end{figure*}

\begin{figure*}
    \centering
    \includegraphics[width=0.48\textwidth]{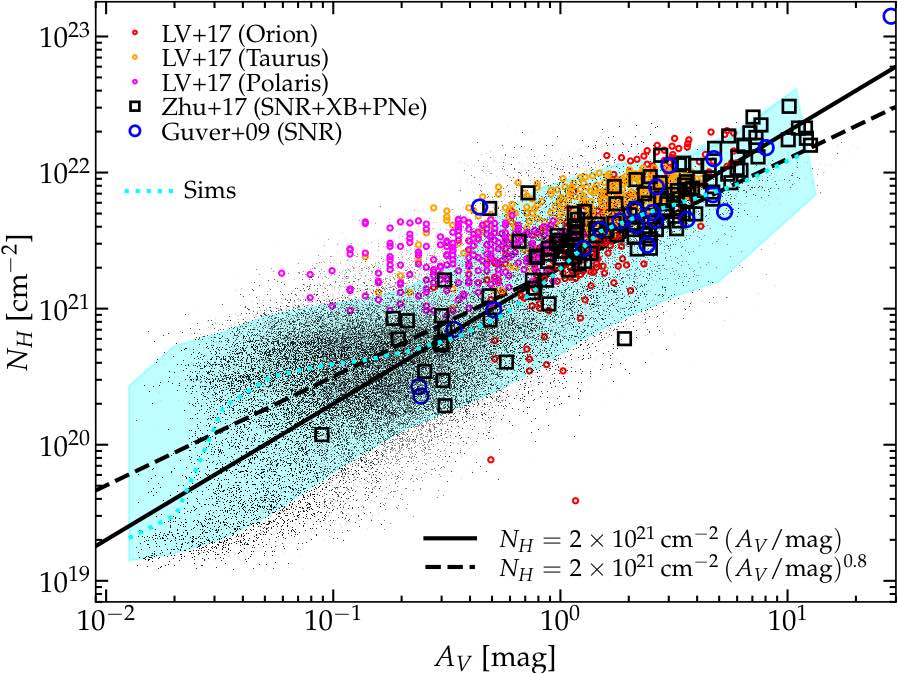}
    \hspace{0.2cm}
    \includegraphics[width=0.49\textwidth]{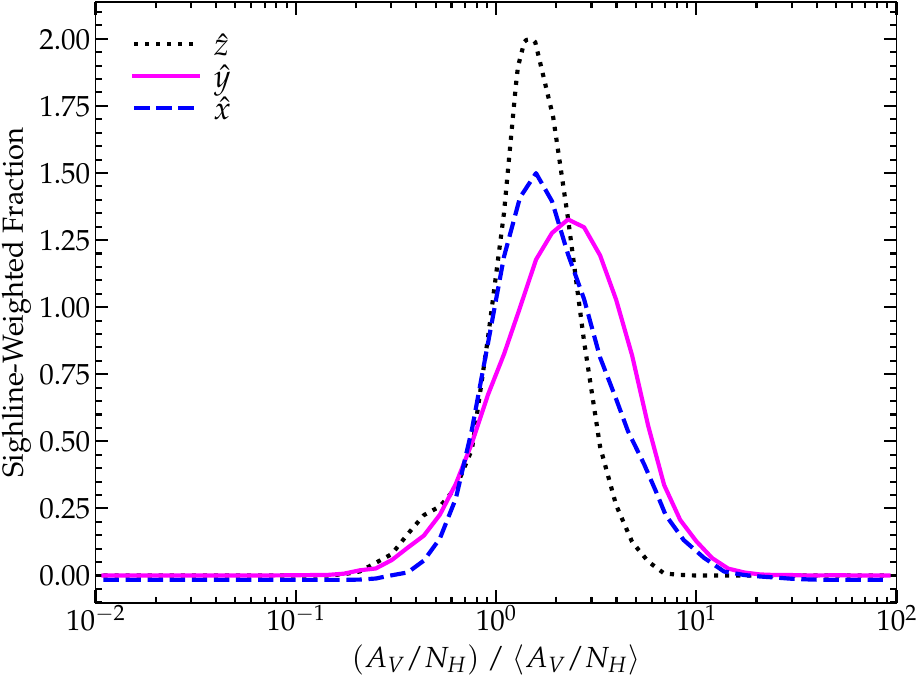}\\
    \includegraphics[width=0.48\textwidth]{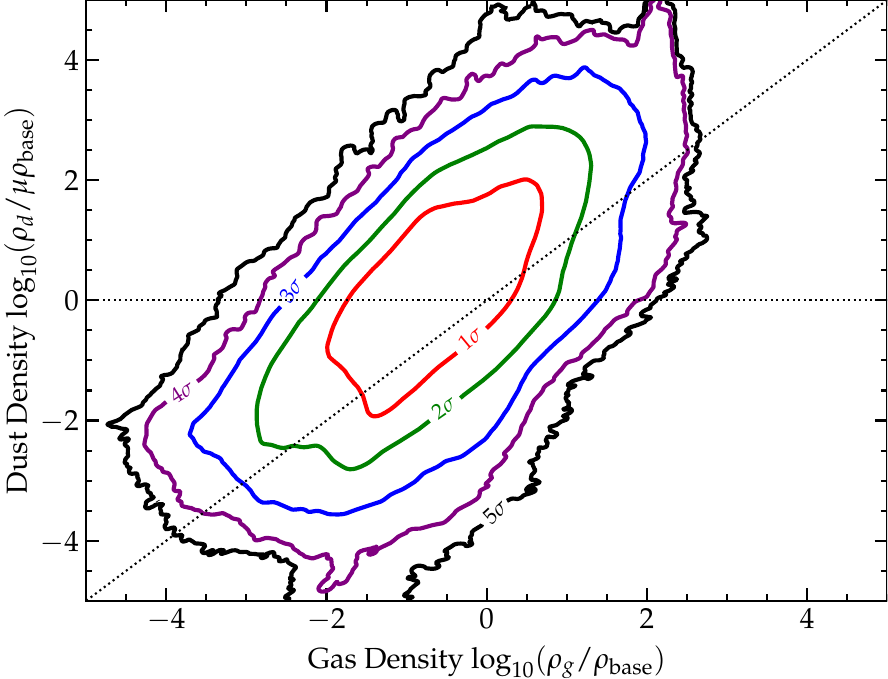}
    \hspace{0.2cm}
    \includegraphics[width=0.49\textwidth]{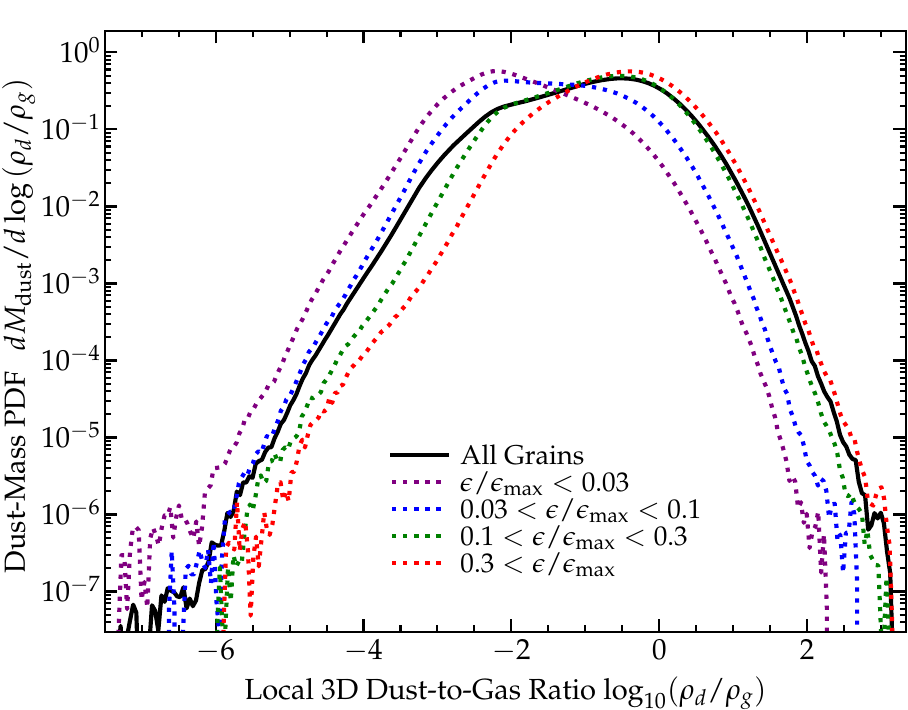}\\
    \vspace{-0.15cm}
    \caption{Distribution of dust and gas densities in {\bf GMC-Q}, both projected ({\em top}) and local 3D ({\em bottom}), averaged over times in the fully-nonlinear regime $t \sim 3-5\,t_{\rm acc}$. 
    {\em Top Left:} $N_{H}$ and $A_{V}$ integrated through random sightlines ({\em black dots}; sampled with pixel size $=0.01\,\Lscale$). Shaded cyan range shows the $5-95\%$ ($\pm2\,\sigma$) inclusion interval at each $A_{V}$, and dotted cyan line shows the median. We compare observations compiled in \citet{guver:2009.av.nh.correlation,lv:2017.dust.gas.ratio.measurements,zhu:2017.av.nh.distrib.milky.way}, and two reference scalings for $N_{H}(A_{V})$. 
    {\em Top Right:} Distribution of $A_{V}/N_{H}$ across sightlines, projected along the $\hat{z}$ (outflow), $\hat{x}$, and $\hat{y}$ axes. The rms dispersion is a similar factor $\sim 1.5-2$ across all sightlines, with slightly smaller dispersion along the outflow (long) axis as more variation is ``integrated out'', and some small differences along $\hat{x}$ vs.\ $\hat{y}$ owing to the bulk magnetic field direction being along $\hat{x}$. 
    {\em Bottom Left:} Bivariate distribution of the local 3D $\dustden$ and $\gasden$, weighted by dust mass (i.e.\ probability of a given $\dustden$ and $\gasden$ {\em around a grain}), at the resolution scale ($\sim (10^{-3}-10^{-2})\,\Lscale$). For reference lines denote uniform dust density and perfect coupling ($\dustden=\dustgas\,\gasden$).
    {\em Bottom Right:} PDF of the local 3D dust-to-gas ratio as Fig.~\ref{fig:dust.gas.pdf}, but now time-averaged.
    The RDIs produce variations in $N_{H}/A_{V}$ comparable to observed; with modest factor $\sim2$ $1\sigma$ scatter for point-source sightlines and slightly shallowed-than-unity slope of $N_{H} \propto A_{V}^{0.7-0.9}$. On small scales the {\em local, dust-weighted} variation in $\dustden/\gasden$ can be much larger, with $\sim 1\,$dex $1\sigma$ scatter and non-Gaussian tails at the $\sim 5\,\sigma$ level spanning from $\dustden/\gasden \lesssim 10^{-7}$ to $\dustden/\gasden \gtrsim 100$ (with the absolute gas-density reaching up to $\sim 10^{5}$ times its initial maximum value, i.e.\ $\sim 1000\,\rhobase$).
    \label{fig:dustgas.pdf.GMC.Q}}
\end{figure*}

\begin{figure}
    \centering
    \includegraphics[width=0.95\columnwidth]{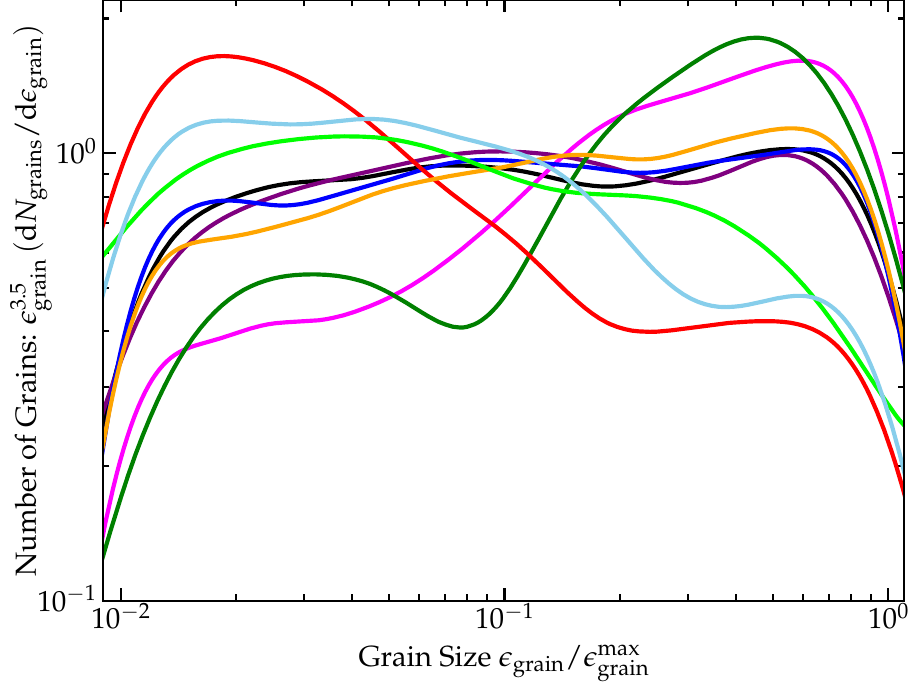}\\
    \includegraphics[width=0.95\columnwidth]{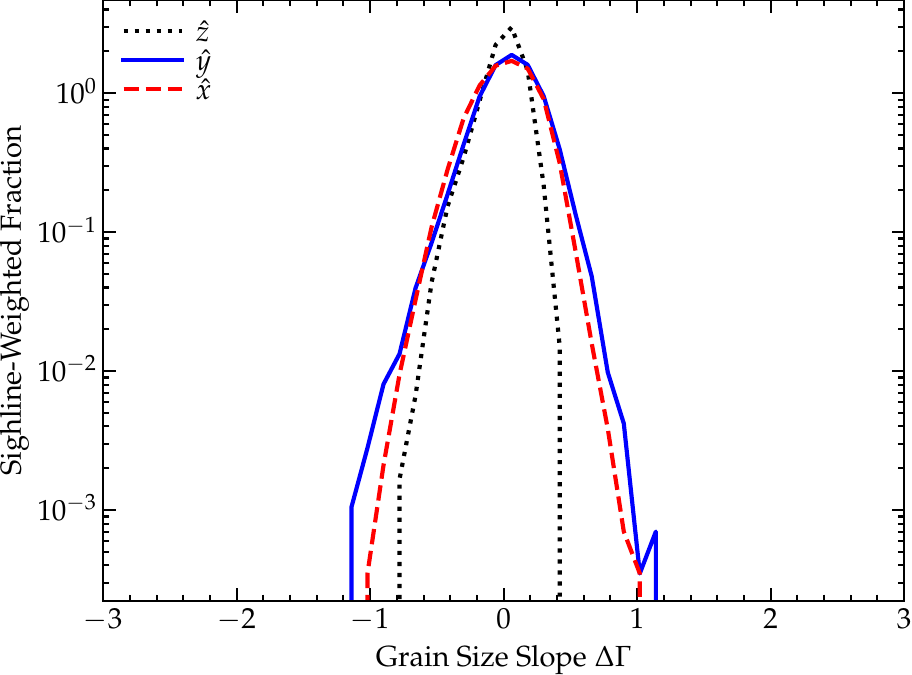}\\
    \caption{{\em Top:} Example of some of the different grain size distributions (GSDs) corresponding to the fluctuations $\Delta\Gamma$ in Fig.~\ref{fig:projection.maps.GMC.Q}, normalized so the box-averaged (MRN) distribution is unity. The cutoffs at $<0.01\,\grainsizemax$ and $\grainsizemax$ are imposed by our initial conditions. 
    {\em Bottom:} Histogram of the $\Delta\Gamma$ values for maps (of the same snapshot) with the line-of-sight along different axes ($\hat{z}$ is outflow direction). 
    Most sightlines are close to ``typical,'' but some have small grain abundances enhanced by factors of $\sim 2-4$ relative to large, or vice versa, comparable to observed variations in GMCs. 
    \label{fig:los.grainsize.hist.demo}}
\end{figure}

\begin{figure*}
    \centering
    \includegraphics[width=0.97\textwidth]{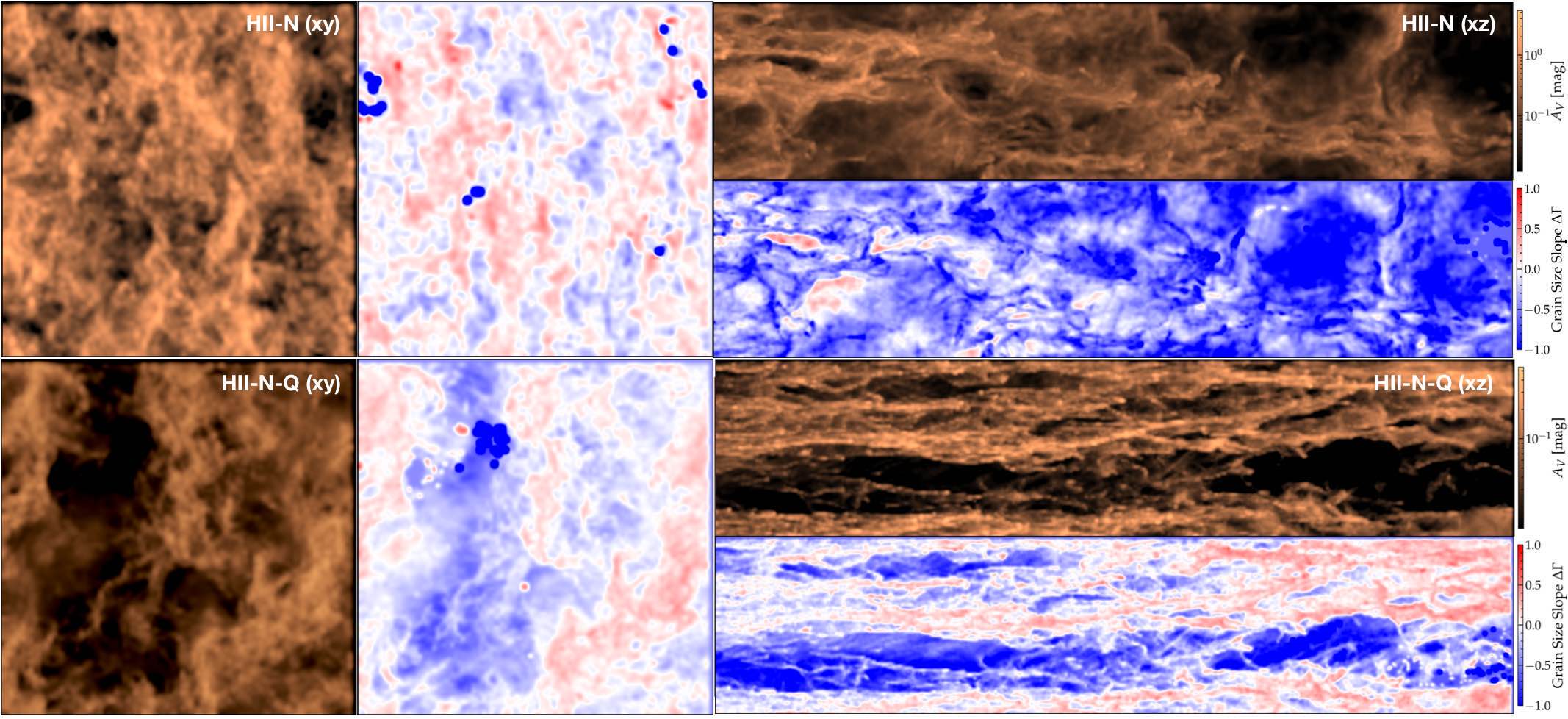}
    \caption{Maps of $A_{V}$ and GSDs as Fig.~\ref{fig:projection.maps.GMC.Q}, for {\bf HII-N} ({\em top}) and {\bf HII-N-Q} ({\em bottom}). For each the squares show the box projected along $\hat{z}$ (size $\Lscale\times\Lscale$) and the rectangles projected along $\hat{y}$ ($\Lscale \times 5\,\Lscale$), zoomed into a region around the median dust+gas position. 
    Although the GSD histograms are similar to those in Fig.~\ref{fig:los.grainsize.hist.demo}, these demonstrate a common trend across our suite. 
    Simulations with $Q\propto \grainsize$ (e.g.\ {\bf HII-N-Q}) exhibit a clear correlation with larger grains relatively more concentrated in regions of high extinction; simulations with $Q\sim\,$constant (e.g.\ {\bf HII-N}) exhibit a much weaker and inverted correlation. Generically the grain sizes which dominate the opacity tend to be the most concentrated (Fig.~\ref{fig:clumping.factors}) and to correlate positively with the extinction.
    \label{fig:los.grainsize}}
\end{figure*}

\begin{figure}
    \centering
    \includegraphics[width=0.5\columnwidth]{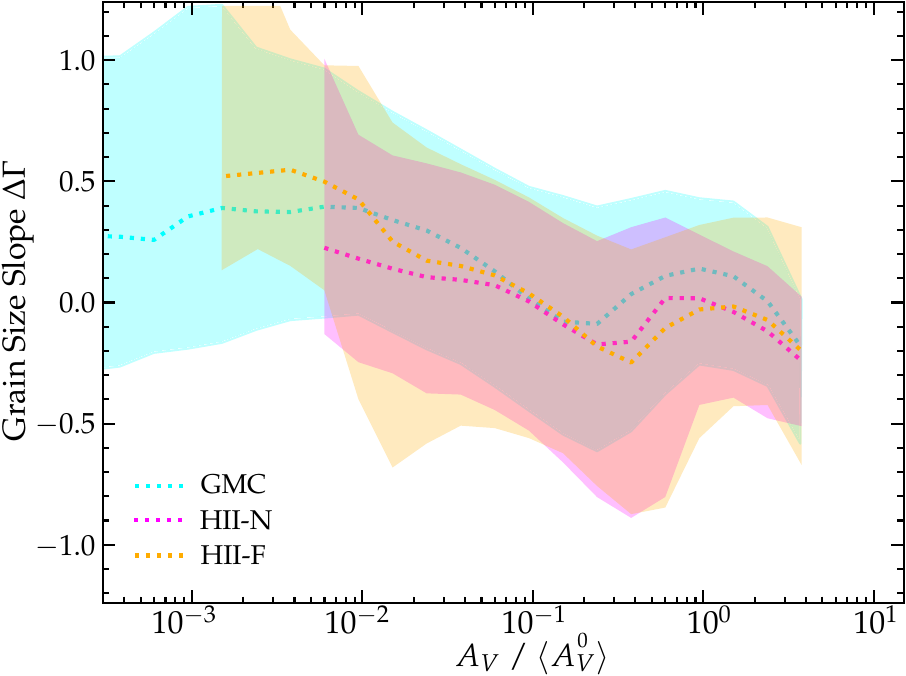}\hspace{-0.1cm}\includegraphics[width=0.5\columnwidth]{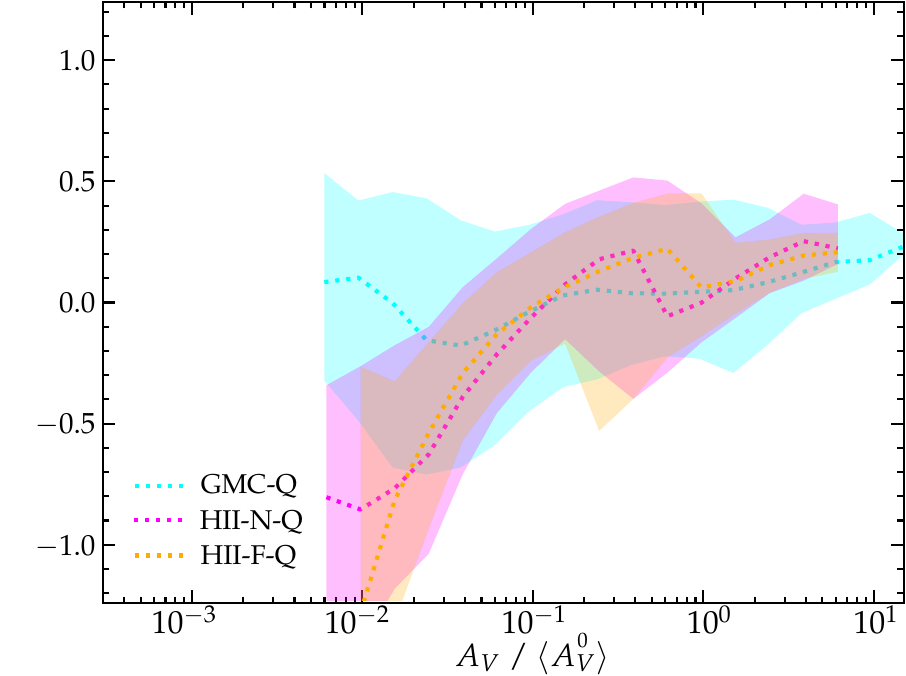}\\
    \includegraphics[width=0.5\columnwidth]{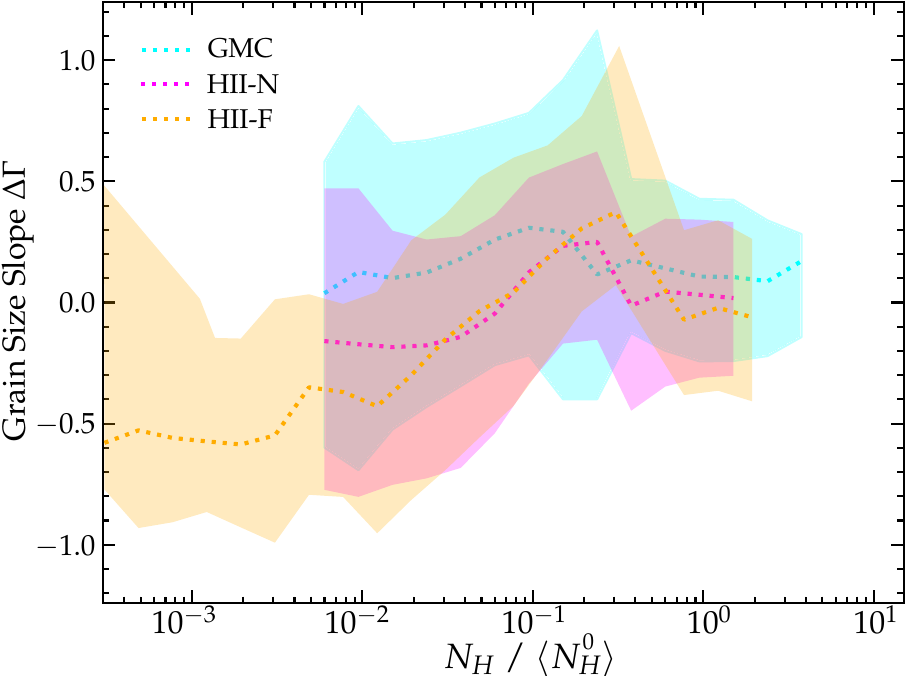}\hspace{-0.1cm}\includegraphics[width=0.5\columnwidth]{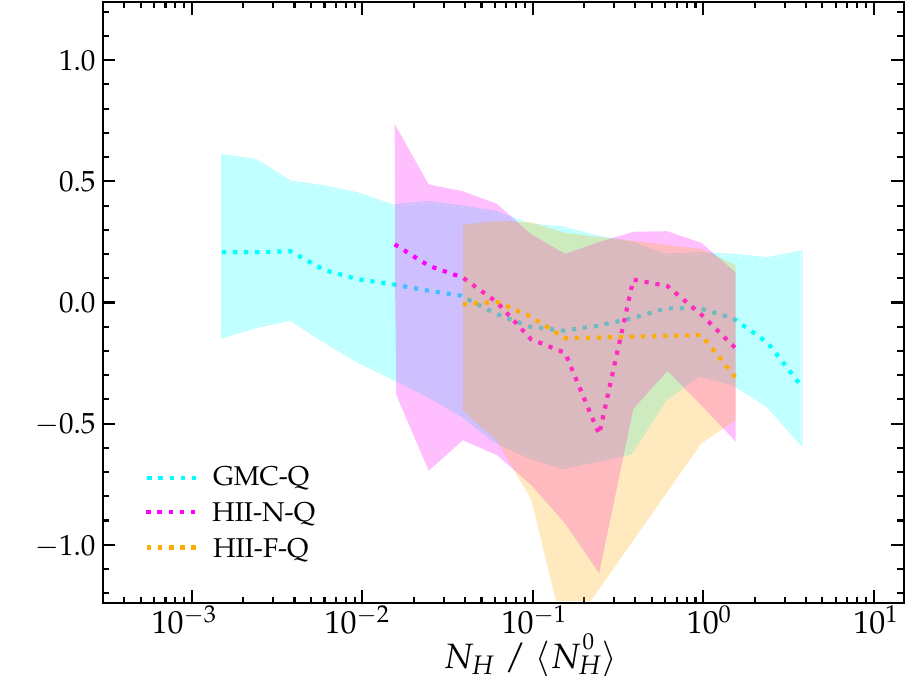}\\
    \vspace{-0.15cm}
    \caption{Distribution of GSDs ($\Delta \Gamma$) as a function of extinction $A_{V}$ ({\em top}) or gas column $N_{H}$ ({\em bottom}) for different fiducial runs (as labeled; with $Q\sim\,$constant {\em left} and $Q\propto \grainsize$ {\rm right}), from maps as Fig.~\ref{fig:los.grainsize} at one instant in time around $t\sim3\,t_{\rm acc}$ over the range of heights containing $\sim 50\%$ of the dust mass viewed from random angles. Dotted lines and shaded interval show median and $5-95\%$ range. 
    $\langle A_{V}^{0}\rangle$ and $\langle N_{H}^{0}\rangle$ refer to the values of $A_{V}$ or $N_{H}$ integrated from the base of the box to infinity in the initial conditions. This quantifies the trend in Fig.~\ref{fig:los.grainsize} for the grain sizes which dominate the opacity to be over-represented in sightlines of higher opacity relative to the local mean. There is no trend (or even a weakly opposite trend) with $N_{H}$.
    \label{fig:los.grainsize.vs.column}}
\end{figure}

Note that we focus here on the characteristic spatial structure/morphology of the simulated systems, as on large (observationally-resolveable) scales in the systems (GMCs and HII regions) of interest the timescales of the global dynamics (timescales for resolved structures to evolve) are long compared to human-observable scales. However in e.g.\ \citet{steinwandel:2021.dust.rdi.variable.stars} we consider the time-resolved dynamics of RDI-driven dust clustering on smaller scales in a different parameter space (there considering dust in cool-star photospheres and outflows) and showed it produces temporal variations roughly corresponding to characteristic growth rates of the different RDI modes on different spatial scales (see \paperone\ and \citealt{hopkins:2018.mhd.rdi} for quantitative expressions for these).

\subsection{Effects of Different Physics}
\label{sec:physics}

\subsubsection{Explicit Charged-Grain Dynamics \&\ RDIs are Essential}
\label{sec:physics:rdis}

We now illustrate the most important physics for the effects here. Fig.~\ref{fig:rdi.physics.demo} compares otherwise-identical variants of run {\bf GMC}. If we assume dust simply traces gas (the ``perfectly-coupled'' limit), then the RDIs and essentially all structure in these outflows vanish. Specifically, we run optically-thin and full RDMHD simulations\footnote{Fig.~\ref{fig:rdi.physics.demo} RDMHD runs choose $Q_{\rm ext,\,0}$ so that the initial total extinction at the wavelengths of the incident radiation is $A_{\rm incident}\approx 1.6\,$mag, with 
albedo $A_{0}=1$, but we vary these below.} where we assume a constant opacity for the gas and apply the radiation forces {\em directly} to the gas in the usual gas radiation-MHD manner (instead of applying the force to the dust and integrating the dust dynamics and back-reaction).\footnote{In the ``dust traces gas'' runs in Fig.~\ref{fig:rdi.physics.demo}, we integrate the dust as a passive scalar using the locally-interpolated gas velocities per \S~\ref{sec:rad}, to confirm that this makes a negligible difference (up to some $\sim1\%$-level integration-error noise) compared to assuming dust exactly follows gas. This, like the explicit tests in e.g.\ \cite{hopkins.2016:dust.gas.molecular.cloud.dynamics.sims,lee:dynamics.charged.dust.gmcs} and \citet{moseley:2018.acoustic.rdi.sims}, verifies that the sort of Lagrangian integration-error effects described in \citet{genel:tracer.particle.method} are negligible.} In the optically-thin (constant-flux) case, this has a trivial exact analytic solution, which we verify our simulations recover up to integration error: the entire dust+gas system simply accelerates exactly with a uniform ${\bf a} = a_{\rm eff}\,\hat{z}$. With explicit radiation transport, the fact that this specific setup is actually moderately optically-thick means the solution has a slightly different vertical profile, but it clearly resembles the optically-thin case, and most important for our purposes, is completely stable and forms no appreciable sub-structure.

Note that if we allow uniform dust drift at the local homogeneous equilibrium drift velocity, but otherwise continue to assume the ``perfectly-coupled'' limit, we obtain nearly-identical results with no appreciable substructure. If we evolve dust dynamics without including the ``back-reaction'' on the gas (momentum transfer from grains to gas), then of course the dust simply unphysically ``ejects'' -- accelerating out of the box uniformly on a very short timescale and leaving the gas entirely behind.

Now allowing for explicit dust dynamics, we consider the case if we ignore Lorentz forces on dust: note that the {\em gas} still obeys MHD (magnetic fields are still present), but we ignore the charge of dust grains. This still leaves the stratified acoustic RDI (see \citealt{hopkins:2017.acoustic.RDI}, Appendix~C). However, the acoustic RDI has a vastly-simpler structure compared to the MHD RDIs, with only one resonance available at each wavenumber $k$, and those resonances all have the same angle at $\cos{\theta_{k}} = \hat{\bf k} \cdot \hat{z} = \cs/|\driftvel|$ independent of $k$ -- in fact we can see this angle traced prominently in the dust. The RDIs for charged grains on the other hand, feature a wide range of modes (with acoustic, cosmic-ray like, \Alf\ and fast/slow magnetosonic MHD-wave, \Alf\ and fast/slow gyro RDIs, and others, with up to $\sim 20$ different resonant angles tracing a complex multi-dimensional structure at a given $k$; see \citealt{hopkins:2018.mhd.rdi}). We clearly see this translate to qualitatively distinct structures.

We next compare our default optically-thin (``Optically-Thin+Dust Dynamics'') and full RDMHD (explicit radiation-dust-MHD; ``RDMHD+Dust Dynamics'') runs (\S~\ref{sec:rad}), which explicitly follow dust dynamics and back reaction. The radiation treatment makes some quantitative differences in detail (discussed below), but the qualitative behavior is identical in all properties described in \S~\ref{sec:results:general}.

\subsubsection{Optical Properties of Grains}
\label{sec:physics:optical}

Although much less dramatic than the effects of removing the MHD RDIs (\S~\ref{sec:physics:rdis}), Figs.~\ref{fig:bulk.props.vs.z.gmc.Q}-\ref{fig:side.view.multi} demonstrate that the optical properties of grains, specifically how $Q$ (and therefore the grain acceleration) scales with $\grainsize$, can have a significant quantitative effect of the resulting behavior (see \S~\ref{sec:results:general}). 

Fig.~\ref{fig:albedo.comparison} extends this by considering the effects of the dust albedo $A_{0}$ as well, in our full RDMHD simulations. Recall, in the optically-thin limit, optical properties beyond $\accsizedep$, such as $A_{0}$ and the normalization of $Q$ do not enter the dynamics individually (only in degenerate combinations, implicit in our dimensionless simulation parameters): they only become non-degenerate and important if the system becomes optically-thick. So we focus on the {\bf GMC} case (instead of {\bf HII-N} or {\bf HII-F}), as this has the highest geometric optical depth (defined as the optical depth if $Q=1$ for all grains), so the effects of different albedo will be most prominent. We consider three variants of {\bf GMC} and {\bf GMC-Q}, with $Q$ normalized so the total extinction through the initial column at the source frequency is $\approx 1.6$\,mag (chosen to be similar to typical GMCs in the Local Group; \citealt{bolatto:2008.gmc.properties}), but albedo (1) $A_{0}=1$ (pure-scattering, appropriate for high-energy source photons such as X-rays or, to $\mathcal{O}(v/c)$, for grey IR absorption and re-emission), (2) $A_{0}=1/2$ (equal scattering and absorption, appropriate for incident radiation with wavelengths of order grain sizes, i.e.\ UV/optical), or (3) $A_{0}=0$ (pure absorption, not physically relevant for the cases here but a useful comparison case). 

The differences are modest -- much smaller than those in Fig.~\ref{fig:rdi.physics.demo} -- but not negligible. As $A_{0}\rightarrow 0$, the grains become more clumped into smaller, denser structures (e.g.\ the black ``globules'' in {\bf GMC} with $A_{0}=0$). As the absorption optical depth $\tau_{\rm abs} = (1-A_{0})\,\opticaldepth$ increases, the outflow requires more time to accelerate (as absorption without re-emission reduces the total photon momentum coupled by a factor $(1-\exp{[-\tau_{\rm abs}]})/\tau_{\rm abs}$), and becomes more ``shell like'' especially in early stages (as the absorption occurs in an increasingly thin shell as $\tau_{\rm abs}\rightarrow \infty$). Of course, in the limit $\tau_{\rm abs} \gg 1$, we should really consider the IR multiple-scattering problem instead; but this is not the regime we focus on in this paper.

\subsubsection{Additional Parameters}
\label{sec:physics:additional}

The simulations in \tref{table:sims.all} survey a large number of additional parameters. Fig.~\ref{fig:parameter.morphology.effects} surveys several of these, including resolution, magnetic field direction, incident flux, magnetic field strength, grain charge, grain size, and strength of gravity. These produce non-negligible quantitative effects, some of which will be studied in future work. However since this is a low-resolution survey, and many of the micro-physical effects of these parameters were studied in more detail in idealized high-resolution simulations in \citet{hopkins:2019.mhd.rdi.periodic.box.sims}, we restrict our comparison here to a brief summary, noting that none of these appear to change any of the {\em qualitative} behaviors seen in \S~\ref{sec:results:general}.

Typically, weaker radiative forcing (smaller $\accparam$) leads to somewhat more coherent but ``wavier'' filaments, while much stronger supersonic forcing produces more vertically-aligned structures, as the RDIs become more supersonic-acoustic-like \citep[see][]{moseley:2018.acoustic.rdi.sims}. Smaller/larger grains (smaller/larger $\sizeparam$) lead to narrower/thicker filaments, corresponding to the change in grain collisional mean free paths or stopping lengths as noted above. Provided gravity remains sub-dominant to the outward force, changing $\gravparam$ has little effect. Lower/higher dust-to-gas ratios $\dustgas$ produce stronger/weaker dust concentration as discussed above. Changing magnetic angles or the background $\beta$ rotates the resonant angles, changing the geometry of some structures \citep[see][]{hopkins:2019.mhd.rdi.periodic.box.sims}. Although accounting for magnetization/charge of the dust is crucial, changing the dust charge-to-mass ratio ($\chargeparam$) by factors of $\sim 100$ in Fig.~\ref{fig:parameter.morphology.effects} produces modest effects, as the dust gyro radii are still much smaller than the scales of the modes of greatest interest -- the results only begin to resemble the ``No Lorentz Forces'' case in Fig.~\ref{fig:rdi.physics.demo} if we lower $\chargeparam$ by factors $\gtrsim 10^{4}$.

\subsubsection{Distinction from Rayleigh-Taylor Instabilities}
\label{sec:physics:rrti}

It is worth briefly noting how the character of the dominant instabilities here (the RDIs) is qualitatively distinct from the radiative Rayleigh-Taylor instability (RRTI), which has been previously studied in simulations which ignore dust dynamics (like the ``dust traces gas'' runs in Fig.~\ref{fig:rdi.physics.demo}; see e.g.\ \citealt{krumholz:2012.rad.pressure.rt.instab,davis:2014.rad.pressure.outflows}). Most obviously, the RRTI is not actually unstable here: it can only grow on scales much larger than the photon mean free path ($\sim 0.7\,\Lscale$, here), with a steep opacity law (e.g.\ $\kappa \propto T_{\rm rad}^{2}$), and the RRTI is stabilized by magnetic fields. The RDIs, in contrast, generally grow {\em faster} when mean-free paths are longer, or in the presence of magnetic fields, are unstable on all wavelengths down to ion gyro radii, and do not depend significantly on the opacity law \citep{squire.hopkins:RDI}. The physics is entirely different: RDIs derive from resonance between natural dust and fluid frequencies (which can be entirely unrelated to any stratification of the medium), and produce mode eigenstructure, fastest-growing wavelengths, and non-linear morphologies (e.g.\ banding/sheets, shell modes, globules) totally unlike the RRTI, and vastly stronger non-linear dust clumping.

\begin{figure}
    \centering
    \includegraphics[width=0.57\columnwidth]{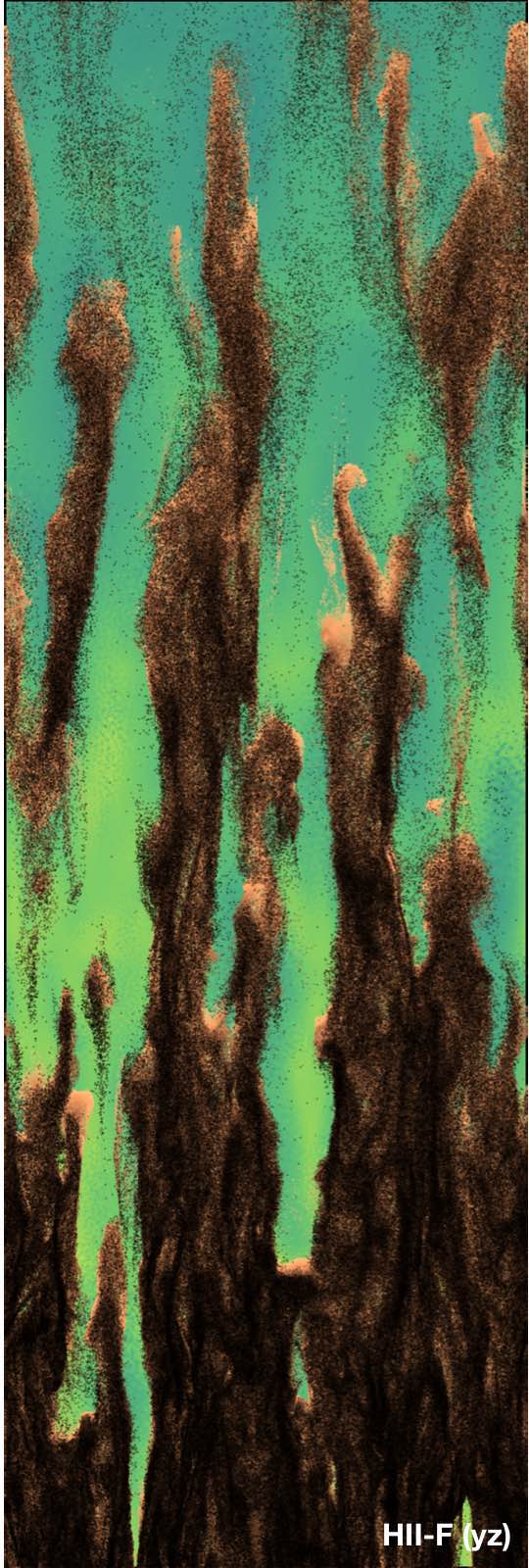}
    \caption{Zoom-in projection (as Fig.~\ref{fig:side.view.multi}) of a random sub-volume of our  {\bf HII-F-Q} simulation (image short-axis size $\sim 0.1\,\Lscale$). This happens to exhibit a ``pillar''-type morphology, similar to many observed structures (see \S~\ref{sec:obs:morph}). We show this to indicate the richness of the detailed morphological structures.
    \label{fig:pillars.of.creation}}
\end{figure}

\subsection{Observable Effects on Dust Structure \&\ Extinction}
\label{sec:obs}

\subsubsection{Morphologies}
\label{sec:obs:morph}

The most immediate observable effect of the RDIs is clearly how they shape the morphology of the dust and gas (\S~\ref{sec:results:general:morphology}). Figs.~\ref{fig:isometric.view.GMC}-\ref{fig:isometric.view.HII} and Fig.~\ref{fig:side.view.multi} show some of the representative morphologies, with the physics driving these discussed in \S~\ref{sec:results:general:morphology} \&\ \S~\ref{sec:physics:rdis}. 

One striking aspect of the dust morphologies is how different they are from the morphologies that arise in simulations of e.g.\ pure MHD-turbulence -- even when those simulations have nearly identical sonic and \Alf{ic} Mach numbers (compare e.g.\ Fig.~1 in \citet{2020ApJ...894L...2B}, which has very similar gas $\mathcal{M}_{s,\,A}$ to our {\bf GMC} and {\bf GMC-Q}). Likewise, the morphologies are totally distinct from those obtained by integrating the trajectories of ``passive'' tracer-particle grains (grains which exert no force on the gas, so cannot drive outflows or RDIs) in MHD gas turbulence (compare Figs.~1 \&\ 2 of \citealt{hopkins.conroy.2015:metal.poor.star.abundances.dust,hopkins.2016:dust.gas.molecular.cloud.dynamics.sims}). Those simulations can produce filamentary structures, but the structures are vastly less coherent and well-aligned and have an obviously distinct distribution of axis ratios from those here, and they tend to be exclusively associated with the locations of strong shocks. Certain morphological features here such as the diffuse cirrus and ``cumulus'' or ``stratocumulus''-like structures simply never occur in ``passive grain'' or pure MHD-turbulence simulations. Others, like some of the knots, pillars or ``horsehead'' or ``mushroom cap'' type structures can form in simulations that include additional gas physics (e.g.\ knots can form in simulations with self-gravity at local points of collapse, pillars and related structure in simulations including ionization fronts as a phase contrast) -- but these necessarily involve different physics from those modeled here, and therefore would occur in different locations with different frequencies. In the passive or ``tracer particle'' dust simulations, larger dust grains are always more diffuse and fail to cluster on small scales, while very small grains are trapped into incredibly narrow ``ridgeline''-type structures (being trapped at local strain maxima at the interstices of vorticity maxima; \citealt{olla:2010.grain.preferential.concentration.randomfield.notes}) -- often completely opposite their behaviors here. 

The filamentary structures predicted here are morphologically remarkably similar to dust filaments in GMCs and massive star-forming region complexes \citep[e.g.][]{apai:2005.HII.region.filamentary.dust.structures,goldsmith:2008.taurus.gmc.mapping,menshchikov:2010.filaments.herschel,2013A&A...550A..38P,andre:2017.filaments.in.gmcs.star.formation.obs.review}.\footnote{To the extent that there is a characteristic scale in the structures, e.g.\ the large grain mean-free-path, this is also suggestive: $\lambda_{\rm mfp}^{\rm grain} \sim \driftvelmag\,\ts \sim \internaldensity\,\grainsize / \gasden \sim 0.1\,{\rm pc}\,(\grainsizemax/0.1\,\mu{\rm m})\,(n_{\rm gas}/100\,{\rm cm^{-3}})^{-1}$, similar to observationally-suggested characteristic scales \citep{koch:2015.filament.magic.size}, but we caution that there are RDIs over a wide hierarchy in scales and similarly the observed spatial power spectra of clouds do not actually show a characteristic scale but a broad distribution, with the appearance of a specific scale in filament-identification more representative of its extremes \citep{panopoulou:characteristic.filament.width.artefact.of.obs.and.power.spectrum}.} This goes well beyond their globally ``filamentary'' structure to include sub-structure and ``feathering'' or ``whisker'' structures, the contrast ratios of edges of the structures, the coherence over very large relative axis ratios, the relative incidence of ``knots,'' and more. On even smaller scales in e.g.\ our HII region-like simulations, we see structures very similar to the whisker fine-structure seen ubiquitously in well-resolved HII regions \citep{odell:2002.pne.knots.review,apai:2005.HII.region.filamentary.dust.structures}, most famously in $\eta$ Carinae \citep{morse:1998.eta.carinae.whiskers.dustlanes.filaments}. We could easily select hundreds of qualitative examples of morphological structures similar to those observed -- for just one example, we note a randomly selected ``zoom-in'' to a subvolume of one of our simulations which happens to produce a ``pillar''-type morphology on these scales in Fig.~\ref{fig:pillars.of.creation}.

We stress that we are not saying only RDIs can form these sorts of structures: turbulence can certainly form filaments with some properties similar to observations \citep{kirk:2015.turb.filaments.vs.obs}, and it is well-established that expanding ionization fronts can produce pillar-type structures \citep{gritschneder:2010.pillar.hii.region.formation,arthur:2011.rhd.hii.region.sims.morph.structure,tremblin:2012.pillar.form.ionization.sims}. However there are also anomalous features in many observed cases, which do not yet have explanations \citep[see e.g.][]{westmoquette:2013.super.star.cluster.pillars.mixing.layer.not.explained.simple.models,roccatagliata:2013.carina.structures.morph.not.well.explained,paron:2017.pillar.hii.region.obs.not.implosion,klaassen:2020.pillars.carina.alma.obs.favor.ion.models}. These include relatively extreme examples, such as pillar-like structures pointing in the ``wrong direction'' from HII regions (i.e.\ away from the nearest massive star/front, which has a natural explanation here as the filaments in dust-driven outflows have this ``head'' structure in both directions), or dust ring/shell structures which are explicitly {\em not} associated with a local similar structure in the gas phase \citep[see][]{topchieva:2017.dust.HII.region.shape.irregular.rings.shells}. 

Of course, the above morphological information is largely qualitative. This motivates the importance of developing {\em quantitative} observables in future work which can distinguish the morphologies dominated by the action of the RDIs as compared to other mechanisms, using e.g.\ higher-order topological characteristics, bi and tri-spectra of the projected densities, and other tools.

\subsubsection{Dust-to-Gas Ratios}
\label{sec:obs:nhav}

We discussed local (micro-scale) variations in e.g.\ the dust-to-gas ratio $(\dustden/\gasden)$ above (\S~\ref{sec:results:general:clumping}), but these are not observable: Fig.~\ref{fig:projection.maps.GMC.Q} attempts to directly construct observable quantities. We take one of our simulations in its non-linear stages, project it along an axis perpendicular to $\hat{z}$, and integrate lines-of-sight convolved into pixels of side-length $\sim 0.01\,\Lscale$, in a box centered on the median location of the dust+gas mass (containing $\sim 30\%$ of the total mass). For consistency, we assume the same scaling of $Q$ with $\grainsize$ assumed in the simulations.\footnote{We focus on {\bf GMC-Q} instead of {\bf GMC} here because if we are interested in $V$-band (wavelength $\lambda_{V} \sim 0.55\,{\rm \mu m}$) and assume $\grainsizemax \sim 0.1\,{\rm \mu m}$, then indeed most grains have $\grainsize \ll \lambda_{\rm rad}$.} To convert to physical units, we assume following e.g.\ \citet{weingartner:2001.dust.size.distrib} that $Q_{\rm ext,\,0}\approx0.2$ for the largest grains ($\grainsize=\grainsizemax$) at $V$-band ($0.55\,{\rm \mu m}$) and $\grainsizemax\,\internaldensity \approx 0.1\,{\rm \mu m\,g\,cm^{-3}}$ (i.e.\ largest grains $\sim 0.1\,{\rm \mu m}$); this fully determines $A_{V}$ and $N_{H}$. 

Immediately, it is striking how closely the main filament morphologically resembles many observed dust filaments (compare e.g.\ Herschel images of Taurus filaments in \citealt{2013A&A...550A..38P}). Most of the key features discussed in \S~\ref{sec:obs:morph} are retained, even in $A_{V}$, although of course finite-resolution effects reduce the ``sharpness'' of some small-scale structures.

It is also immediately visually obvious that dust traces gas on large scales. We show this quantitatively in Fig.~\ref{fig:dustgas.pdf.GMC.Q}, where we plot the distribution of $A_{V}$ versus $N_{H}$, and PDF of $A_{V}/N_{H}$ aggregating different snapshots and projections. To first order, $N_{H} \sim 2-3\times10^{21}\,{\rm cm^{-2}}\,(A_{V}/{\rm mag})$, with factor $\sim 2$ or smaller $1\sigma$ log-normal scatter. This is in excellent agreement with the observed mean trend and well within observational bounds on the intrinsic scatter in $N_{H}/A_{V}$ \citep[see][]{guver:2009.av.nh.correlation,zafar:2011.grb.extinction.curves.av.nh,galliano:2011.lmc.regional.dust.to.gas.variations,lv:2017.dust.gas.ratio.measurements,zhu:2017.av.nh.distrib.milky.way}, including the variations inferred within a given star-forming complex \citep{roman-duval.2014:smc.lmc.dust.to.gas.ratio.fluctuations.variations,lv:2017.dust.gas.ratio.measurements}. The result does not depend strongly on time (provided we are in the non-linear stages) or projection angle (though projecting directly along $\hat{z}$ always produces somewhat smaller scatter,  as we integrate through the ``entire'' outflow, not just a portion). At second-order, there are some weak trends: the correlation is slightly sub-linear ($N_{H} \propto A_{V}^{\alpha}$ with $\alpha \sim 0.8$, if we perform a simple least-squares fit), i.e.\ high-column regions have slightly higher $A_{V}/N_{H}$, on average, and slightly reduced scatter. These trends have also been observed suggested by observations, though with less certainty (see references above, and e.g.\ \citealt{draine:2003.dust.review,apai:dust.review}). They arise naturally here because (1) the dust is clumped more strongly than the gas, and (2) the dust-gas coupling is stronger in higher-density regions. 

The robustness of $A_{V}/N_{H}$ might at first appear contradictory to the enormous fluctuations in $(\dustden/\gasden) \sim 10^{-7} - 10^{3}$ (Fig.~\ref{fig:dust.gas.pdf}) in these simulations. But it is essential to recall that the latter is defined as a {\em local} 3D quantity at (ideally) infinitesimal scales, while the former is a line-of-sight integral (and of course, we must distinguish between the extreme tails and typical rms width of the distributions). Moreover, independent of what drives the fluctuations in $\dustden/\gasden$, \citet{hopkins:2012.intermittent.turb.density.pdfs,squire.hopkins:turb.density.pdf} note how mass conservation requires that different small-scale line-of-sight fluctuations must be correlated in a manner such that the line-of-sight-integrated PDF must always converge to the mean {\em faster} than e.g.\ the central limit theorem would imply. 

There are still some outliers where large-scale modes in the sky plane create large fluctuations in $A_{V}/N_{H}$. But these are observed as well. Low density regions with very little dust would not be detected in most observations in Fig.~\ref{fig:dustgas.pdf.GMC.Q}, but such regions exist and are usually simply assumed to have been dust-depleted \citep{galliano:2011.lmc.regional.dust.to.gas.variations}. Conversely, there are a number of well known examples in e.g.\ HII regions of dense dust ``knots'' or filaments which do not appear coincident with gas-phase density enhancement on small scales \citep[see e.g.][]{garnett:2001.ring.nebula.dusty.knots.offset.emission.line.no.gas.overdensity}, which the RDIs here can naturally explain.

\subsubsection{Extinction Curve Variations}
\label{sec:obs:extcurve}

Going beyond $A_{V}$, the RDIs should also imprint sightline-to-sightline variations in extinction curve shape. Modeling this in detail requires radiation-transport calculations including anisotropic scattering, with a detailed model for the grain chemistry/optical properties \citep[see e.g.][]{seon.draine:extinction.curve.vs.radiative.transfer.turbulent.sims}, which (while an important future question) is beyond the scope of our study here. We can however immediately (without adding additional assumptions) predict the first-order important quantity for the extinction curve: the grain size distribution (GSD) along a given sightline. Recall we always begin from an MRN GSD which is universal everywhere in the box: $d N_{\grainsuff} / d \grainsize \propto \grainsize^{-3.5}$, from $\grainsizemin=0.01\,\grainsizemax$ to $\grainsizemax$. The individual grain sizes are conserved (we do not model collisions/growth/destruction) -- so any variation in the GSD must arise from differential grain dynamics of e.g.\ large-vs-small grains. 

Fig.~\ref{fig:projection.maps.GMC.Q} shows an example of this in projection, and Figs.~\ref{fig:los.grainsize.hist.demo}, \ref{fig:los.grainsize}, \&\ \ref{fig:los.grainsize.vs.column} show more detailed statistics. Although we can quantify the full GSD for each sightline, the very detailed structure here is (a) cumbersome to analyze statistically, and (b) prone to noise, given our finite numerical dust-element resolution and binning into very narrow sightlines (each pixel contains only $\sim 10^{-5}$ of the total dust mass). We can reduce the GSD to a single statistic by considering e.g.\ the mean grain size $\grainsizebar$ along each sightline (weighted by e.g.\ contribution to grain mass, or area, or $V$-band extinction), or by fitting a power-law $d N_{\grainsuff} / d \grainsize \propto \grainsize^{-3.5+\Delta \Gamma}$ to the set of discrete grain sizes.\footnote{We use the method from \citet{bauke:2007.maximum.likelihood.methods.for.discrete.power.law.distributed.data} to robustly fit a maximum-likelihood power-law GSD over a finite interval: $d N_{\grainsuff} / d \grainsize \propto \grainsize^{-3.5+\Delta \Gamma}$ from $\grainsizemin=0.01\,\grainsizemax$ to $\grainsizemax$ directly to the un-binned set of grain sizes sampled along each sightline.} Smaller $\grainsizebar$ ($\Delta\Gamma < 0$) correspond to ``steeper'' UV extinction curves, while larger $\grainsizebar$ ($\Delta\Gamma>0$) correspond to ``flatter'' or more ``grey'' extinction at short wavelengths.\footnote{For a toy model with $Q=Q_{0}\,{\rm MIN}(\grainsize/c\,\lambda_{\rm rad},\,1)$ with $\grainsizemin \ll c\,\lambda_{\rm rad} \ll \grainsizemax$, the slope $\alpha$ of the extinction curve $A_{\lambda} \propto \lambda^{\alpha}$ in this intermediate range of wavelengths is modified by $\Delta \alpha \sim 0.4\,\Delta\Gamma$, while at $c\,\lambda_{\rm rad} \gg \grainsizemax$, $\alpha$ is un-modified.} 

We see in Fig.~\ref{fig:projection.maps.GMC.Q}, and quantify in Figs.~\ref{fig:los.grainsize.hist.demo}-\ref{fig:los.grainsize.vs.column}, that there can be significant small-scale variation in the GSD even within a single cloud/GMC/HII region at a given time, with slope variations from $\Delta\Gamma \sim -1$ to $+1$, corresponding to $\grainsizebar$ varying from $\sim 1/3$ to $\sim 3$ times its value for an MRN GSD, though the $1\sigma$ scatter is quite a bit smaller: $\sigma(\Delta\Gamma) \sim 0.3$ (corresponding to UV extinction curve slope variations of just $\Delta \alpha \sim \pm 0.1$). These are well within the range of ``effective'' GSDs fit to different LMC/SMC/MW regions in e.g.\ \citet{weingartner:2001.dust.size.distrib,fitzpatrick:2007.extinction.curves.vs.Rv,fitzpatrick:2009.extinction.law.variation.with.Rv}, or different sightlines within the diffuse Galactic ISM \citep{ysard:2005.dust.extinction.curve.variations.diffuse.ism,schlafly:galactic.extinction.curve.variations,wang:2017.extinction.law.variation.in.diffuse.ism} -- let alone the variation observed across different galaxies \citep{pei92:reddening.curves,calzetti:1994.dust.distributions,hopkins:dust,kriek:2013.dust.extinction.curve.vs.galaxy.type,salim:2018.extinction.curve.variations.by.galaxy}. But more importantly they are within or broadly similar to the range of extinction curve slopes or inferred GSDs observed across sightlines {\em within a single star forming complex} in Galactic or SMC/LMC regions \citep[see e.g.][]{gordon:2003.large.variations.extinction.curves.in.lmc.smc.mw.sightlines,bernard:2008.30.dor.grain.size.modifications,gosling:2009.variable.nir.extinction.towards.mw.nucleus,demarchi:2014.extinction.law.variation.within.30dor}.

From Fig.~\ref{fig:los.grainsize}, quantified in Fig.~\ref{fig:los.grainsize.vs.column}, one can see a second-order correlation in {\bf HII-N-Q} but representative of all our simulations with $Q \propto \grainsize$, for larger grains to be over-represented ($\Delta\Gamma>0$, i.e.\ flatter extinction or higher $R_{V}$) in high-density (high-$N_{H}$) regions in the filaments (where large grains have shorter mean-free paths), and correspondingly for smaller grains (steeper or lower-$R_{V}$) to be (fractionally) over-represented in lower-column regions. Comparing our other simulations in Fig.~\ref{fig:los.grainsize.vs.column} (also Fig.~\ref{fig:projection.maps.GMC.Q}), that trend is much weaker or even inverted in simulations with $Q\sim\,$constant. This follows from two simple considerations: first, in Fig.~\ref{fig:clumping.factors}, we showed the grains that dominate the opacity (hence absorption hence bulk acceleration) clump most strongly. And second, more obviously, if there is a mixture of grain sizes, the grains which dominate the opacity will be best-correlated with the total extinction -- this appears to be the dominant effect in our simulations, as demonstrated by the fact that in Fig.~\ref{fig:los.grainsize.vs.column} there is a much weaker (or even slightly opposite) correlation between $\Delta \Gamma$ and $N_{H}$, compared to the more significant correlation between $\Delta \Gamma$ and $A_{V}$.  This sort of qualitative trend of $R_{V}$ with $A_{V}$ has been known observationally for decades \citep[see][and references therein]{fitzpatrick:2009.extinction.law.variation.with.Rv,gordon:2009.fuse.extinction.curve.slope.vs.rv}, and has traditionally been interpreted in terms of grain growth/chemistry (e.g.\ grain growth in dense regions or destruction in {the} diffuse ISM; see \citealt{schnee:2014.mm.sized.grains.in.star.forming.regions,hirashita:2014.model.growth.in.dense.clouds,wang:2014.ism.extinction.variations.grain.growth.model}). Those processes certainly occur, but a quantitative model for their relative effects would require including those processes and modeling more diffuse regions. Our point here is simply that these trends can {\em also} arise entirely owing to dust dynamics.

\begin{figure}
    \centering
    \includegraphics[width=0.483\columnwidth]{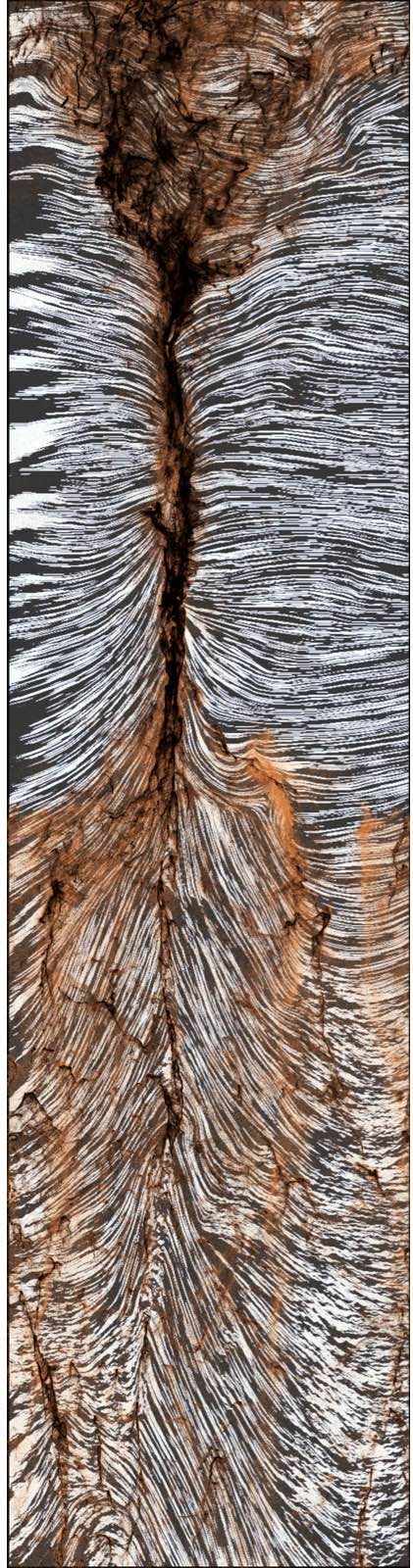}
    \includegraphics[width=0.457\columnwidth]{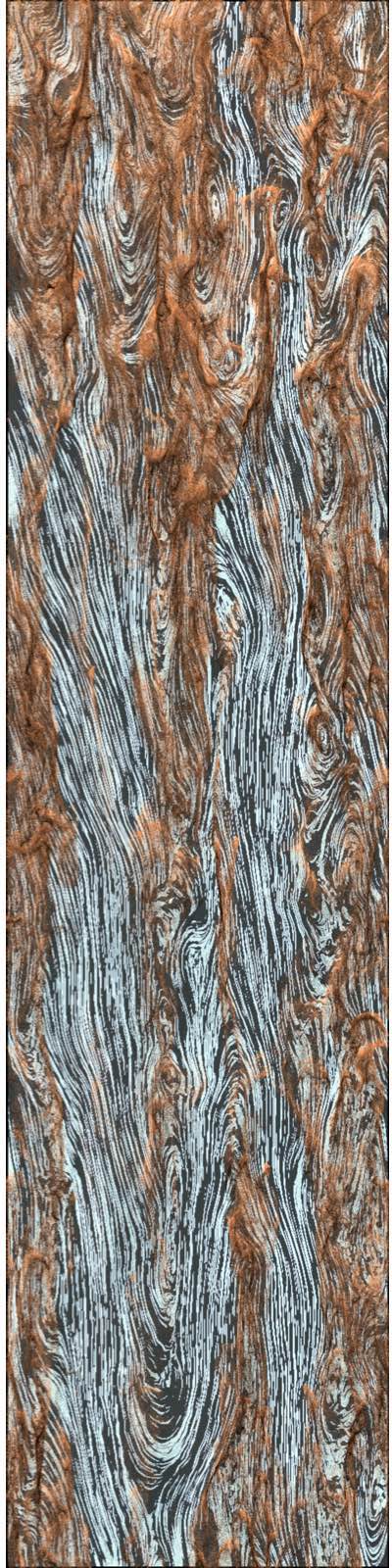}
    \caption{Example of the magnetic field line structure in fully-nonlinear stages of evolution (see \S~\ref{sec:obs:bfield}). Grains are shown in brown/black as Figs.~\ref{fig:isometric.view.GMC}-\ref{fig:pillars.of.creation} above (short axis size $\Lscale$); for gas we show a line-integral-convolution tracing the magnetic field lines (black lines). Images are taken from a {\bf GMC} run with initial ${\bf B} = B_{0}\,\hat{x}$ perpendicular to the outflow direction, at times similar to the previous images. 
    {\em Left:} An $xz$ projection, at slightly earlier times. In the upper half of the image, collapse of dust dragging gas along $\hat{\bf B}$ in one of the aligned-RDI modes creates a dense filament perpendicular to $\hat{\bf B}$. In the lower half, upward dust motion in a more diffuse structure surrounding a thinner dust filament is bending field lines to align with the  structure.
    {\em Right:} A $yz$ projection at later times. The filaments here have almost-fully aligned the magnetic field with their preferred direction (the outflow direction $\hat{z}$), despite being initially perpendicular.
    \label{fig:Bfield.morphology.demo}}
\end{figure}

\subsubsection{Magnetic Field Structure}
\label{sec:obs:bfield}

Fig.~\ref{fig:Bfield.morphology.demo} shows a typical example of the field line structure. We select a case where the initial field direction $\bhat$ is nearly perpendicular to the stratification direction $\hat{z}$ ($|\cos{\theta_{B}^{0}}|\ll 1$). As the outflows go non-linear, the fields become increasingly re-oriented to point along $\hat{z}$: non-linear structures with locally-higher $(\dustden/\gasden)$ are accelerated in $\hat{z}$ more rapidly (see \S~\ref{sec:results:general:morphology}), forming the filamentary structure and dragging gas collisionally with the dust, which (being flux-frozen since we assume ideal MHD for the gas) drag and bend the field lines. However, at some dense ``nodes,'' one can still see the initially-perpendicular fields (example in Fig.~\ref{fig:Bfield.morphology.demo}). Here we see the behavior described in \citet{hopkins:2019.mhd.rdi.periodic.box.sims} \S~5.4: the dominant mode at scales $\sim \Lscale$ (here, from linear theory, the ``quasi-sound'' mode) is one of the ``aligned,'' compressible modes which can grow rapidly and has $\hat{\bf k} \approx \bhat$  when $\tauparam \equiv \ts/\tL \gtrsim 1$ (and the drift is trans-sonic or faster), causing the dust to collapse along $\bhat$ into sheet-like structures.

In summary, we see that dense dust structures forming relatively early can, viewed edge-on, appear as filaments with $\bhat$ perpendicular to their axis; while more generally the filaments late in the non-linear evolution of the outflow at large distances have $\bhat$ parallel to the filament and outflow direction. This is tantalizingly similar to observational suggestions of parallel alignment between parallel fields and filaments along lower-density filaments and perpendicular alignment for high-dust-density structures \citep{clark:2014.magnetic.alignment.filaments,planck:2016.magnetic.field.alignment.filaments.gmcs}, although there are many other viable physical explanations for the observed behavior \citep{nakamura:2008.magnetic.alignment.filaments.from.sims}, and a number of recent studies have questioned the statistical and physical (3D) significance of those correlations \citep{planck:2016.magnetic.field.corr.structure.weak.except.highest.densities,alina:2019.magnetic.field.alignment.analysis.evidence.weak}. 

Magnetic fields can also be amplified by the RDIs, generally following scalings for a turbulent dynamo with the trans-magnetosonic velocities seen here. This is studied in \paperone, for cases with weak initial fields, but since we generally begin here initial conditions with from already-large field strengths, the amplification effects in our fiducial simulations are modest.

\section{Conclusions}
\label{sec:discussion}

We have presented the first simulations of ``radiation-dust-driven outflows'' which explicitly integrate the dust motion and dust-gas coupling, accounting for drag and Lorentz forces on grains. Specifically we simulate radiation interacting with a realistic spectrum of dust grain sizes and grain charge-to-mass ratios, which in turn interact with gas via collisional (drag) and electrodynamic forces, in a stratified inhomogeneous medium, with initial conditions chosen to resemble dusty gas in HII regions, GMCs, and star-forming regions in the Local Group. In these systems, the dust mean-free paths and gyro radii are much smaller than global scale-lengths, but not nearly as small as those of ions, so the dust cannot in fact be treated as a ``tightly coupled'' fluid. In fact, the dust is unstable to a broad spectrum of RDIs, with growth timescales even on global length scales  shorter than other large-scale flow times. These can only be captured in simulations that explicitly follow grain dynamics. Our main conclusions include:

\begin{enumerate}

\item{\bf RDIs:} The RDIs do, in fact, grow rapidly, and saturate at large non-linear amplitudes, completely changing the structure and dynamics of the outflows. Stratification, gravity, explicit radiation-hydrodynamics, grain size spectra, magnetization/charge, {and} different gas equations-of-state do not eliminate the instabilities. Ignoring the dust dynamics (treating dust as a ``tightly coupled fluid'' moving with gas, or as a constant gas opacity), the outflows here (with modest optical depths $A_{V} \lesssim$\,a few) are completely different, and form essentially no structure. Ignoring the grain interactions with magnetic fields (grain charge \&\ electromagnetic forces) leads to qualitatively different RDIs with little structure. Once explicit dust drag and electromagnetic interactions are included, the qualitative results are robust, though details depend significantly on optical properties of the grains.

\item{\bf Outflows \&\ Turbulence:} Despite the strong dust clustering induced by the RDIs, the simulations robustly launch outflows. Grains are not ``spit out'' leaving gas behind, and the inhomogeneity does not dramatically reduce the efficiency of radiation coupling. ``Leakage'' of e.g.\ UV photons should be dramatically enhanced as optically-thin channels are created, and a non-negligible fraction of the initial mass (tens of percent) can be left behind or ``sink'' as other material is accelerated, owing to inhomogeneity. The RDIs provide yet another mechanism to drive small-scale turbulence within outflows, driving trans-magnetosonic ($|\delta \gasvel| \sim v_{\rm fast}$) turbulence on scales $\sim \Lscale$ (the global outflow scale-length), with gas density fluctuations following the usual trans-sonic MHD scalings.

\item{\bf Dust Clustering:} The RDIs can drive strong micro-scale dust-dust clustering. Most dust grains locally ``see'' a median dust density of order the gas density (despite a volume-averaged dust-to-gas ratio $\sim 0.01$) -- i.e.\ the typical dust-dust clumping factor $C_{d d} \equiv \langle \dustden^{2} \rangle/\langle \dustden \rangle^{2} \sim 1/\dustgas$, for a volume-averaged dust-to-gas-ratio $\dustgas$, and the micro-scale dust-to-gas ratio can span factors $\sim 10^{8}$ at the $\pm 5\,\sigma$ level. Grain-grain relative velocities are also typically much smaller than canonical turbulence models ignoring the RDIs \citep[e.g.][]{ormel:2007.closed.form.grain.rel.velocities} would predict. Gas-gas clumping and gas-dust-cross-clumping are much weaker, more consistent with standard MHD turbulence. This can have dramatic implications for grain collisions \&\ coagulation, enhancing their rates by orders of magnitude.

\item{\bf Extinction Curve \&\ Dust-to-Gas Ratio Variation:} Despite the enormous variation in the {\em local} dust-to-gas ratio, the sightline-integrated $A_{V}/N_{H}$ varies by a modest factor $\sim2$, as the micro-scale variations are integrated out. However the fact that different grain sizes cluster differently can produce variations in the extinction curve broadly similar in magnitude and shape to those observed in many well-studied clouds. Observed second-order correlations such as a slightly non-linear $A_{V}-N_{H}$ relation or correlation of $R_{V}$ with $A_{V}$ appear in many of our simulations. These do not have to come from dust chemistry, but can arise purely from dust-dynamical processes.

\item{\bf Morphologies:} The instabilities studied drive the dust+gas morphology into filamentary structures with sub-structure including whiskers, knots, pillars, and more remarkably similar in visual morphology to observed structures in GMCs and large massive star-forming complexes. These morphologies are qualitatively distinct from simpler simulations that ignore dust dynamics entirely or ignore the Lorentz forces on grains. They are also visually distinct in a number of ways from e.g.\ radiative Rayleigh-Taylor instabilities, driven MHD turbulence, shock fronts and I-fronts, and other commonly-invoked explanations for structure in GMCs, and can explain some observed features that these phenomena cannot. The filamentary structures can collapse along magnetic field lines but also strongly re-shape the fields as they differentially accelerate, bending fields into alignment; we therefore find a heuristic mix of parallel-and-perpendicular field-filament geometries similar to recent observational suggestions.

\end{enumerate}

We stress that the simulations here are still intentionally idealized in terms of chemistry and dynamics (they are far from ``full physics'' star formation \&\ GMC dynamics models). However this has allowed us to identify the most important physics in various regimes and isolate the role of the RDIs. In future work, it will be interesting and important to explore the effects of additional physics, such as ionization and radiation pressure on gas, more realistic geometries, more detailed dust optical properties (and corresponding charge and acceleration laws and size distributions). It is also important to explore very different regimes, where the relevant limits of the RDIs or radiation could be quite different -- for example, dust-driven outflows around AGN, or in cool-star photospheres, or planetary atmospheres, where the dominant modes are qualitatively distinct and relevant values of some of the key parameters here can be several-orders-of-magnitude different. Another essential goal for future work will be to explore more quantitative metrics to compare the morphology of the structures predicted here to observations: the striking visual similarity and contrast from e.g. pure driven-MHD turbulence simulations should be testable with the ongoing development of novel quantitative morphological and topological measures.

\acknowledgments{Support for PFH was provided by NSF Research Grants 1911233 \&\ 20009234, NSF CAREER grant 1455342, NASA grants 80NSSC18K0562, HST-AR-15800.001-A. Numerical calculations were run on the Caltech compute cluster ``Wheeler,'' allocations FTA-Hopkins supported by the NSF and TACC, and NASA HEC SMD-16-7592. Support for ALR was provided by the Institute for Theory \& Computation at Harvard University. GVP acknowledges support by NASA through the NASA Hubble Fellowship grant \#HST-HF2-51444.001-A.}

\datastatement{The data supporting the plots within this article are available on reasonable request to the corresponding author. A public version of the GIZMO code is available at \gizmourl.}

\bibliography{ms_extracted}

\begin{thebibliography}{}
\makeatletter
\relax
\def\mn@urlcharsother{\let\do\@makeother \do\$\do\&\do\#\do\^\do\_\do\%\do\~}
\def\mn@doi{\begingroup\mn@urlcharsother \@ifnextchar [ {\mn@doi@}
  {\mn@doi@[]}}
\def\mn@doi@[#1]#2{\def\@tempa{#1}\ifx\@tempa\@empty \href
  {http://dx.doi.org/#2} {doi:#2}\else \href {http://dx.doi.org/#2} {#1}\fi
  \endgroup}
\def\mn@eprint#1#2{\mn@eprint@#1:#2::\@nil}
\def\mn@eprint@arXiv#1{\href {http://arxiv.org/abs/#1} {{\tt arXiv:#1}}}
\def\mn@eprint@dblp#1{\href {http://dblp.uni-trier.de/rec/bibtex/#1.xml}
  {dblp:#1}}
\def\mn@eprint@#1:#2:#3:#4\@nil{\def\@tempa {#1}\def\@tempb {#2}\def\@tempc
  {#3}\ifx \@tempc \@empty \let \@tempc \@tempb \let \@tempb \@tempa \fi \ifx
  \@tempb \@empty \def\@tempb {arXiv}\fi \@ifundefined
  {mn@eprint@\@tempb}{\@tempb:\@tempc}{\expandafter \expandafter \csname
  mn@eprint@\@tempb\endcsname \expandafter{\@tempc}}}

\bibitem[\protect\citeauthoryear{{Abergel} et~al.,}{{Abergel}
  et~al.}{2002}]{abergel:2002.size.segregation.effects.seen.in.orion.small.dust.abundances}
{Abergel} A.,  et~al., 2002, \mn@doi [\aap] {10.1051/0004-6361:20020324}, \href
  {https://ui.adsabs.harvard.edu/abs/2002A&A...389..239A} {389, 239}

\bibitem[\protect\citeauthoryear{{Akimkin}, {Kirsanova}, {Pavlyuchenkov}  \&
  {Wiebe}}{{Akimkin}
  et~al.}{2017}]{akimkin:2017.dustgasHII.expansion.cavity.density.metallicity}
{Akimkin} V.~V.,  {Kirsanova} M.~S.,  {Pavlyuchenkov} Y.~N.,   {Wiebe} D.~S.,
  2017, \mn@doi [\mnras] {10.1093/mnras/stx797}, \href
  {http://adsabs.harvard.edu/abs/2017MNRAS.469..630A} {469, 630}

\bibitem[\protect\citeauthoryear{{Alina}, {Ristorcelli}, {Montier},
  {Abdikamalov}, {Juvela}, {Ferri{\`e}re}, {Bernard}  \& {Micelotta}}{{Alina}
  et~al.}{2019}]{alina:2019.magnetic.field.alignment.analysis.evidence.weak}
{Alina} D.,  {Ristorcelli} I.,  {Montier} L.,  {Abdikamalov} E.,  {Juvela} M.,
  {Ferri{\`e}re} K.,  {Bernard} J.~P.,   {Micelotta} E.~R.,  2019, \mn@doi
  [\mnras] {10.1093/mnras/stz508}, \href
  {https://ui.adsabs.harvard.edu/abs/2019MNRAS.485.2825A} {485, 2825}

\bibitem[\protect\citeauthoryear{{Anderson} et~al.,}{{Anderson}
  et~al.}{2010}]{anderson:2010.dust.HII.leakage.clumps.induced.sf.filaments}
{Anderson} L.~D.,  et~al., 2010, \mn@doi [\aap] {10.1051/0004-6361/201014657},
  \href {http://adsabs.harvard.edu/abs/2010A%26A...518L..99A} {518, L99}

\bibitem[\protect\citeauthoryear{{Andr{\'e}}}{{Andr{\'e}}}{2017}]{andre:2017.filaments.in.gmcs.star.formation.obs.review}
{Andr{\'e}} P.,  2017, \mn@doi [Comptes Rendus Geoscience]
  {10.1016/j.crte.2017.07.002}, \href
  {https://ui.adsabs.harvard.edu/abs/2017CRGeo.349..187A} {349, 187}

\bibitem[\protect\citeauthoryear{{Apai} \& {Lauretta}}{{Apai} \&
  {Lauretta}}{2010}]{apai:dust.review}
{Apai} D.~A.,  {Lauretta} D.~S.,  2010, {Protoplanetary Dust: Astrophysical and
  Cosmochemical Perspectives}.
Cambridge University Press, Cambridge, UK; eds.: D. Apai, D. S. Lauretta

\bibitem[\protect\citeauthoryear{{Apai}, {Linz}, {Henning}  \&
  {Stecklum}}{{Apai}
  et~al.}{2005}]{apai:2005.HII.region.filamentary.dust.structures}
{Apai} D.,  {Linz} H.,  {Henning} T.,   {Stecklum} B.,  2005, \mn@doi [\aap]
  {10.1051/0004-6361:20035890}, \href
  {http://adsabs.harvard.edu/abs/2005A%26A...434..987A} {434, 987}

\bibitem[\protect\citeauthoryear{{Arthur}, {Henney}, {Mellema}, {de Colle}  \&
  {V{\'a}zquez-Semadeni}}{{Arthur}
  et~al.}{2011}]{arthur:2011.rhd.hii.region.sims.morph.structure}
{Arthur} S.~J.,  {Henney} W.~J.,  {Mellema} G.,  {de Colle} F.,
  {V{\'a}zquez-Semadeni} E.,  2011, \mn@doi [\mnras]
  {10.1111/j.1365-2966.2011.18507.x}, \href
  {https://ui.adsabs.harvard.edu/abs/2011MNRAS.414.1747A} {414, 1747}

\bibitem[\protect\citeauthoryear{{Bai} \& {Stone}}{{Bai} \&
  {Stone}}{2010}]{bai:2010.grain.streaming.vs.diskparams}
{Bai} X.-N.,  {Stone} J.~M.,  2010, \mn@doi [\apjl]
  {10.1088/2041-8205/722/2/L220}, \href
  {http://adsabs.harvard.edu/abs/2010ApJ...722L.220B} {722, L220}

\bibitem[\protect\citeauthoryear{{Bauke}}{{Bauke}}{2007}]{bauke:2007.maximum.likelihood.methods.for.discrete.power.law.distributed.data}
{Bauke} H.,  2007, \mn@doi [European Physical Journal B]
  {10.1140/epjb/e2007-00219-y}, \href
  {https://ui.adsabs.harvard.edu/abs/2007EPJB...58..167B} {58, 167}

\bibitem[\protect\citeauthoryear{{Bec}, {Biferale}, {Cencini}, {Lanotte}  \&
  {Toschi}}{{Bec}
  et~al.}{2009}]{bec:2009.caustics.intermittency.key.to.largegrain.clustering}
{Bec} J.,  {Biferale} L.,  {Cencini} M.,  {Lanotte} A.~S.,   {Toschi} F.,
  2009, eprint arxiv:0905.1192, \href
  {http://adsabs.harvard.edu/abs/2009arXiv0905.1192B} {}

\bibitem[\protect\citeauthoryear{{Bell}}{{Bell}}{2004}]{bell.2004.cosmic.rays}
{Bell} A.~R.,  2004, \mn@doi [\mnras] {10.1111/j.1365-2966.2004.08097.x}, \href
  {http://adsabs.harvard.edu/abs/2004MNRAS.353..550B} {353, 550}

\bibitem[\protect\citeauthoryear{{Benincasa} et~al.,}{{Benincasa}
  et~al.}{2020}]{benincasa:2020.gmc.lifetimes.fire}
{Benincasa} S.~M.,  et~al., 2020, \mn@doi [\mnras] {10.1093/mnras/staa2116},
  \href {https://ui.adsabs.harvard.edu/abs/2020MNRAS.497.3993B} {497, 3993}

\bibitem[\protect\citeauthoryear{{Bernard} et~al.,}{{Bernard}
  et~al.}{2008}]{bernard:2008.30.dor.grain.size.modifications}
{Bernard} J.-P.,  et~al., 2008, \mn@doi [\aj] {10.1088/0004-6256/136/3/919},
  \href {https://ui.adsabs.harvard.edu/abs/2008AJ....136..919B} {136, 919}

\bibitem[\protect\citeauthoryear{{Berruyer}}{{Berruyer}}{1991}]{berruyer:dust.wind.unstable.pressure.gradient}
{Berruyer} N.,  1991, \aap, \href
  {http://adsabs.harvard.edu/abs/1991A%26A...249..181B} {249, 181}

\bibitem[\protect\citeauthoryear{{Bialy} \& {Burkhart}}{{Bialy} \&
  {Burkhart}}{2020}]{2020ApJ...894L...2B}
{Bialy} S.,  {Burkhart} B.,  2020, \mn@doi [\apjl] {10.3847/2041-8213/ab8a32},
  \href {https://ui.adsabs.harvard.edu/abs/2020ApJ...894L...2B} {894, L2}

\bibitem[\protect\citeauthoryear{{Bolatto}, {Leroy}, {Rosolowsky}, {Walter}  \&
  {Blitz}}{{Bolatto} et~al.}{2008}]{bolatto:2008.gmc.properties}
{Bolatto} A.~D.,  {Leroy} A.~K.,  {Rosolowsky} E.,  {Walter} F.,   {Blitz} L.,
  2008, \mn@doi [\apj] {10.1086/591513}, \href
  {http://adsabs.harvard.edu/abs/2008ApJ...686..948B} {686, 948}

\bibitem[\protect\citeauthoryear{{Calzetti}, {Kinney}  \&
  {Storchi-Bergmann}}{{Calzetti}
  et~al.}{1994}]{calzetti:1994.dust.distributions}
{Calzetti} D.,  {Kinney} A.~L.,   {Storchi-Bergmann} T.,  1994, \mn@doi [\apj]
  {10.1086/174346}, \href {http://adsabs.harvard.edu/abs/1994ApJ...429..582C}
  {429, 582}

\bibitem[\protect\citeauthoryear{{Carballido}, {Stone}  \&
  {Turner}}{{Carballido}
  et~al.}{2008}]{carballido:2008.grain.streaming.instab.sims}
{Carballido} A.,  {Stone} J.~M.,   {Turner} N.~J.,  2008, \mn@doi [\mnras]
  {10.1111/j.1365-2966.2008.13014.x}, \href
  {http://adsabs.harvard.edu/abs/2008MNRAS.386..145C} {386, 145}

\bibitem[\protect\citeauthoryear{{Chang}, {Schiano}  \& {Wolfe}}{{Chang}
  et~al.}{1987}]{chang:qso.rad.feedback}
{Chang} C.~A.,  {Schiano} A.~V.~R.,   {Wolfe} A.~M.,  1987, \mn@doi [\apj]
  {10.1086/165714}, \href {http://adsabs.harvard.edu/abs/1987ApJ...322..180C}
  {322, 180}

\bibitem[\protect\citeauthoryear{{Clark}, {Peek}  \& {Putman}}{{Clark}
  et~al.}{2014}]{clark:2014.magnetic.alignment.filaments}
{Clark} S.~E.,  {Peek} J.~E.~G.,   {Putman} M.~E.,  2014, \mn@doi [\apj]
  {10.1088/0004-637X/789/1/82}, \href
  {https://ui.adsabs.harvard.edu/abs/2014ApJ...789...82C} {789, 82}

\bibitem[\protect\citeauthoryear{{Colbrook}, {Ma}, {Hopkins}  \&
  {Squire}}{{Colbrook} et~al.}{2017}]{colbrook:passive.scalar.scalings}
{Colbrook} M.~J.,  {Ma} X.,  {Hopkins} P.~F.,   {Squire} J.,  2017, \mn@doi
  [\mnras] {10.1093/mnras/stx261}, \href
  {http://adsabs.harvard.edu/abs/2017MNRAS.467.2421C} {467, 2421}

\bibitem[\protect\citeauthoryear{{Crutcher}, {Wandelt}, {Heiles}, {Falgarone}
  \& {Troland}}{{Crutcher} et~al.}{2010}]{crutcher:cloud.b.fields}
{Crutcher} R.~M.,  {Wandelt} B.,  {Heiles} C.,  {Falgarone} E.,   {Troland}
  T.~H.,  2010, \mn@doi [\apj] {10.1088/0004-637X/725/1/466}, \href
  {http://adsabs.harvard.edu/abs/2010ApJ...725..466C} {725, 466}

\bibitem[\protect\citeauthoryear{{Cuzzi}, {Hogan}, {Paque}  \&
  {Dobrovolskis}}{{Cuzzi}
  et~al.}{2001}]{cuzzi:2001.grain.concentration.chondrules}
{Cuzzi} J.~N.,  {Hogan} R.~C.,  {Paque} J.~M.,   {Dobrovolskis} A.~R.,  2001,
  \mn@doi [\apj] {10.1086/318233}, \href
  {http://adsabs.harvard.edu/abs/2001ApJ...546..496C} {546, 496}

\bibitem[\protect\citeauthoryear{{Davis}, {Jiang}, {Stone}  \&
  {Murray}}{{Davis} et~al.}{2014}]{davis:2014.rad.pressure.outflows}
{Davis} S.~W.,  {Jiang} Y.-F.,  {Stone} J.~M.,   {Murray} N.,  2014, \mn@doi
  [\apj] {10.1088/0004-637X/796/2/107}, \href
  {http://adsabs.harvard.edu/abs/2014ApJ...796..107D} {796, 107}

\bibitem[\protect\citeauthoryear{{De Marchi} \& {Panagia}}{{De Marchi} \&
  {Panagia}}{2014}]{demarchi:2014.extinction.law.variation.within.30dor}
{De Marchi} G.,  {Panagia} N.,  2014, \mn@doi [\mnras] {10.1093/mnras/stu1694},
  \href {https://ui.adsabs.harvard.edu/abs/2014MNRAS.445...93D} {445, 93}

\bibitem[\protect\citeauthoryear{{Dorschner}}{{Dorschner}}{2003}]{dorschner:dust.mineralogy.review}
{Dorschner} J.,  2003, in {Henning} T.~K.,  ed.,  Lecture Notes in Physics,
  Berlin Springer Verlag Vol. 609, Astromineralogy; University Observatory
  Schillerg{\"a}sschen 3, D-07745 Jena, Germany. pp 1--54

\bibitem[\protect\citeauthoryear{{Draine}}{{Draine}}{2003}]{draine:2003.dust.review}
{Draine} B.~T.,  2003, \mn@doi [\araa]
  {10.1146/annurev.astro.41.011802.094840}, \href
  {http://adsabs.harvard.edu/abs/2003ARA%26A..41..241D} {41, 241}

\bibitem[\protect\citeauthoryear{{Draine} \& {Sutin}}{{Draine} \&
  {Sutin}}{1987}]{draine:1987.grain.charging}
{Draine} B.~T.,  {Sutin} B.,  1987, \mn@doi [\apj] {10.1086/165596}, \href
  {http://adsabs.harvard.edu/abs/1987ApJ...320..803D} {320, 803}

\bibitem[\protect\citeauthoryear{{Fitzpatrick} \& {Massa}}{{Fitzpatrick} \&
  {Massa}}{2007}]{fitzpatrick:2007.extinction.curves.vs.Rv}
{Fitzpatrick} E.~L.,  {Massa} D.,  2007, \mn@doi [\apj] {10.1086/518158}, \href
  {https://ui.adsabs.harvard.edu/abs/2007ApJ...663..320F} {663, 320}

\bibitem[\protect\citeauthoryear{{Fitzpatrick} \& {Massa}}{{Fitzpatrick} \&
  {Massa}}{2009}]{fitzpatrick:2009.extinction.law.variation.with.Rv}
{Fitzpatrick} E.~L.,  {Massa} D.,  2009, \mn@doi [\apj]
  {10.1088/0004-637X/699/2/1209}, \href
  {https://ui.adsabs.harvard.edu/abs/2009ApJ...699.1209F} {699, 1209}

\bibitem[\protect\citeauthoryear{{Franco}, {Ferrini}, {Barsella}  \&
  {Ferrara}}{{Franco}
  et~al.}{1991}]{franco:dust.rad.pressure.galactic.fountain}
{Franco} J.,  {Ferrini} F.,  {Barsella} B.,   {Ferrara} A.,  1991, \mn@doi
  [\apj] {10.1086/169578}, \href
  {http://adsabs.harvard.edu/abs/1991ApJ...366..443F} {366, 443}

\bibitem[\protect\citeauthoryear{{Galliano} et~al.,}{{Galliano}
  et~al.}{2011}]{galliano:2011.lmc.regional.dust.to.gas.variations}
{Galliano} F.,  et~al., 2011, \mn@doi [\aap] {10.1051/0004-6361/201117952},
  \href {https://ui.adsabs.harvard.edu/abs/2011A&A...536A..88G} {536, A88}

\bibitem[\protect\citeauthoryear{{Garnett} \& {Dinerstein}}{{Garnett} \&
  {Dinerstein}}{2001}]{garnett:2001.ring.nebula.dusty.knots.offset.emission.line.no.gas.overdensity}
{Garnett} D.~R.,  {Dinerstein} H.~L.,  2001, \mn@doi [\apj] {10.1086/322452},
  \href {http://adsabs.harvard.edu/abs/2001ApJ...558..145G} {558, 145}

\bibitem[\protect\citeauthoryear{{Genel}, {Vogelsberger}, {Nelson}, {Sijacki},
  {Springel}  \& {Hernquist}}{{Genel}
  et~al.}{2013}]{genel:tracer.particle.method}
{Genel} S.,  {Vogelsberger} M.,  {Nelson} D.,  {Sijacki} D.,  {Springel} V.,
  {Hernquist} L.,  2013, \mn@doi [\mnras] {10.1093/mnras/stt1383}, \href
  {http://adsabs.harvard.edu/abs/2013MNRAS.435.1426G} {435, 1426}

\bibitem[\protect\citeauthoryear{{Goldsmith}, {Heyer}, {Narayanan}, {Snell},
  {Li}  \& {Brunt}}{{Goldsmith}
  et~al.}{2008}]{goldsmith:2008.taurus.gmc.mapping}
{Goldsmith} P.~F.,  {Heyer} M.,  {Narayanan} G.,  {Snell} R.,  {Li} D.,
  {Brunt} C.,  2008, \mn@doi [\apj] {10.1086/587166}, \href
  {http://adsabs.harvard.edu/abs/2008ApJ...680..428G} {680, 428}

\bibitem[\protect\citeauthoryear{{Gordon}, {Clayton}, {Misselt}, {Landolt}  \&
  {Wolff}}{{Gordon}
  et~al.}{2003}]{gordon:2003.large.variations.extinction.curves.in.lmc.smc.mw.sightlines}
{Gordon} K.~D.,  {Clayton} G.~C.,  {Misselt} K.~A.,  {Landolt} A.~U.,   {Wolff}
  M.~J.,  2003, \mn@doi [\apj] {10.1086/376774}, \href
  {http://adsabs.harvard.edu/abs/2003ApJ...594..279G} {594, 279}

\bibitem[\protect\citeauthoryear{{Gordon}, {Cartledge}  \& {Clayton}}{{Gordon}
  et~al.}{2009}]{gordon:2009.fuse.extinction.curve.slope.vs.rv}
{Gordon} K.~D.,  {Cartledge} S.,   {Clayton} G.~C.,  2009, \mn@doi [\apj]
  {10.1088/0004-637X/705/2/1320}, \href
  {https://ui.adsabs.harvard.edu/abs/2009ApJ...705.1320G} {705, 1320}

\bibitem[\protect\citeauthoryear{{Gosling}, {Bandyopadhyay}  \&
  {Blundell}}{{Gosling}
  et~al.}{2009}]{gosling:2009.variable.nir.extinction.towards.mw.nucleus}
{Gosling} A.~J.,  {Bandyopadhyay} R.~M.,   {Blundell} K.~M.,  2009, \mn@doi
  [\mnras] {10.1111/j.1365-2966.2009.14493.x}, \href
  {https://ui.adsabs.harvard.edu/abs/2009MNRAS.394.2247G} {394, 2247}

\bibitem[\protect\citeauthoryear{{Gritschneder}, {Burkert}, {Naab}  \&
  {Walch}}{{Gritschneder}
  et~al.}{2010}]{gritschneder:2010.pillar.hii.region.formation}
{Gritschneder} M.,  {Burkert} A.,  {Naab} T.,   {Walch} S.,  2010, \mn@doi
  [\apj] {10.1088/0004-637X/723/2/971}, \href
  {https://ui.adsabs.harvard.edu/abs/2010ApJ...723..971G} {723, 971}

\bibitem[\protect\citeauthoryear{{Grudi{\'c}}, {Hopkins},
  {Faucher-Gigu{\`e}re}, {Quataert}, {Murray}  \& {Kere{\v s}}}{{Grudi{\'c}}
  et~al.}{2018}]{grudic:sfe.cluster.form.surface.density}
{Grudi{\'c}} M.~Y.,  {Hopkins} P.~F.,  {Faucher-Gigu{\`e}re} C.-A.,  {Quataert}
  E.,  {Murray} N.,   {Kere{\v s}} D.,  2018, \mn@doi [\mnras]
  {10.1093/mnras/sty035}, \href
  {http://adsabs.harvard.edu/abs/2018MNRAS.475.3511G} {475, 3511}

\bibitem[\protect\citeauthoryear{{Grudi{\'c}}, {Hopkins}, {Lee}, {Murray},
  {Faucher-Gigu{\`e}re}  \& {Johnson}}{{Grudi{\'c}}
  et~al.}{2019}]{grudic:sfe.gmcs.vs.obs}
{Grudi{\'c}} M.~Y.,  {Hopkins} P.~F.,  {Lee} E.~J.,  {Murray} N.,
  {Faucher-Gigu{\`e}re} C.-A.,   {Johnson} L.~C.,  2019, \mn@doi [\mnras]
  {10.1093/mnras/stz1758}, \href
  {https://ui.adsabs.harvard.edu/abs/2019MNRAS.488.1501G} {488, 1501}

\bibitem[\protect\citeauthoryear{{Grudi{\'c}}, {Guszejnov}, {Hopkins}, {Offner}
   \& {Faucher-Gigu{\`e}re}}{{Grudi{\'c}}
  et~al.}{2020}]{grudic:starforge.methods}
{Grudi{\'c}} M.~Y.,  {Guszejnov} D.,  {Hopkins} P.~F.,  {Offner} S. S.~R.,
  {Faucher-Gigu{\`e}re} C.-A.,  2020, MNRAS, submitted, arXiv:2010.11254, \href
  {https://ui.adsabs.harvard.edu/abs/2020arXiv201011254G} {p. arXiv:2010.11254}

\bibitem[\protect\citeauthoryear{{Guszejnov}, {Hopkins}  \&
  {Graus}}{{Guszejnov}
  et~al.}{2019}]{guszejnov:2019.imf.variation.vs.galaxy.props.not.variable}
{Guszejnov} D.,  {Hopkins} P.~F.,   {Graus} A.~S.,  2019, \mn@doi [\mnras]
  {10.1093/mnras/stz736}, \href
  {https://ui.adsabs.harvard.edu/abs/2019MNRAS.485.4852G} {485, 4852}

\bibitem[\protect\citeauthoryear{{Guszejnov}, {Grudi{\'c}}, {Offner},
  {Boylan-Kolchin}, {Faucher-Gigu{\`e}re}, {Wetzel}, {Benincasa}  \&
  {Loebman}}{{Guszejnov} et~al.}{2020}]{guszejnov:fire.gmc.props.vs.z}
{Guszejnov} D.,  {Grudi{\'c}} M.~Y.,  {Offner} S. S.~R.,  {Boylan-Kolchin} M.,
  {Faucher-Gigu{\`e}re} C.-A.,  {Wetzel} A.,  {Benincasa} S.~M.,   {Loebman}
  S.,  2020, \mn@doi [\mnras] {10.1093/mnras/stz3527}, \href
  {https://ui.adsabs.harvard.edu/abs/2020MNRAS.492..488G} {492, 488}

\bibitem[\protect\citeauthoryear{{G{\"u}ver} \& {{\"O}zel}}{{G{\"u}ver} \&
  {{\"O}zel}}{2009}]{guver:2009.av.nh.correlation}
{G{\"u}ver} T.,  {{\"O}zel} F.,  2009, \mn@doi [\mnras]
  {10.1111/j.1365-2966.2009.15598.x}, \href
  {https://ui.adsabs.harvard.edu/abs/2009MNRAS.400.2050G} {400, 2050}

\bibitem[\protect\citeauthoryear{{Hartquist} \& {Havnes}}{{Hartquist} \&
  {Havnes}}{1994}]{hartquist:bfield.dust.coupling.cool.star.winds}
{Hartquist} T.~W.,  {Havnes} O.,  1994, \mn@doi [Astrophysics and Space
  Science] {10.1007/BF00658063}, \href
  {http://adsabs.harvard.edu/abs/1994Ap%26SS.218...23H} {218, 23}

\bibitem[\protect\citeauthoryear{{Heckman}, {Armus}  \& {Miley}}{{Heckman}
  et~al.}{1990}]{heckman:1990.sb.superwinds}
{Heckman} T.~M.,  {Armus} L.,   {Miley} G.~K.,  1990, \mn@doi [\apjs]
  {10.1086/191522}, \href {http://adsabs.harvard.edu/abs/1990ApJS...74..833H}
  {74, 833}

\bibitem[\protect\citeauthoryear{{Hirashita} \& {Voshchinnikov}}{{Hirashita} \&
  {Voshchinnikov}}{2014}]{hirashita:2014.model.growth.in.dense.clouds}
{Hirashita} H.,  {Voshchinnikov} N.~V.,  2014, \mn@doi [\mnras]
  {10.1093/mnras/stt1997}, \href
  {https://ui.adsabs.harvard.edu/abs/2014MNRAS.437.1636H} {437, 1636}

\bibitem[\protect\citeauthoryear{{H{\"o}fner} \& {Olofsson}}{{H{\"o}fner} \&
  {Olofsson}}{2018}]{2018A&ARv..26....1H}
{H{\"o}fner} S.,  {Olofsson} H.,  2018, \mn@doi [\aapr]
  {10.1007/s00159-017-0106-5}, \href
  {https://ui.adsabs.harvard.edu/abs/2018A&ARv..26....1H} {26, 1}

\bibitem[\protect\citeauthoryear{{Hopkins}}{{Hopkins}}{2013a}]{hopkins:frag.theory}
{Hopkins} P.~F.,  2013a, \mn@doi [\mnras] {10.1093/mnras/sts704}, \href
  {http://adsabs.harvard.edu/abs/2013MNRAS.430.1653H} {430, 1653}

\bibitem[\protect\citeauthoryear{{Hopkins}}{{Hopkins}}{2013b}]{hopkins:2012.intermittent.turb.density.pdfs}
{Hopkins} P.~F.,  2013b, \mn@doi [\mnras] {10.1093/mnras/stt010}, \href
  {http://adsabs.harvard.edu/abs/2013MNRAS.430.1880H} {430, 1880}

\bibitem[\protect\citeauthoryear{{Hopkins}}{{Hopkins}}{2014}]{hopkins:totally.metal.stars}
{Hopkins} P.~F.,  2014, \mn@doi [\apj] {10.1088/0004-637X/797/1/59}, \href
  {http://adsabs.harvard.edu/abs/2014ApJ...797...59H} {797, 59}

\bibitem[\protect\citeauthoryear{{Hopkins}}{{Hopkins}}{2015}]{hopkins:gizmo}
{Hopkins} P.~F.,  2015, \mn@doi [\mnras] {10.1093/mnras/stv195}, \href
  {http://adsabs.harvard.edu/abs/2015MNRAS.450...53H} {450, 53}

\bibitem[\protect\citeauthoryear{{Hopkins}}{{Hopkins}}{2016a}]{hopkins:2013.grain.clustering}
{Hopkins} P.~F.,  2016a, \mn@doi [\mnras] {10.1093/mnras/stv2226}, \href
  {http://adsabs.harvard.edu/abs/2016MNRAS.455...89H} {455, 89}

\bibitem[\protect\citeauthoryear{{Hopkins}}{{Hopkins}}{2016b}]{hopkins:cg.mhd.gizmo}
{Hopkins} P.~F.,  2016b, \mn@doi [\mnras] {10.1093/mnras/stw1578}, \href
  {http://adsabs.harvard.edu/abs/2016MNRAS.462..576H} {462, 576}

\bibitem[\protect\citeauthoryear{{Hopkins}}{{Hopkins}}{2017}]{hopkins:gizmo.diffusion}
{Hopkins} P.~F.,  2017, \mn@doi [\mnras] {10.1093/mnras/stw3306}, \href
  {http://adsabs.harvard.edu/abs/2017MNRAS.466.3387H} {466, 3387}

\bibitem[\protect\citeauthoryear{{Hopkins} \& {Conroy}}{{Hopkins} \&
  {Conroy}}{2017}]{hopkins.conroy.2015:metal.poor.star.abundances.dust}
{Hopkins} P.~F.,  {Conroy} C.,  2017, \mn@doi [\apj]
  {10.3847/1538-4357/835/2/154}, \href
  {http://adsabs.harvard.edu/abs/2017ApJ...835..154H} {835, 154}

\bibitem[\protect\citeauthoryear{{Hopkins} \& {Grudi{\'c}}}{{Hopkins} \&
  {Grudi{\'c}}}{2019}]{hopkins:2019.grudic.photon.momentum.rad.pressure.coupling}
{Hopkins} P.~F.,  {Grudi{\'c}} M.~Y.,  2019, \mn@doi [\mnras]
  {10.1093/mnras/sty3089}, \href
  {https://ui.adsabs.harvard.edu/abs/2019MNRAS.483.4187H} {483, 4187}

\bibitem[\protect\citeauthoryear{{Hopkins} \& {Lee}}{{Hopkins} \&
  {Lee}}{2016}]{hopkins.2016:dust.gas.molecular.cloud.dynamics.sims}
{Hopkins} P.~F.,  {Lee} H.,  2016, \mn@doi [\mnras] {10.1093/mnras/stv2745},
  \href {http://adsabs.harvard.edu/abs/2016MNRAS.456.4174H} {456, 4174}

\bibitem[\protect\citeauthoryear{{Hopkins} \& {Raives}}{{Hopkins} \&
  {Raives}}{2016}]{hopkins:mhd.gizmo}
{Hopkins} P.~F.,  {Raives} M.~J.,  2016, \mn@doi [\mnras]
  {10.1093/mnras/stv2180}, \href
  {http://adsabs.harvard.edu/abs/2016MNRAS.455...51H} {455, 51}

\bibitem[\protect\citeauthoryear{{Hopkins} \& {Squire}}{{Hopkins} \&
  {Squire}}{2018a}]{hopkins:2018.mhd.rdi}
{Hopkins} P.~F.,  {Squire} J.,  2018a, \mn@doi [\mnras]
  {10.1093/mnras/sty1604}, \href
  {http://adsabs.harvard.edu/abs/2018MNRAS.479.4681H} {479, 4681}

\bibitem[\protect\citeauthoryear{{Hopkins} \& {Squire}}{{Hopkins} \&
  {Squire}}{2018b}]{hopkins:2017.acoustic.RDI}
{Hopkins} P.~F.,  {Squire} J.,  2018b, \mn@doi [\mnras]
  {10.1093/mnras/sty1982}, \href
  {http://adsabs.harvard.edu/abs/2018MNRAS.480.2813H} {480, 2813}

\bibitem[\protect\citeauthoryear{{Hopkins} et~al.,}{{Hopkins}
  et~al.}{2004}]{hopkins:dust}
{Hopkins} P.~F.,  et~al., 2004, \mn@doi [\aj] {10.1086/423291}, \href
  {http://adsabs.harvard.edu/abs/2004AJ....128.1112H} {128, 1112}

\bibitem[\protect\citeauthoryear{{Hopkins}, {Grudi{\'c}}, {Wetzel},
  {Kere{\v{s}}}, {Faucher-Gigu{\`e}re}, {Ma}, {Murray}  \& {Butcher}}{{Hopkins}
  et~al.}{2020a}]{hopkins:radiation.methods}
{Hopkins} P.~F.,  {Grudi{\'c}} M.~Y.,  {Wetzel} A.,  {Kere{\v{s}}} D.,
  {Faucher-Gigu{\`e}re} C.-A.,  {Ma} X.,  {Murray} N.,   {Butcher} N.,  2020a,
  \mn@doi [\mnras] {10.1093/mnras/stz3129}, \href
  {https://ui.adsabs.harvard.edu/abs/2020MNRAS.491.3702H} {491, 3702}

\bibitem[\protect\citeauthoryear{{Hopkins}, {Squire}  \& {Seligman}}{{Hopkins}
  et~al.}{2020b}]{hopkins:2019.mhd.rdi.periodic.box.sims}
{Hopkins} P.~F.,  {Squire} J.,   {Seligman} D.,  2020b, \mn@doi [\mnras]
  {10.1093/mnras/staa1046}, \href
  {https://ui.adsabs.harvard.edu/abs/2020MNRAS.496.2123H} {496, 2123}

\bibitem[\protect\citeauthoryear{{Johansen}, {Youdin}  \& {Mac Low}}{{Johansen}
  et~al.}{2009}]{johansen:2009.particle.clumping.metallicity.dependence}
{Johansen} A.,  {Youdin} A.,   {Mac Low} M.-M.,  2009, \mn@doi [\apjl]
  {10.1088/0004-637X/704/2/L75}, \href
  {http://adsabs.harvard.edu/abs/2009ApJ...704L..75J} {704, L75}

\bibitem[\protect\citeauthoryear{{Kim}, {Kim}  \& {Ostriker}}{{Kim}
  et~al.}{2018}]{2018ApJ...859...68K}
{Kim} J.-G.,  {Kim} W.-T.,   {Ostriker} E.~C.,  2018, \mn@doi [\apj]
  {10.3847/1538-4357/aabe27}, \href
  {https://ui.adsabs.harvard.edu/abs/2018ApJ...859...68K} {859, 68}

\bibitem[\protect\citeauthoryear{{Kirk}, {Klassen}, {Pudritz}  \&
  {Pillsworth}}{{Kirk} et~al.}{2015}]{kirk:2015.turb.filaments.vs.obs}
{Kirk} H.,  {Klassen} M.,  {Pudritz} R.,   {Pillsworth} S.,  2015, \mn@doi
  [\apj] {10.1088/0004-637X/802/2/75}, \href
  {https://ui.adsabs.harvard.edu/abs/2015ApJ...802...75K} {802, 75}

\bibitem[\protect\citeauthoryear{{Klaassen}, {Reiter}, {McLeod}, {Mottram},
  {Dale}  \& {Gritschneder}}{{Klaassen}
  et~al.}{2020}]{klaassen:2020.pillars.carina.alma.obs.favor.ion.models}
{Klaassen} P.~D.,  {Reiter} M.~R.,  {McLeod} A.~F.,  {Mottram} J.~C.,  {Dale}
  J.~E.,   {Gritschneder} M.,  2020, \mn@doi [\mnras] {10.1093/mnras/stz3012},
  \href {https://ui.adsabs.harvard.edu/abs/2020MNRAS.491..178K} {491, 178}

\bibitem[\protect\citeauthoryear{{Koch} \& {Rosolowsky}}{{Koch} \&
  {Rosolowsky}}{2015}]{koch:2015.filament.magic.size}
{Koch} E.~W.,  {Rosolowsky} E.~W.,  2015, \mn@doi [\mnras]
  {10.1093/mnras/stv1521}, \href
  {https://ui.adsabs.harvard.edu/abs/2015MNRAS.452.3435K} {452, 3435}

\bibitem[\protect\citeauthoryear{{Konstandin}, {Girichidis}, {Federrath}  \&
  {Klessen}}{{Konstandin} et~al.}{2012}]{konstantin:mach.compressive.relation}
{Konstandin} L.,  {Girichidis} P.,  {Federrath} C.,   {Klessen} R.~S.,  2012,
  \mn@doi [\apj] {10.1088/0004-637X/761/2/149}, \href
  {http://adsabs.harvard.edu/abs/2012arXiv1206.4524K} {761, 149}

\bibitem[\protect\citeauthoryear{{Kriek} \& {Conroy}}{{Kriek} \&
  {Conroy}}{2013}]{kriek:2013.dust.extinction.curve.vs.galaxy.type}
{Kriek} M.,  {Conroy} C.,  2013, \mn@doi [\apjl] {10.1088/2041-8205/775/1/L16},
  \href {https://ui.adsabs.harvard.edu/abs/2013ApJ...775L..16K} {775, L16}

\bibitem[\protect\citeauthoryear{{Kruijssen} et~al.,}{{Kruijssen}
  et~al.}{2019}]{2019Natur.569..519K}
{Kruijssen} J.~M.~D.,  et~al., 2019, \mn@doi [\nat]
  {10.1038/s41586-019-1194-3}, \href
  {https://ui.adsabs.harvard.edu/abs/2019Natur.569..519K} {569, 519}

\bibitem[\protect\citeauthoryear{{Krumholz} \& {Thompson}}{{Krumholz} \&
  {Thompson}}{2012}]{krumholz:2012.rad.pressure.rt.instab}
{Krumholz} M.~R.,  {Thompson} T.~A.,  2012, \mn@doi [\apj]
  {10.1088/0004-637X/760/2/155}, \href
  {http://adsabs.harvard.edu/abs/2012arXiv1203.2926K} {760, 155}

\bibitem[\protect\citeauthoryear{{Kuiper}, {Klahr}, {Beuther}  \&
  {Henning}}{{Kuiper}
  et~al.}{2012}]{kuiper:2012.rad.pressure.outflow.vs.rt.method}
{Kuiper} R.,  {Klahr} H.,  {Beuther} H.,   {Henning} T.,  2012, \mn@doi [\aap]
  {10.1051/0004-6361/201117808}, \href
  {http://adsabs.harvard.edu/abs/2012A%26A...537A.122K} {537, A122}

\bibitem[\protect\citeauthoryear{{Lamers} \& {Cassinelli}}{{Lamers} \&
  {Cassinelli}}{1999}]{1999isw..book.....L}
{Lamers} H. J.~G.~L.~M.,  {Cassinelli} J.~P.,  1999, {Introduction to Stellar
  Winds}.
Cambridge, UK: Cambridge University Press

\bibitem[\protect\citeauthoryear{{Lee} \& {Hopkins}}{{Lee} \&
  {Hopkins}}{2020}]{lee:2020.hopkins.stars.planets.born.intense.rad.fields}
{Lee} E.~J.,  {Hopkins} P.~F.,  2020, \mn@doi [\mnras]
  {10.1093/mnrasl/slaa050}, \href
  {https://ui.adsabs.harvard.edu/abs/2020MNRAS.495L..86L} {495, L86}

\bibitem[\protect\citeauthoryear{{Lee}, {Hopkins}  \& {Squire}}{{Lee}
  et~al.}{2017}]{lee:dynamics.charged.dust.gmcs}
{Lee} H.,  {Hopkins} P.~F.,   {Squire} J.,  2017, \mn@doi [\mnras]
  {10.1093/mnras/stx1097}, \href
  {http://adsabs.harvard.edu/abs/2017MNRAS.469.3532L} {469, 3532}

\bibitem[\protect\citeauthoryear{{Levermore}}{{Levermore}}{1984}]{levermore:1984.FLD.M1}
{Levermore} C.~D.,  1984, \mn@doi [Journal of Quantitative Spectroscopy and
  Radiative Transfer] {10.1016/0022-4073(84)90112-2}, \href
  {http://adsabs.harvard.edu/abs/1984JQSRT..31..149L} {31, 149}

\bibitem[\protect\citeauthoryear{{Lowrie}, {Morel}  \& {Hittinger}}{{Lowrie}
  et~al.}{1999}]{lowrie:1999.radiation.hydro.coupling}
{Lowrie} R.~B.,  {Morel} J.~E.,   {Hittinger} J.~A.,  1999, \mn@doi [\apj]
  {10.1086/307515}, \href {http://adsabs.harvard.edu/abs/1999ApJ...521..432L}
  {521, 432}

\bibitem[\protect\citeauthoryear{{Lupi}, {Volonteri}  \& {Silk}}{{Lupi}
  et~al.}{2017}]{lupi:2017.gizmo.galaxy.form.methods}
{Lupi} A.,  {Volonteri} M.,   {Silk} J.,  2017, \mn@doi [\mnras]
  {10.1093/mnras/stx1313}, \href
  {https://ui.adsabs.harvard.edu/abs/2017MNRAS.470.1673L} {470, 1673}

\bibitem[\protect\citeauthoryear{{Lupi}, {Bovino}, {Capelo}, {Volonteri}  \&
  {Silk}}{{Lupi} et~al.}{2018}]{lupi:2018.h2.sfr.rhd.gizmo.methods}
{Lupi} A.,  {Bovino} S.,  {Capelo} P.~R.,  {Volonteri} M.,   {Silk} J.,  2018,
  \mn@doi [\mnras] {10.1093/mnras/stx2874}, \href
  {http://adsabs.harvard.edu/abs/2018MNRAS.474.2884L} {474, 2884}

\bibitem[\protect\citeauthoryear{{Lv}, {Jiang}  \& {Li}}{{Lv}
  et~al.}{2017}]{lv:2017.dust.gas.ratio.measurements}
{Lv} Z.~P.,  {Jiang} B.~W.,   {Li} J.,  2017, Acta Astronomica Sinica, \href
  {https://ui.adsabs.harvard.edu/abs/2017AcASn..58...11L} {58, 11}

\bibitem[\protect\citeauthoryear{{Ma}, {Kasen}, {Hopkins},
  {Faucher-Gigu{\`e}re}, {Quataert}, {Kere{\v s}}  \& {Murray}}{{Ma}
  et~al.}{2015}]{ma:2015.fire.escape.fractions}
{Ma} X.,  {Kasen} D.,  {Hopkins} P.~F.,  {Faucher-Gigu{\`e}re} C.-A.,
  {Quataert} E.,  {Kere{\v s}} D.,   {Murray} N.,  2015, \mn@doi [\mnras]
  {10.1093/mnras/stv1679}, \href
  {http://adsabs.harvard.edu/abs/2015MNRAS.453..960M} {453, 960}

\bibitem[\protect\citeauthoryear{{Ma}, {Hopkins}, {Kasen}, {Quataert},
  {Faucher-Gigu{\`e}re}, {Kere{\v s}}, {Murray}  \& {Strom}}{{Ma}
  et~al.}{2016}]{ma.2016:binary.star.escape.fraction.effects}
{Ma} X.,  {Hopkins} P.~F.,  {Kasen} D.,  {Quataert} E.,  {Faucher-Gigu{\`e}re}
  C.-A.,  {Kere{\v s}} D.,  {Murray} N.,   {Strom} A.,  2016, \mn@doi [\mnras]
  {10.1093/mnras/stw941}, \href
  {http://adsabs.harvard.edu/abs/2016MNRAS.459.3614M} {459, 3614}

\bibitem[\protect\citeauthoryear{{Ma}, {Quataert}, {Wetzel}, {Hopkins},
  {Faucher-Gigu{\`e}re}  \& {Kere{\v{s}}}}{{Ma}
  et~al.}{2020}]{ma:2020.no.missing.photons.for.reion.supershells}
{Ma} X.,  {Quataert} E.,  {Wetzel} A.,  {Hopkins} P.~F.,  {Faucher-Gigu{\`e}re}
  C.-A.,   {Kere{\v{s}}} D.,  2020, \mn@doi [\mnras] {10.1093/mnras/staa2404},
  \href {https://ui.adsabs.harvard.edu/abs/2020MNRAS.498.2001M} {498, 2001}

\bibitem[\protect\citeauthoryear{{Mathis}, {Rumpl}  \& {Nordsieck}}{{Mathis}
  et~al.}{1977}]{mathis:1977.grain.sizes}
{Mathis} J.~S.,  {Rumpl} W.,   {Nordsieck} K.~H.,  1977, \mn@doi [\apj]
  {10.1086/155591}, \href {http://adsabs.harvard.edu/abs/1977ApJ...217..425M}
  {217, 425}

\bibitem[\protect\citeauthoryear{{McKinnon}, {Vogelsberger}, {Torrey},
  {Marinacci}  \& {Kannan}}{{McKinnon} et~al.}{2018}]{2018MNRAS.478.2851M}
{McKinnon} R.,  {Vogelsberger} M.,  {Torrey} P.,  {Marinacci} F.,   {Kannan}
  R.,  2018, \mn@doi [\mnras] {10.1093/mnras/sty1248}, \href
  {https://ui.adsabs.harvard.edu/abs/2018MNRAS.478.2851M} {478, 2851}

\bibitem[\protect\citeauthoryear{{Men'shchikov} et~al.,}{{Men'shchikov}
  et~al.}{2010}]{menshchikov:2010.filaments.herschel}
{Men'shchikov} A.,  et~al., 2010, \mn@doi [\aap] {10.1051/0004-6361/201014668},
  \href {https://ui.adsabs.harvard.edu/abs/2010A&A...518L.103M} {518, L103}

\bibitem[\protect\citeauthoryear{{Mihalas} \& {Mihalas}}{{Mihalas} \&
  {Mihalas}}{1984}]{mihalas:1984oup..book.....M}
{Mihalas} D.,  {Mihalas} B.~W.,  eds, 1984, {Foundations of radiation
  hydrodynamics}.
New York, Oxford University Press, 731 p.

\bibitem[\protect\citeauthoryear{{Miville-Desch{\^e}nes}, {Boulanger}, {Joncas}
   \& {Falgarone}}{{Miville-Desch{\^e}nes}
  et~al.}{2002}]{miville-deschenes:2002.large.fluct.in.small.grain.abundances}
{Miville-Desch{\^e}nes} M.-A.,  {Boulanger} F.,  {Joncas} G.,   {Falgarone} E.,
   2002, \mn@doi [\aap] {10.1051/0004-6361:20011074}, \href
  {http://adsabs.harvard.edu/abs/2002A%26A...381..209M} {381, 209}

\bibitem[\protect\citeauthoryear{Monchaux, Bourgoin  \& Cartellier}{Monchaux
  et~al.}{2010}]{monchaux:2010.grain.concentration.experiments.voronoi}
Monchaux R.,  Bourgoin M.,   Cartellier A.,  2010, \mn@doi [Physics of Fluids]
  {10.1063/1.3489987}, 22, 103304

\bibitem[\protect\citeauthoryear{{Monchaux}, {Bourgoin}  \&
  {Cartellier}}{{Monchaux}
  et~al.}{2012}]{monchaux:2012.grain.concentration.experiment.review}
{Monchaux} R.,  {Bourgoin} M.,   {Cartellier} A.,  2012, \mn@doi [International
  Journal of Multiphase Flow] {10.1016/j.ijmultiphaseflow.2011.12.001}, 40, 1

\bibitem[\protect\citeauthoryear{{Morse}, {Davidson}, {Bally}, {Ebbets},
  {Balick}  \& {Frank}}{{Morse}
  et~al.}{1998}]{morse:1998.eta.carinae.whiskers.dustlanes.filaments}
{Morse} J.~A.,  {Davidson} K.,  {Bally} J.,  {Ebbets} D.,  {Balick} B.,
  {Frank} A.,  1998, \mn@doi [\aj] {10.1086/300581}, \href
  {http://adsabs.harvard.edu/abs/1998AJ....116.2443M} {116, 2443}

\bibitem[\protect\citeauthoryear{{Moseley}, {Squire}  \& {Hopkins}}{{Moseley}
  et~al.}{2019}]{moseley:2018.acoustic.rdi.sims}
{Moseley} E.~R.,  {Squire} J.,   {Hopkins} P.~F.,  2019, \mn@doi [\mnras]
  {10.1093/mnras/stz2128}, \href
  {https://ui.adsabs.harvard.edu/abs/2019MNRAS.489..325M} {489, 325}

\bibitem[\protect\citeauthoryear{{Nakamura} \& {Li}}{{Nakamura} \&
  {Li}}{2008}]{nakamura:2008.magnetic.alignment.filaments.from.sims}
{Nakamura} F.,  {Li} Z.-Y.,  2008, \mn@doi [\apj] {10.1086/591641}, \href
  {https://ui.adsabs.harvard.edu/abs/2008ApJ...687..354N} {687, 354}

\bibitem[\protect\citeauthoryear{{Nyland} et~al.,}{{Nyland}
  et~al.}{2013}]{nyland:2013.radio.core.ngc1266}
{Nyland} K.,  et~al., 2013, \mn@doi [\apj] {10.1088/0004-637X/779/2/173}, \href
  {https://ui.adsabs.harvard.edu/abs/2013ApJ...779..173N} {779, 173}

\bibitem[\protect\citeauthoryear{{O'Dell}, {Balick}, {Hajian}, {Henney}  \&
  {Burkert}}{{O'Dell} et~al.}{2002}]{odell:2002.pne.knots.review}
{O'Dell} C.~R.,  {Balick} B.,  {Hajian} A.~R.,  {Henney} W.~J.,   {Burkert} A.,
   2002, \mn@doi [\aj] {10.1086/340726}, \href
  {http://adsabs.harvard.edu/abs/2002AJ....123.3329O} {123, 3329}

\bibitem[\protect\citeauthoryear{{Olla}}{{Olla}}{2010}]{olla:2010.grain.preferential.concentration.randomfield.notes}
{Olla} P.,  2010, \mn@doi [Phys. Rev. E] {10.1103/PhysRevE.81.016305}, \href
  {http://adsabs.harvard.edu/abs/2010PhRvE..81a6305O} {81, 016305}

\bibitem[\protect\citeauthoryear{{Ormel} \& {Cuzzi}}{{Ormel} \&
  {Cuzzi}}{2007}]{ormel:2007.closed.form.grain.rel.velocities}
{Ormel} C.~W.,  {Cuzzi} J.~N.,  2007, \mn@doi [\aap]
  {10.1051/0004-6361:20066899}, \href
  {http://adsabs.harvard.edu/abs/2007A%26A...466..413O} {466, 413}

\bibitem[\protect\citeauthoryear{{Padoan}, {Cambr{\'e}sy}, {Juvela}, {Kritsuk},
  {Langer}  \& {Norman}}{{Padoan}
  et~al.}{2006}]{padoan:dust.fluct.taurus.vs.sims}
{Padoan} P.,  {Cambr{\'e}sy} L.,  {Juvela} M.,  {Kritsuk} A.,  {Langer} W.~D.,
   {Norman} M.~L.,  2006, \mn@doi [\apj] {10.1086/507068}, \href
  {http://adsabs.harvard.edu/abs/2006ApJ...649..807P} {649, 807}

\bibitem[\protect\citeauthoryear{{Palmeirim} et~al.,}{{Palmeirim}
  et~al.}{2013}]{2013A&A...550A..38P}
{Palmeirim} P.,  et~al., 2013, \mn@doi [\aap] {10.1051/0004-6361/201220500},
  \href {https://ui.adsabs.harvard.edu/abs/2013A&A...550A..38P} {550, A38}

\bibitem[\protect\citeauthoryear{{Pan} \& {Padoan}}{{Pan} \&
  {Padoan}}{2010}]{pan:2010.grain.velocity.sims}
{Pan} L.,  {Padoan} P.,  2010, \mn@doi [Journal of Fluid Mechanics]
  {10.1017/S0022112010002855}, \href
  {http://adsabs.harvard.edu/abs/2010JFM...661...73P} {661, 73}

\bibitem[\protect\citeauthoryear{{Pan} \& {Padoan}}{{Pan} \&
  {Padoan}}{2013}]{pan:2013.grain.relative.velocity.calc}
{Pan} L.,  {Padoan} P.,  2013, \apj, in press, arXiv:1305.0307, \href
  {http://adsabs.harvard.edu/abs/2013arXiv1305.0307P} {}

\bibitem[\protect\citeauthoryear{{Pan}, {Padoan}, {Scalo}, {Kritsuk}  \&
  {Norman}}{{Pan} et~al.}{2011}]{pan:2011.grain.clustering.midstokes.sims}
{Pan} L.,  {Padoan} P.,  {Scalo} J.,  {Kritsuk} A.~G.,   {Norman} M.~L.,  2011,
  \mn@doi [\apj] {10.1088/0004-637X/740/1/6}, \href
  {http://adsabs.harvard.edu/abs/2011ApJ...740....6P} {740, 6}

\bibitem[\protect\citeauthoryear{{Panopoulou}, {Psaradaki}, {Skalidis},
  {Tassis}  \& {Andrews}}{{Panopoulou}
  et~al.}{2017}]{panopoulou:characteristic.filament.width.artefact.of.obs.and.power.spectrum}
{Panopoulou} G.~V.,  {Psaradaki} I.,  {Skalidis} R.,  {Tassis} K.,   {Andrews}
  J.~J.,  2017, \mn@doi [\mnras] {10.1093/mnras/stw3060}, \href
  {https://ui.adsabs.harvard.edu/abs/2017MNRAS.466.2529P} {466, 2529}

\bibitem[\protect\citeauthoryear{{Paron}, {Celis Pe{\~n}a}, {Ortega},
  {Fari{\~n}a}, {Petriella}, {Rubio}  \& {Ashley}}{{Paron}
  et~al.}{2017}]{paron:2017.pillar.hii.region.obs.not.implosion}
{Paron} S.,  {Celis Pe{\~n}a} M.,  {Ortega} M.~E.,  {Fari{\~n}a} C.,
  {Petriella} A.,  {Rubio} M.,   {Ashley} R.~P.,  2017, \mn@doi [\mnras]
  {10.1093/mnras/stx1486}, \href
  {https://ui.adsabs.harvard.edu/abs/2017MNRAS.470.4662P} {470, 4662}

\bibitem[\protect\citeauthoryear{{Pei}}{{Pei}}{1992}]{pei92:reddening.curves}
{Pei} Y.~C.,  1992, \mn@doi [\apj] {10.1086/171637}, \href
  {http://adsabs.harvard.edu/abs/1992ApJ...395..130P} {395, 130}

\bibitem[\protect\citeauthoryear{{Pellegrini} et~al.,}{{Pellegrini}
  et~al.}{2013}]{pellegrini:2013.ngc.1266.shocked.molecules}
{Pellegrini} E.~W.,  et~al., 2013, \mn@doi [\apjl]
  {10.1088/2041-8205/779/2/L19}, \href
  {https://ui.adsabs.harvard.edu/abs/2013ApJ...779L..19P} {779, L19}

\bibitem[\protect\citeauthoryear{{Pineda}, {Goldsmith}, {Chapman}, {Snell},
  {Li}, {Cambr{\'e}sy}  \& {Brunt}}{{Pineda}
  et~al.}{2010}]{pineda:2010.taurus.large.extinction.variations}
{Pineda} J.~L.,  {Goldsmith} P.~F.,  {Chapman} N.,  {Snell} R.~L.,  {Li} D.,
  {Cambr{\'e}sy} L.,   {Brunt} C.,  2010, \mn@doi [\apj]
  {10.1088/0004-637X/721/1/686}, \href
  {http://adsabs.harvard.edu/abs/2010ApJ...721..686P} {721, 686}

\bibitem[\protect\citeauthoryear{{Planck Collaboration} et~al.,}{{Planck
  Collaboration}
  et~al.}{2016a}]{planck:2016.magnetic.field.corr.structure.weak.except.highest.densities}
{Planck Collaboration} et~al., 2016a, \mn@doi [\aap]
  {10.1051/0004-6361/201425044}, \href
  {https://ui.adsabs.harvard.edu/abs/2016A&A...586A.135P} {586, A135}

\bibitem[\protect\citeauthoryear{{Planck Collaboration} et~al.,}{{Planck
  Collaboration}
  et~al.}{2016b}]{planck:2016.magnetic.field.alignment.filaments.gmcs}
{Planck Collaboration} et~al., 2016b, \mn@doi [\aap]
  {10.1051/0004-6361/201525896}, \href
  {https://ui.adsabs.harvard.edu/abs/2016A&A...586A.138P} {586, A138}

\bibitem[\protect\citeauthoryear{{Raskutti}, {Ostriker}  \&
  {Skinner}}{{Raskutti} et~al.}{2016}]{raskutti:2016.m1.cloud.sims}
{Raskutti} S.,  {Ostriker} E.~C.,   {Skinner} M.~A.,  2016, \mnras, submitted,
  arXiv:1608.04469, \href {http://adsabs.harvard.edu/abs/2016arXiv160804469R}
  {}

\bibitem[\protect\citeauthoryear{{Rice}, {Goodman}, {Bergin}, {Beaumont}  \&
  {Dame}}{{Rice} et~al.}{2016}]{rice:2016.gmc.mw.catalogue}
{Rice} T.~S.,  {Goodman} A.~A.,  {Bergin} E.~A.,  {Beaumont} C.,   {Dame}
  T.~M.,  2016, \mn@doi [\apj] {10.3847/0004-637X/822/1/52}, \href
  {http://adsabs.harvard.edu/abs/2016ApJ...822...52R} {822, 52}

\bibitem[\protect\citeauthoryear{{Roccatagliata}, {Preibisch}, {Ratzka}  \&
  {Gaczkowski}}{{Roccatagliata}
  et~al.}{2013}]{roccatagliata:2013.carina.structures.morph.not.well.explained}
{Roccatagliata} V.,  {Preibisch} T.,  {Ratzka} T.,   {Gaczkowski} B.,  2013,
  \mn@doi [\aap] {10.1051/0004-6361/201321081}, \href
  {https://ui.adsabs.harvard.edu/abs/2013A&A...554A...6R} {554, A6}

\bibitem[\protect\citeauthoryear{{Roman-Duval} et~al.,}{{Roman-Duval}
  et~al.}{2014}]{roman-duval.2014:smc.lmc.dust.to.gas.ratio.fluctuations.variations}
{Roman-Duval} J.,  et~al., 2014, \mn@doi [\apj] {10.1088/0004-637X/797/2/86},
  \href {http://adsabs.harvard.edu/abs/2014ApJ...797...86R} {797, 86}

\bibitem[\protect\citeauthoryear{{Rosen}, {Krumholz}, {McKee}  \&
  {Klein}}{{Rosen} et~al.}{2016}]{rosen:massive.sf.rhd}
{Rosen} A.~L.,  {Krumholz} M.~R.,  {McKee} C.~F.,   {Klein} R.~I.,  2016,
  \mn@doi [\mnras] {10.1093/mnras/stw2153}, \href
  {http://adsabs.harvard.edu/abs/2016MNRAS.463.2553R} {463, 2553}

\bibitem[\protect\citeauthoryear{{Salim}, {Boquien}  \& {Lee}}{{Salim}
  et~al.}{2018}]{salim:2018.extinction.curve.variations.by.galaxy}
{Salim} S.,  {Boquien} M.,   {Lee} J.~C.,  2018, \mn@doi [\apj]
  {10.3847/1538-4357/aabf3c}, \href
  {https://ui.adsabs.harvard.edu/abs/2018ApJ...859...11S} {859, 11}

\bibitem[\protect\citeauthoryear{{Sandford}, {Whitaker}  \& {Klein}}{{Sandford}
  et~al.}{1984}]{sandford:radiatively.driven.dust.bounded.gmc.globules}
{Sandford} II M.~T.,  {Whitaker} R.~W.,   {Klein} R.~I.,  1984, \mn@doi [\apj]
  {10.1086/162189}, \href {http://adsabs.harvard.edu/abs/1984ApJ...282..178S}
  {282, 178}

\bibitem[\protect\citeauthoryear{{Scalo}, {Vazquez-Semadeni}, {Chappell}  \&
  {Passot}}{{Scalo} et~al.}{1998}]{scalo:1998.turb.density.pdf}
{Scalo} J.,  {Vazquez-Semadeni} E.,  {Chappell} D.,   {Passot} T.,  1998,
  \mn@doi [\apj] {10.1086/306099}, \href
  {http://adsabs.harvard.edu/abs/1998ApJ...504..835S} {504, 835}

\bibitem[\protect\citeauthoryear{{Schlafly} et~al.,}{{Schlafly}
  et~al.}{2016}]{schlafly:galactic.extinction.curve.variations}
{Schlafly} E.~F.,  et~al., 2016, \mn@doi [\apj] {10.3847/0004-637X/821/2/78},
  \href {http://adsabs.harvard.edu/abs/2016ApJ...821...78S} {821, 78}

\bibitem[\protect\citeauthoryear{{Schnee}, {Mason}, {Di Francesco}, {Friesen},
  {Li}, {Sadavoy}  \& {Stanke}}{{Schnee}
  et~al.}{2014}]{schnee:2014.mm.sized.grains.in.star.forming.regions}
{Schnee} S.,  {Mason} B.,  {Di Francesco} J.,  {Friesen} R.,  {Li} D.,
  {Sadavoy} S.,   {Stanke} T.,  2014, \mnras, in press, arxiv:1408.5429, \href
  {http://adsabs.harvard.edu/abs/2014arXiv1408.5429S} {}

\bibitem[\protect\citeauthoryear{{Scoville}, {Polletta}, {Ewald}, {Stolovy},
  {Thompson}  \& {Rieke}}{{Scoville}
  et~al.}{2001}]{scoville:2001.dust.pressure.in.sb.regions}
{Scoville} N.~Z.,  {Polletta} M.,  {Ewald} S.,  {Stolovy} S.~R.,  {Thompson}
  R.,   {Rieke} M.,  2001, \mn@doi [\aj] {10.1086/323445}, \href
  {http://adsabs.harvard.edu/abs/2001AJ....122.3017S} {122, 3017}

\bibitem[\protect\citeauthoryear{{Seligman}, {Hopkins}  \& {Squire}}{{Seligman}
  et~al.}{2019}]{seligman:2018.mhd.rdi.sims}
{Seligman} D.,  {Hopkins} P.~F.,   {Squire} J.,  2019, \mn@doi [\mnras]
  {10.1093/mnras/stz666}, \href
  {https://ui.adsabs.harvard.edu/abs/2019MNRAS.485.3991S} {485, 3991}

\bibitem[\protect\citeauthoryear{{Seon} \& {Draine}}{{Seon} \&
  {Draine}}{2016}]{seon.draine:extinction.curve.vs.radiative.transfer.turbulent.sims}
{Seon} K.-I.,  {Draine} B.~T.,  2016, \mn@doi [\apj]
  {10.3847/1538-4357/833/2/201}, \href
  {http://adsabs.harvard.edu/abs/2016ApJ...833..201S} {833, 201}

\bibitem[\protect\citeauthoryear{{Shields} \& {Kennicutt}}{{Shields} \&
  {Kennicutt}}{1995}]{shields:1995.dust.HII.region.chem.rad.fx}
{Shields} J.~C.,  {Kennicutt} Jr. R.~C.,  1995, \mn@doi [\apj]
  {10.1086/176533}, \href {http://adsabs.harvard.edu/abs/1995ApJ...454..807S}
  {454, 807}

\bibitem[\protect\citeauthoryear{{Skilling}}{{Skilling}}{1975}]{1975MNRAS.172..557S}
{Skilling} J.,  1975, \mn@doi [\mnras] {10.1093/mnras/172.3.557}, \href
  {http://adsabs.harvard.edu/abs/1975MNRAS.172..557S} {172, 557}

\bibitem[\protect\citeauthoryear{{Squire} \& {Hopkins}}{{Squire} \&
  {Hopkins}}{2017}]{squire.hopkins:turb.density.pdf}
{Squire} J.,  {Hopkins} P.~F.,  2017, \mn@doi [\mnras] {10.1093/mnras/stx1817},
  \href {http://adsabs.harvard.edu/abs/2017MNRAS.471.3753S} {471, 3753}

\bibitem[\protect\citeauthoryear{{Squire} \& {Hopkins}}{{Squire} \&
  {Hopkins}}{2018a}]{squire:rdi.ppd}
{Squire} J.,  {Hopkins} P.~F.,  2018a, \mn@doi [\mnras] {10.1093/mnras/sty854},
  \href {http://adsabs.harvard.edu/abs/2018MNRAS.477.5011S} {477, 5011}

\bibitem[\protect\citeauthoryear{{Squire} \& {Hopkins}}{{Squire} \&
  {Hopkins}}{2018b}]{squire.hopkins:RDI}
{Squire} J.,  {Hopkins} P.~F.,  2018b, \mn@doi [\apjl]
  {10.3847/2041-8213/aab54d}, \href
  {http://adsabs.harvard.edu/abs/2018ApJ...856L..15S} {856, L15}

\bibitem[\protect\citeauthoryear{{Squire}, {Moroianu}  \& {Hopkins}}{{Squire}
  et~al.}{2022}]{squire:2022.acoustic.rdi.size.spectrum}
{Squire} J.,  {Moroianu} S.,   {Hopkins} P.~F.,  2022, \mn@doi [\mnras]
  {10.1093/mnras/stab3377}, \href
  {https://ui.adsabs.harvard.edu/abs/2022MNRAS.510..110S} {510, 110}

\bibitem[\protect\citeauthoryear{{Steinwandel}, {Kaurov}, {Hopkins}  \&
  {Squire}}{{Steinwandel}
  et~al.}{2021}]{steinwandel:2021.dust.rdi.variable.stars}
{Steinwandel} U.~P.,  {Kaurov} A.~A.,  {Hopkins} P.~F.,   {Squire} J.,  2021,
  arXiv e-prints, \href {https://ui.adsabs.harvard.edu/abs/2021arXiv211109335S}
  {p. arXiv:2111.09335}

\bibitem[\protect\citeauthoryear{{Su}, {Hopkins}, {Hayward},
  {Faucher-Gigu{\`e}re}, {Kere{\v s}}, {Ma}  \& {Robles}}{{Su}
  et~al.}{2017}]{su:2016.weak.mhd.cond.visc.turbdiff.fx}
{Su} K.-Y.,  {Hopkins} P.~F.,  {Hayward} C.~C.,  {Faucher-Gigu{\`e}re} C.-A.,
  {Kere{\v s}} D.,  {Ma} X.,   {Robles} V.~H.,  2017, \mn@doi [\mnras]
  {10.1093/mnras/stx1463}, \href
  {http://adsabs.harvard.edu/abs/2017MNRAS.471..144S} {471, 144}

\bibitem[\protect\citeauthoryear{{Thompson}, {Quataert}  \&
  {Murray}}{{Thompson} et~al.}{2005}]{thompson:rad.pressure}
{Thompson} T.~A.,  {Quataert} E.,   {Murray} N.,  2005, \mn@doi [\apj]
  {10.1086/431923}, \href {http://adsabs.harvard.edu/abs/2005ApJ...630..167T}
  {630, 167}

\bibitem[\protect\citeauthoryear{{Thoraval}, {Boisse}  \& {Duvert}}{{Thoraval}
  et~al.}{1997}]{thoraval:1997.sub.0pt04pc.no.cloud.extinction.fluct.but.are.on.larger.scales}
{Thoraval} S.,  {Boisse} P.,   {Duvert} G.,  1997, \aap, \href
  {http://adsabs.harvard.edu/abs/1997A%26A...319..948T} {319, 948}

\bibitem[\protect\citeauthoryear{{Thoraval}, {Boiss{\'e}}  \&
  {Duvert}}{{Thoraval}
  et~al.}{1999}]{thoraval:1999.small.scale.dust.to.gas.density.fluctuations}
{Thoraval} S.,  {Boiss{\'e}} P.,   {Duvert} G.,  1999, \aap, \href
  {http://adsabs.harvard.edu/abs/1999A%26A...351.1051T} {351, 1051}

\bibitem[\protect\citeauthoryear{{Tielens}}{{Tielens}}{1998}]{tielens:1998.dust.is.amorphous.iron.poor.silicates}
{Tielens} A.~G.~G.~M.,  1998, \mn@doi [\apj] {10.1086/305640}, \href
  {http://adsabs.harvard.edu/abs/1998ApJ...499..267T} {499, 267}

\bibitem[\protect\citeauthoryear{{Tielens}}{{Tielens}}{2005}]{tielens:2005.book}
{Tielens} A.~G.~G.~M.,  2005, {The Physics and Chemistry of the Interstellar
  Medium}.
Cambridge, UK: Cambridge University Press

\bibitem[\protect\citeauthoryear{{Tielens}, {Waters}, {Molster}  \&
  {Justtanont}}{{Tielens} et~al.}{1998}]{tielens:silicate.dust.composition}
{Tielens} A.~G.~G.~M.,  {Waters} L.~B.~F.~M.,  {Molster} F.~J.,   {Justtanont}
  K.,  1998, \mn@doi [Astrophysics and Space Science]
  {10.1023/A:1001585120472}, \href
  {http://adsabs.harvard.edu/abs/1998Ap%26SS.255..415T} {255, 415}

\bibitem[\protect\citeauthoryear{{Topchieva}, {Wiebe}, {Kirsanova}  \&
  {Krushinsky}}{{Topchieva}
  et~al.}{2017}]{topchieva:2017.dust.HII.region.shape.irregular.rings.shells}
{Topchieva} A.,  {Wiebe} D.,  {Kirsanova} M.,   {Krushinsky} V.,  2017, in
  {Balega} Y.~Y.,  {Kudryavtsev} D.~O.,  {Romanyuk} I.~I.,   {Yakunin} I.~A.,
  eds,  Astronomical Society of the Pacific Conference Series Vol. 510, Stars:
  From Collapse to Collapse, San Francisco: Astronomical Society of the
  Pacific. p.~98

\bibitem[\protect\citeauthoryear{{Tremblin}, {Audit}, {Minier}, {Schmidt}  \&
  {Schneider}}{{Tremblin}
  et~al.}{2012}]{tremblin:2012.pillar.form.ionization.sims}
{Tremblin} P.,  {Audit} E.,  {Minier} V.,  {Schmidt} W.,   {Schneider} N.,
  2012, \mn@doi [\aap] {10.1051/0004-6361/201219224}, \href
  {https://ui.adsabs.harvard.edu/abs/2012A&A...546A..33T} {546, A33}

\bibitem[\protect\citeauthoryear{{Tsang} \& {Milosavljevi{\'c}}}{{Tsang} \&
  {Milosavljevi{\'c}}}{2015}]{tsang:monte.carlo.rhd.dusty.wind}
{Tsang} B.~T.-H.,  {Milosavljevi{\'c}} M.,  2015, \mn@doi [\mnras]
  {10.1093/mnras/stv1707}, \href
  {http://adsabs.harvard.edu/abs/2015MNRAS.453.1108T} {453, 1108}

\bibitem[\protect\citeauthoryear{{Vazquez-Semadeni}}{{Vazquez-Semadeni}}{1994}]{vazquez-semadeni:1994.turb.density.pdf}
{Vazquez-Semadeni} E.,  1994, \mn@doi [\apj] {10.1086/173847}, \href
  {http://adsabs.harvard.edu/abs/1994ApJ...423..681V} {423, 681}

\bibitem[\protect\citeauthoryear{{Wang}, {Li}  \& {Jiang}}{{Wang}
  et~al.}{2014}]{wang:2014.ism.extinction.variations.grain.growth.model}
{Wang} S.,  {Li} A.,   {Jiang} B.~W.,  2014, \mn@doi [\planss]
  {10.1016/j.pss.2014.03.018}, \href
  {https://ui.adsabs.harvard.edu/abs/2014P&SS..100...32W} {100, 32}

\bibitem[\protect\citeauthoryear{{Wang}, {Jiang}, {Zhao}, {Chen}  \& {de
  Grijs}}{{Wang}
  et~al.}{2017}]{wang:2017.extinction.law.variation.in.diffuse.ism}
{Wang} S.,  {Jiang} B.~W.,  {Zhao} H.,  {Chen} X.,   {de Grijs} R.,  2017,
  \mn@doi [\apj] {10.3847/1538-4357/aa8db7}, \href
  {https://ui.adsabs.harvard.edu/abs/2017ApJ...848..106W} {848, 106}

\bibitem[\protect\citeauthoryear{{Weingartner} \& {Draine}}{{Weingartner} \&
  {Draine}}{2001a}]{weingartner:2001.grain.charging.photoelectric}
{Weingartner} J.~C.,  {Draine} B.~T.,  2001a, \mn@doi [\apjs] {10.1086/320852},
  \href {http://adsabs.harvard.edu/abs/2001ApJS..134..263W} {134, 263}

\bibitem[\protect\citeauthoryear{{Weingartner} \& {Draine}}{{Weingartner} \&
  {Draine}}{2001b}]{weingartner:2001.dust.size.distrib}
{Weingartner} J.~C.,  {Draine} B.~T.,  2001b, \mn@doi [\apj] {10.1086/318651},
  \href {http://adsabs.harvard.edu/abs/2001ApJ...548..296W} {548, 296}

\bibitem[\protect\citeauthoryear{{Weingartner} \& {Draine}}{{Weingartner} \&
  {Draine}}{2001c}]{weingertner.draine:photo.forces.on.dust.hard.ism.rad}
{Weingartner} J.~C.,  {Draine} B.~T.,  2001c, \mn@doi [\apj] {10.1086/320963},
  \href {http://adsabs.harvard.edu/abs/2001ApJ...553..581W} {553, 581}

\bibitem[\protect\citeauthoryear{{Westmoquette}, {Dale}, {Ercolano}  \&
  {Smith}}{{Westmoquette}
  et~al.}{2013}]{westmoquette:2013.super.star.cluster.pillars.mixing.layer.not.explained.simple.models}
{Westmoquette} M.~S.,  {Dale} J.~E.,  {Ercolano} B.,   {Smith} L.~J.,  2013,
  \mn@doi [\mnras] {10.1093/mnras/stt1172}, \href
  {https://ui.adsabs.harvard.edu/abs/2013MNRAS.435...30W} {435, 30}

\bibitem[\protect\citeauthoryear{{Wise}, {Abel}, {Turk}, {Norman}  \&
  {Smith}}{{Wise} et~al.}{2012}]{wise:2012.rad.pressure.effects}
{Wise} J.~H.,  {Abel} T.,  {Turk} M.~J.,  {Norman} M.~L.,   {Smith} B.~D.,
  2012, \mn@doi [\mnras] {10.1111/j.1365-2966.2012.21809.x}, \href
  {http://adsabs.harvard.edu/abs/2012MNRAS.427..311W} {427, 311}

\bibitem[\protect\citeauthoryear{{Yakhot}}{{Yakhot}}{1997}]{yakhot:1997.scalar.field.turb.pdfs}
{Yakhot} V.,  1997, \mn@doi [Physical Review E] {10.1103/PhysRevE.55.329},
  \href {http://adsabs.harvard.edu/abs/1997PhRvE..55..329Y} {55, 329}

\bibitem[\protect\citeauthoryear{{Yan}, {Lazarian}  \& {Draine}}{{Yan}
  et~al.}{2004}]{yan.2004:lorentz.forces.drag.dust.ism.analytic}
{Yan} H.,  {Lazarian} A.,   {Draine} B.~T.,  2004, \mn@doi [\apj]
  {10.1086/425111}, \href {http://adsabs.harvard.edu/abs/2004ApJ...616..895Y}
  {616, 895}

\bibitem[\protect\citeauthoryear{Yoshimoto \& Goto}{Yoshimoto \&
  Goto}{2007}]{yoshimoto:2007.grain.clustering.selfsimilar.inertial.range}
Yoshimoto H.,  Goto S.,  2007, \mn@doi [Journal of Fluid Mechanics]
  {10.1017/S0022112007004946}, 577, 275

\bibitem[\protect\citeauthoryear{{Ysard}, {K{\"o}hler}, {Jones},
  {Miville-Desch{\^e}nes}, {Abergel}  \& {Fanciullo}}{{Ysard}
  et~al.}{2015}]{ysard:2005.dust.extinction.curve.variations.diffuse.ism}
{Ysard} N.,  {K{\"o}hler} M.,  {Jones} A.,  {Miville-Desch{\^e}nes} M.~A.,
  {Abergel} A.,   {Fanciullo} L.,  2015, \mn@doi [\aap]
  {10.1051/0004-6361/201425523}, \href
  {https://ui.adsabs.harvard.edu/abs/2015A&A...577A.110Y} {577, A110}

\bibitem[\protect\citeauthoryear{{Zafar}, {Watson}, {Fynbo}, {Malesani},
  {Jakobsson}  \& {de Ugarte Postigo}}{{Zafar}
  et~al.}{2011}]{zafar:2011.grb.extinction.curves.av.nh}
{Zafar} T.,  {Watson} D.,  {Fynbo} J.~P.~U.,  {Malesani} D.,  {Jakobsson} P.,
  {de Ugarte Postigo} A.,  2011, \mn@doi [\aap] {10.1051/0004-6361/201116663},
  \href {https://ui.adsabs.harvard.edu/abs/2011A&A...532A.143Z} {532, A143}

\bibitem[\protect\citeauthoryear{{Zel'dovich}}{{Zel'dovich}}{1970}]{zeldovich:1970.pancakes}
{Zel'dovich} Y.~B.,  1970, \aap, \href
  {http://adsabs.harvard.edu/abs/1970A%26A.....5...84Z} {5, 84}

\bibitem[\protect\citeauthoryear{{Zhu}, {Tian}, {Li}  \& {Zhang}}{{Zhu}
  et~al.}{2017}]{zhu:2017.av.nh.distrib.milky.way}
{Zhu} H.,  {Tian} W.,  {Li} A.,   {Zhang} M.,  2017, \mn@doi [\mnras]
  {10.1093/mnras/stx1580}, \href
  {https://ui.adsabs.harvard.edu/abs/2017MNRAS.471.3494Z} {471, 3494}

\makeatother
\end{thebibliography}

\begin{appendix}

\section{Complete List of Simulations}
\label{sec:appendix:simslist}

\tref{table:sims.all} presents the full list of simulations we have run, to extend \tref{table:sims} in the main text.

%
%
\begin{table*}
\begin{center}
 \begin{tabular}{| l c c c c c c c r |}
 \hline
Name & $\accparammax$ $\left( \frac{|\initvalupper{ \driftvel }|}{\initvalupper{ \cs }} \right)$ & $\sizeparammax$ $\left( \frac{\initvalupper{ \cs}\, \initvalupper{\ts }}{\Lscale} \right)$ & $\chargeparammax$ ($\tauparam$) & $\beta_{0}$ & $\cos{\Bangle^{0}}$ & $\gravparam$ ($\lambda_{\rm Edd}$)  & $\accsizedep$ & Notes \\
 \hline\hline
{\bf GMC-Q} & 65 (4e-2 - 3) & 1e-3 (6e-6 - 4e-4) & 400 (1e3 - 2e5) & 0.01 & 0.1 & 110 (6) &  0 & ``default'': $Q\propto \grainsize^{1}$ \\ 
{\bf GMC} & -- (3) & -- (4e-6 - 4e-4) & -- (1e3 - 1e5) & -- & -- & -- (6) &  1 & ``default'': $Q \propto \grainsize^{0}$ \\ 
{GMC-lr-Q} & -- & -- & -- & -- & -- & -- &  0 & lower-resolution ``default'' \\ 
{GMC-lr} & -- (3) & -- (4e-6 - 4e-4) & -- (1e3 - 1e5) & -- & -- & -- (6) &  1 & lower-resolution ``default'' \\ 
{GMC-lr-Q-a} & 650 (0.4 - 10) & -- (6e-6 - 1e-4) & 40 (50 - 2e4) & -- & -- & -- (60) &  0 & higher $F_{\rm rad}$, lower $Z_{\rm grain}$ \\ 
{GMC-lr-Q-b} & 650 (0.4 - 10) & -- (6e-6 - 1e-4) & -- (5e2 - 2e5) & -- & -- & -- (60) &  0 & higher flux $F_{\rm rad}$  \\ 
{GMC-lr-Q-c} & 650 (0.4 - 10) & -- (6e-6 - 1e-4) & -- (5e2 - 2e5) & -- & -- & -- & 0 &  lower dust-to-gas $\dustgas=0.001$ \\ 
{GMC-lr-Q-d} & 6.5 (4e-4 - 0.4) & -- (6e-6 - 6e-4) & -- (2e3 - 2e5) & -- & -- & -- (0.6) &  0 & lower $F_{\rm rad}$  \\ 
{GMC-lr-Q-e} & 650 (0.4 - 10) & -- (6e-6 - 1e-4) & 4e4 (5e4 - 2e7) & -- & -- & -- (60) &  0 &  much higher charge $Z_{\rm grain}$ \\ 
{GMC-lr-Q-f} & 650 (0.4 - 20) & -- (6e-6 - 8e-5) & 4 (3 - 2e3) & -- & -- & -- (60) &  0 & much lower $Z_{\rm grain}$  \\ 
{GMC-lr-Q-g} & 650 (0.4 - 10) & -- (6e-6 - 1e-4) & -- (5e2 - 2e5) & -- & -- & (600) & 0 &  higher dust-to-gas $\dustgas=0.1$ \\ 
{GMC-lr-Q-h} & -- (2e-6 - 2e-2) & -- (6e-6 - 6e-4) & -- (2e3 - 2e5) & -- & 0 & -- &  0 & perpendicular initial ${\bf B}$ \\ 
{GMC-lr-Q-i} & -- (0.3-8) & -- (6e-6 - 2e-4) & -- (6e2 - 6e4) & -- & 0.7 & -- & 0 &  $45^{\circ}$ initial ${\bf B}$ \\ 
{GMC-lr-Q-j} & -- (0.3-8) & -- (6e-6 - 2e-4) & -- (6e2 - 6e4) & -- & 0.7 & -- & 0 &  ICs turbulent ${\bf B}$, ${\bf v}$ \\ 
{GMC-lr-Q-k} & 650 (4e-1 - 30) & 1e-2 (6e-5 - 4e-4) & -- & -- & -- & -- & 0 &  larger grain size $\grainsize$ \\ 
{GMC-lr-Q-l} & 6.5 (4e-3 - 0.3) & 1e-4 (6e-7 - 4e-5) & -- & -- & -- & -- & 0 &  smaller grain size $\grainsize$ \\ 
{\bf GMC-U-Q-M} & -- & 0.1 (4e-4 - 4e-2) & --  & -- & -- & U &  0 & unstratified ``zoom-in'' \\ 
{\bf GMC-U-M} & -- (3) & 0.1 (6e-4 - 4e-2) & -- (1e3 - 1e5) & -- & -- & U &  1 & unstratified ``zoom-in'' \\ 
{\bf GMC-U-S} & -- (3) & 10 (6e-2 - 4) & -- (1e3 - 1e5) & -- & -- & U &  1 & unstratified ``zoom-in'' \\ 
{GMC-R-A0} & -- (3) & -- (4e-6 - 4e-4) & -- (1e3 - 1e5) & -- & -- & -- &  1 & RDMHD ($A_{0}=0$) \\ 
{GMC-R-A0.5} & -- (3) & -- (4e-6 - 4e-4) & -- (1e3 - 1e5) & -- & -- & -- &  1 & RDMHD ($A_{0}=0.5$) \\ 
{GMC-R-A0-Lo} & -- (3) & -- (4e-6 - 4e-4) & --  (1e3 - 1e5) & -- & -- & -- &  1 & RDMHD ($A_{0}=1$), $A_{v} \times 0.01$ \\ 
{GMC-R-A0} & -- (3) & -- (4e-6 - 4e-4) & -- (1e3 - 1e5) & -- & -- & -- &  1 & RDMHD ($A_{0}=0$) \\ 
{GMC-Q-R-A0} & -- & -- & -- & -- & -- & -- &  0 & RDMHD ($A_{0}=0$) \\ 
{GMC-Q-R-A0.5} & -- & --  & -- & -- & -- & -- &  0 & RDMHD ($A_{0}=0.5$) \\ 
{GMC-Q-R-A1} & -- & -- & -- & -- & -- & -- &  0 & RDMHD ($A_{0}=1$) \\ 
%
%
%
\hline
{\bf HII-N} & 4.5 (0.3) & 3e-3 (2e-5 - 2e-3) & 44 (14 - 1e3) & 4 & 0.1 & 0.01 (1500) &  1 & ``default'': $Q \propto \grainsize^{0}$ \\ 
{\bf HII-N-Q} & -- (3e-3 - 0.3) & -- & -- & -- & -- & -- &  0 & ``default'': $Q \propto \grainsize^{1}$  \\ 
{\bf HII-N-45} & 4.5 (0.3) & 3e-3 (2e-5 - 2e-3) & 44 (14 - 1e3) & 4 & 0.5 & 0.01 (1500) &  1 & $45^{\circ}$ initial ${\bf B}$; $Q \propto \grainsize^{0}$ \\ 
{\bf HII-N-45-Q} & -- (3e-3 - 0.3) & -- & -- & -- & 0.5 & -- &  0 & $45^{\circ}$ initial ${\bf B}$; $Q \propto \grainsize^{1}$  \\ 
{HII-N-lr} & -- & -- & -- & -- & -- & -- &  1 & lower-resolution ``default''  \\ 
{HII-N-lr-a} & 0.45 (3e-2) & -- & -- & -- & -- & -- (150) & 1 & lower $F_{\rm rad}$ \\ 
{HII-N-lr-b} & -- (2) & -- & 1 (5e-2 - 6) & 80 & 0.5 & -- & 1 &  lower field strength $|{\bf B}|$ \\ 
{HII-N-lr-c} & -- & -- & 1000 (300 - 3e4) & -- & -- & -- & 1 &   higher $Z_{\rm grain}$ \\ 
{HII-N-lr-d} & 0.45 (3e-2) & 3e-4 (2e-6 - 2e-4) & -- & -- & -- & -- & 1 &  smaller grain size $\grainsize$ \\ 
{HII-N-lr-e} & --  & -- & -- & -- & -- & 1 (15) & 1 &  stronger gravity ${\bf g}$ \\ 
{HII-N-lr-Q-f} & -- (2e-3 - 0.2) & -- & -- & -- & 0 & -- & 0 & perpendicular initial ${\bf B}$  \\ 
{HII-N-lr-g} & -- (0.2) & -- & -- & -- & 0 & -- & 1 & perpendicular initial ${\bf B}$ \\ 
{HII-N-lr-h} & -- (1) & -- & -- & -- & 0.7 & -- &   1 & $45^{\circ}$ initial ${\bf B}$ \\ 
{\bf HII-N-U-L} & 40 & 7e-3 & 38 & 10 &  0.7 & U & 1 & unstratified ``zoom-in'' \\ 
{\bf HII-N-U-S} & 40 & 780 & 38 & 10 &  0.7 & U & 1 & unstratified ``zoom-in'' \\ 
{\bf HII-N-U-Q-L} & 40 & 7e-3 & 38 & 10 &  0.7 & U & 0 & unstratified ``zoom-in'' \\ 
{\bf HII-N-U-Q-M} & 40 & 2.6 & 38 & 10 &  0.7 & U & 0 & unstratified ``zoom-in'' \\ 
{\bf HII-N-U-Q-S} & 40 & 780 & 38 & 10 &  0.7 & U & 0 & unstratified ``zoom-in'' \\ 
{HII-N-R-A0.5} & -- & -- & -- & -- & -- & -- &  1 & RDHMHD ($A_{0}=0.5$)  \\ 
%
%
%
\hline
{\bf HII-F} & 4.8 (0.3) & 3e-2 (2e-4 - 2e-2) & 440 (140 - 1e4) & 4 & 0.1 & 0.001 (1600) &  1 & ``default'': $Q \propto \grainsize^{0}$ \\ 
{\bf HII-F-Q} & 4.8 (3e-3 - 0.3) & -- & -- & -- & -- & -- & 0 &  ``default'': $Q \propto \grainsize^{1}$  \\ 
{HII-F-lr} & -- & -- & -- & -- & -- & -- & 1 &  lower-resolution ``default''  \\ 
{HII-F-lr-Q} & 4.8 (3e-3 - 0.3) & -- & -- & -- & -- & -- & 0 &  lower-resolution ``default''  \\ 
{HII-F-lr-a} & 0.48 (3e-2) & -- & -- & -- & -- & -- (160) &  1 & lower $F_{\rm rad}$ \\ 
{HII-F-lr-b} & -- (1) & --  (2e-4 - 1e-2) & 10 (0.5 - 60) & 80 & 0.5 & -- &  1 &  lower $|{\bf B}|$ \\ 
{HII-F-lr-c} & -- & -- & 1e4 (3e3 - 3e5) & -- & -- & -- &  1 &  higher $Z_{\rm grain}$ \\ 
{HII-F-lr-d} & 0.48 (3e-2) & 3e-3 (2e-5 - 2e-3) & -- & -- & -- & -- &  1 & smaller $\grainsize$ \\ 
{HII-F-lr-e} & 48 (30) & 0.3 (2e-3 - 2e-1) & -- & -- & -- & -- &  1 & larger $\grainsize$ \\ 
{HII-F-lr-f} & -- & --& -- & -- & -- & 1 (1.6) &  1 &  stronger gravity ${\bf g}$ \\ 
{HII-F-lr-Q-g} & -- (2e-6 - 2e-2) & -- & -- & -- & 0 & -- &  0 & perpendicular initial ${\bf B}$   \\ 
{HII-F-lr-h} & -- (2e-2) & -- & -- & -- & 0 & -- &  1 &  perpendicular initial ${\bf B}$   \\ 
{HII-F-lr-i} & -- (1) & -- & -- & -- & 0.7 & -- &  1 &   $45^{\circ}$ initial ${\bf B}$ \\ 
{\bf HII-F-U-Q} & 0.45 & 5e-3 & 16 & 10 &  0.7 & U & 0 & unstratified ``zoom-in'' \\ 
{\bf HII-F-U-$\tau$-Q} & 0.9 & 1e-2 & 25 & 10 &  0.7 & U & 0 & unstratified ``zoom-in'' \\ 
{\bf HII-F-U-$\tau$} & 0.9 & 1e-2 & 25 & 10 &  0.7 & U &  1 &  unstratified ``zoom-in'' \\ 
{HII-F-R-A0.5} & -- & -- & -- & -- & -- & -- &  1 & RDHMHD ($A_{0}=0.5$)  \\ 
%
%
%
%
\hline
\end{tabular}
\end{center}\vspace{-0.25cm}
\caption{Initial conditions for all simulations, as \tref{table:sims}. Runs in {\bf bold} are fiducial-resolution, other parameter-survey runs use 8 times lower resolution. Boxes marked ``U'' are unstratified and periodic, so are invariant to any value of $\gravparam$ (these are run to ``zoom in'' to effectively higher-resolution on a small patch, see \S~\ref{sec:ics}). Runs marked ``R'' use full radiation-dust-magnetohydrodynamics (RDMHD; \S~\ref{sec:rad}), with the specified normalization for the albedo $A_{0}$ and absorption efficiency $Q_{\rm ext,\,0}$ set to give total extinction $A_{\rm ext}=1.6$ (for GMC runs, except GMC-Q-R-A0-Lo which uses $0.016$) or $0.5$ (HII-N) or $0.05$ (HII-F) for the incident/driving radiation. Entries marked ``--'' use the same value for the given parameter as the ``default'' run of the same group ({\bf GMC-Q}, {\bf HII-N}, {\bf HII-F}). 
\label{table:sims.all}\vspace{-0.25cm}}
\end{table*}

\section{Stokes and Coulomb Drag}
\label{sec:appendix:drag}

Our simulations can technically interpolate between Epstein drag (collisional drag when the physical size of a dust grain is smaller than the gas mean-free-path and/or the dust is moving super-sonically)+Coulomb drag (electrostatic drag forces) and Stokes drag (viscous drag dominant when the grain is moving sub-sonically and has size much larger than the gas collisional mean-free-path $\lambda_{\rm mfp}^{\rm gas}$) regimes. 

In the Stokes regime (when $\grainsize \gtrsim (9/4)\,\lambda_{\rm mfp}^{\rm gas}$), the drag law is just the Epstein drag $\ts$ (Eq.~\ref{eq:ts}) multiplied by $(4\,\grainsize)/(9\,\lambda_{\rm mfp}^{\rm gas})$, but this is never relevant here, as we expect $\lambda^{\rm gas}_{\rm mfp} \gtrsim 10^{10}\,{\rm cm}$ in e.g.\ HII regions and GMCs, with $\grainsize \lesssim 10^{-4}\,{\rm cm}$. 

The stopping time for Coulomb drag scales as $\ts^{\rm Coulomb} = (\pi\,\gamma/2)^{1/2}\,[(\internaldensity\,\grainsize)/(f_{\rm ion}\,\gasden\,\cs\,\ln{\Lambda})]\,(k_{B}\,T/z_{i}\,e\,U)^{2}\,(1+(|\driftvel|^{3}/\cs^{3})\,\sqrt{2\gamma^{3}/9\pi})$, where $z_{i}\sim1$ is the mean charge of gas ions, $f_{\rm ion}$ is the ionized number fraction of the gas, $\ln{\Lambda}$ is a Coulomb logarithm, and $U\sim Z_{\rm grain}\,e/\grainsize$ is the grain electrostatic potential. As discussed in \paperone\ and \citet{hopkins:2018.mhd.rdi}, inserting the relevant scalings for these terms, the ratio of Coulomb drag force to Epstein drag force is given by $\ts^{\rm Epstein} / \ts^{\rm Coulomb}$ which is just $\sim 10\,f_{\rm ion}$ when the drift is sub-sonic ($|\driftvel| \lesssim \cs$), and is suppressed by a power of $\cs^{4}/|\driftvel|^{4}$ when $|\driftvel| \gtrsim \cs$. Thus Coulomb drag forces scale identically (modulo a normalization constant, given that we assume a homogenous gas ionization state) to Epstein drag in the sub-sonic drift limit, and are negligible in the super-sonic drift limit. For conditions in the neutral ISM, i.e.\ the warm neutral or cold neutral or molecular medium, relevant for e.g.\ our GMC or HII-F (outside the Stromgren radius) simulations, we expect $f_{\rm ion} \ll 1$, so Coulomb drag is always a negligible correction and we can safely ignore it. For an ionized HII region as modeled in e.g. our HII-N simulations, we expect (and assume) $f_{\rm ion}\approx 1$, and the drift we predict is entirely sub-sonic (see \tref{table:sims.all}), so the Coulomb drag terms can be entirely subsumed into the normalization of the Epstein drag scaling (multiplying Eq.~\ref{eq:ts} by a constant) or $\sizeparam$.

\end{appendix}

\end{document}